\documentclass[english,cover]{fitthesis} 

\usepackage[latin2]{inputenc}
\usepackage[T1, IL2]{fontenc}
\usepackage{url}
\DeclareUrlCommand\url{\def\UrlLeft{<}\def\UrlRight{>} \urlstyle{tt}}

\usepackage{amsthm}
\usepackage{tikz}
\usepackage{enumerate}
\usepackage[colorinlistoftodos]{todonotes}
\usepackage{verse}
\usepackage[linesnumbered,ruled,longend]{algorithm2e}

\ifWis
\ifx\pdfoutput\undefined 
\else
  \usepackage{color}
  \usepackage[unicode,hyperindex,plainpages=false]{hyperref}
  \definecolor{links}{rgb}{0.4,0.5,0}
  \definecolor{anchors}{rgb}{1,0,0}

\fi
\fi

\projectinfo{
  project=DP,            
  year=2010,             
  date=26. kv\v{e}tna 2010,    
  title.cs={Efektivní knihovna pro práci s konečnými stromovými automaty},  
  title.en={An Efficient Finite Tree Automata Library}, 
  author={Ondřej Lengál},   
  author.title.p=Bc., 
  department=UITS, 
  supervisor=Tomáš Vojnar, 
  supervisor.title.p={doc.\ Ing.},   
  supervisor.title.a={Ph.D.},    
  keywords.cs={stromové automaty, formální verifikace, abstraktní regulární stromový model checking, binární rozhodovací diagramy, multiterminálové binární rozhodovací diagramy}, 
  keywords.en={tree automata, formal verification, abstract regular tree model checking, binary decision diagrams, multi-terminal binary decision diagrams}, 
    abstract.cs={
      Mnoho současných počítačových systémů používá dynamické datové či řídicí struktury předem neomezené velikosti.
      Tyto datové struktury mají často charakter stromů nebo se dají zakódovat jako stromy s některými dodatečnými ukazateli nad stromovou kostrou.
      Této skutečnosti využívají některé v současné době intenzivně studované techniky formální verifikace, které reprezentují nekonečně mnoho stavů konečným stromovým automatem.
      Nicméně v současnosti neexistuje efektivní a flexibilní implementace knihovny pro stromové automaty, která by byla pro tyto techniky vhodná.
      Cílem této diplomové práce je takovouto knihovnu poskytnout.
      Předložený text nejdříve popisuje základy teorie konečných stromových automatů a regulárních stromových jazyků.
      Dále jsou pro\-zkou\-má\-ny existující implementace knihoven pro stromové automaty a různé verifikační techniky pro systémy se stromovou strukturou.
      Poté se text zaobírá návrhem reprezentace stromového automatu a algoritmů provádějících standardní jazykové operace nad touto reprezentací, načež následuje popis implementace knihovny.
      Prostřednictvím provedených experimentů ukazujeme, že knihovna může konkurovat ostatním dostupným knihovnám pro práci se stromovými automaty, přičemž její výkon v určitých oblastech je řádově vyšší.
      }, 
    abstract.en={
      Numerous computer systems use dynamic control and data structures of unbounded size.
      These data structures have often the character of trees or they can be encoded as trees with some additional pointers.
      This is exploited by some currently intensively studied techniques of formal verification that represent an infinite number of states using a finite tree automaton.
      However, currently there is no tree automata library implementation that would provide an efficient and flexible support for such methods.
      Thus the aim of this Master's Thesis is to provide such a library.
      The present paper first describes the theoretical background of finite tree automata and regular tree languages.
      Then it surveys the current implementations of tree automata libraries and studies various verification techniques, outlining requirements for the library.
      Representation of a finite tree automaton and algorithms that perform standard language operations on this representation are proposed in the next part, which is followed by description of library implementation.
      Through a series of experiments it is shown that the library can compete with other available tree automata libraries, in certain areas being even significantly superior to them.
      }, 
  declaration={
    Prohlašuji, že jsem tuto diplomovou práci vypracoval samostatně pod vedením \linebreak
    doc.~Ing.~Tomáše Vojnara, Ph.D. Uvedl jsem všechny literární prameny a publikace, ze kterých jsem čerpal.
    },
  acknowledgment={Děkuji touto cestou vedoucímu práce, doc.~Ing.~Tomáši Vojnarovi, Ph.\ D., za odborný přínos, ochotu ke konzultacím, motivaci a trpělivost.
    Dále bych chtěl poděkovat Mgr.~Lukáši Holíkovi, Ing.~Jiřímu Šimáčkovi a Mgr.~Adamu Rogalewiczovi, Ph.\ D., za poskytnuté konzultace.
    Mé díky patří i všem, kteří mě podporovali: Honzovi, Liborovi, Marcelce, Markovi, Petrovi, Martinovi a hlavně mé rodině.
    } 
}

\begin{document}
  \maketitle


~
\vfill

\hfill
\begin{minipage}{10cm}
\begin{verse}
I think that I shall never see \\
A poem lovely as a tree.\\

A tree whose hungry mouth is prest \\
Against the sweet earth's flowing breast; \\

A tree that looks at God all day, \\
And lifts her leafy arms to pray; \\

A tree that may in summer wear \\
A nest of robins in her hair; \\

Upon whose bosom snow has lain; \\
Who intimately lives with rain. \\

Poems are made by fools like me, \\
But only God can make a tree. \\
\end{verse}
\hfill--- Joyce Kilmer, \textit{Trees}
\end{minipage}

\vfill

  \tableofcontents
  




\SetKwInput{KwGlobalData}{Global data}

\SetKwFunction{Enqueue}{enqueue}
\SetKwFunction{Dequeue}{dequeue}

\SetKw{Break}{break}
\SetKw{Continue}{continue}

\SetKw{True}{true}
\SetKw{False}{false}


\newcommand{\rankalph}{\mathcal{F}}

\newcommand{\variables}{\mathcal{X}}

\newcommand{\position}[1]{\mathcal{P}os(#1)}

\newcommand{\automaton}{\mathcal{A}}

\newcommand{\transducer}{\tau}

\newcommand{\lang}{\mathcal{L}}

\newcommand{\arity}{\mathit{Arity}}

\newcommand{\natnums}{\mathbb{N}}

\newcommand{\terms}{T(\rankalph, \variables)}

\newcommand{\groundterms}{T(\rankalph)}

\newcommand{\contexts}{\mathcal{C}}

\newcommand{\examplegt}{t = f(f(a, a), g(a))}

\newcommand{\setdiff}{\setminus}

\newcommand{\ifandonlyif}{\Leftrightarrow}

\newcommand{\congruence}{\equiv}

\newcommand{\suchthat}{\;|\;}

\newcommand{\move}[2][]{\ifinner{}\to_{#2}^{#1}\else\xrightarrow[#2]{#1}\fi}

\newcommand{\binaryset}{\{0, 1\}}

\newcommand{\apply}{\mathit{Apply}}

\newcommand{\monadicapply}{\mathit{MonadicApply}}

\newcommand{\createmtbdd}{\mathit{CreateMTBDD}}

\newcommand{\createprojection}{\mathit{CreateProjection}}

\newcommand{\project}{\mathit{Project}}

\newcommand{\trimvariables}{\mathit{TrimVariables}}

\newcommand{\renamevariables}{\mathit{RenameVariables}}

\newcommand{\asgn}{\,:=\,}

\newcommand{\transfunc}{\Delta}

\newcommand{\transfuncmtbdd}{\transfunc^{\bullet}}

\newcommand{\issimulatedby}{\preceq}

\newcommand{\bottom}{\bot}


\newcommand{\horbar}{\noindent \hrulefill}

\newcommand{\seq}[2][n]{{#2}_1, \dots, {#2}_#1}

\newcommand{\complexclass}[1]{\textbf{#1}}

\newcommand{\latin}[1]{\textit{#1}}

\newcommand{\libsfta}{libSFTA}
\newcommand{\Libsfta}{LibSFTA}

\newcommand{\superstate}{super-state}
\newcommand{\macrostate}{macrostate}

\newcommand{\cuddsharedmtbdd}{\texttt{CUDDSharedMTBDD}}
\newcommand{\cuddfacade}{\texttt{CUDDFacade}}
\newcommand{\applyfunctor}{\texttt{AbstractApplyFunctorType}}
\newcommand{\mtbddtransitionfunction}{\texttt{MTBDDTransitionFunction}}
\newcommand{\tabuildingdirector}{\texttt{TABuildingDirector}}
\newcommand{\abstracttabuilder}{\texttt{AbstractTABuilder}}
\newcommand{\timbuktabuilder}{\texttt{TimbukTABuilder}}
\newcommand{\treeautomatonclass}{\texttt{TreeAutomaton}}
\newcommand{\mtbddoperation}{\texttt{MTBDDOperation}}

\newcommand{\RootType}{\texttt{RootType}}
\newcommand{\LeafType}{\texttt{LeafType}}
\newcommand{\VariableAssignmentType}{\texttt{VariableAssignmentType}}
\newcommand{\RootAllocator}{\texttt{RootAllocator}}
\newcommand{\LeafAllocator}{\texttt{LeafAllocator}}

\newcommand{\unsigned}{\texttt{unsigned}}


\newcommand{\desc}[1]{\textnormal{#1}}

\newcommand{\vari}[1]{$\mathit{#1}$} 


\theoremstyle{definition}
\newtheorem{example}{Example}
\theoremstyle{plain}

\newtheorem{theorem}{Theorem}
\newtheorem*{myhillnerodetheorem}{Myhill-Nerode Theorem for Tree Languages}


\chapter{Introduction}

Donald Knuth, the pioneer of the analysis of algorithms, says that computer scientists love \emph{trees} more than anybody else~\cite{knuth97}. Indeed, trees play a crucial role in computer science. They recur in many of its fields, from the representation of programs in the form of abstract syntax trees~\cite{meduna_automatas}, through the use for fast data retrieval in search trees~\cite{knuth_taocp_3}, to tree topologies of computer networks. It is not surprising that trees are often a natural way to represent a model of many types of systems including safety-critical systems

Software errors in safety-critical systems may cause severe losses of money and, in the worst case, even human lives (the Ariane 5 failure is perhaps the best-known case of an expensive software failure~\cite{nuseibeh97}). There are several means which help to avoid software bugs in such systems, one of them being verification based on formal mathematical methods, \emph{formal verification}.

Formal verification of computer systems has gained in popularity in recent years. One of the reasons of this increased interest is the fact that testing of systems, which do not need to be very large, can never cover 100\,\% of cases in an acceptable time (even for systems with finite state spaces; infinite state systems \latin{per se} cannot be completely tested). A popular approach to formal verification is \emph{model checking} (introduced in early 1980s by E.~M.~Clarke, E.~A.~Emerson, J.~P.~Queille and J.~Sifakis), a method based on checking whether a given system conforms to given specification by systematically searching the state space of the system. However, in the real world, there exist systems with state spaces that are infinite, though they often have regular structure, e.g.\ systems with unbounded queues or stacks. As one of the approaches to handle infinite-state systems where states have a linear (or effectively linearizable) structure, \emph{regular model checking} has been proposed~\cite{bouajjani00}.
Regular model checking is based on the following ideas: configurations of the systems being verified are represented as finite words over finite alphabet, transitions are represented as relations over words. Then finite (word) automata over the alphabet can be used to represent sets of configurations of the system and finite (word) transducers can be used to express the transition relation.

However, there are also systems that do not have a linear structure which would enable natural encoding of their configuration into finite words.
A special case of these are systems with tree-like structure, such as parametrised tree networks or heaps.
Moreover, it turns out that many more general graph structures that cannot be easily linearized can be effectively encoded using trees (see e.g.~\cite{vojnar06}).
For such cases, it is convenient to generalise the method to \emph{regular tree model checking}~\cite{abdulla02}, where finite tree automata, a generalisation of finite automata to trees, and finite tree transducers are used.

Nonetheless when used for reachability analysis, regular model checking in general may suffer from problems with infinite number of configurations as the transducers may generate ever new configurations.
Therefore various acceleration techniques that ensure finiteness of the method for many real world problems have been proposed.
These methods may need to perform sophisticated operations upon finite tree automata (transducers).
In order to conduct the operations in verification of non-trivial systems in an acceptable time, smart data structures and algorithms must be used. 
However, currently there is no efficient tree automata library that would be suitable for such operations (although MONA, which is discussed in Section~\ref{sec_mona}, includes a fairly sophisticated deterministic finite tree automata implementation).

The aim of this work is to design an efficient library that would be suitable for sophisticated tree model checking techniques while being flexible enough to be used even for methods which have not yet been developed.
The library focuses on an efficient representation of finite tree automata that work with large alphabets.
Unlike most other tree automata libraries, we use symbolic representation to encode transition functions of tree automata.
An exception in this sense is MONA which also uses symbolic representation.
Moreover, unlike MONA, our library allows to handle \emph{nondeterministic} finite tree automata, which turns out to be crucial for the efficiency of many verification approaches.
We have developed a set of algorithms that conduct standard language operations on symbolically represented nondeterministic finite tree automata, as well as algorithms that perform several non-standard operations, such as reduction according to \emph{downward simulation} or inclusion checking based on \emph{antichains}.
A prototype of the library has been implemented and evaluated through a series of experiments.

The text is divided into several chapters.
Chapter~\ref{chap_theoretical_background} introduces terms, trees, finite tree automata and regular tree languages, while Chapter~\ref{chap_related_work} discusses available libraries that support work with tree automata.
In Chapter~\ref{chap_analysis}, various formal verification techniques using tree automata are studied and requirements for the library are outlined.
Chapter~\ref{chap_design} describes the proposed tree automata representation and algorithms for standard operations that work with this representation.
This is followed by a description of the implementation of the library in Chapter~\ref{chap_implementation}.
Chapter~\ref{chap_evaluation} gives experimental results.
Finally, Chapter~\ref{chap_conclusion} summarizes the work and outlines its possible further development.


\chapter{Theoretical Background} \label{chap_theoretical_background}

This chapter introduces standard definitions, which were taken from \cite{tata2007}.
Theorems are presented without proofs as they can be found in the same source.
First, terms over a ranked alphabet and trees are defined, followed by a description of tree automata and an analysis of closure properties of regular tree languages.
Then the concept of tree automata minimisation is introduced, and decision problems for tree automata languages are discussed.
Finally, a definition of tree transducers concludes the chapter.

\section{Terms and Trees}

This section introduces terms over ranked alphabet and trees.

\subsection{Terms}

A \emph{ranked alphabet} is a couple $(\rankalph, \arity)$ where $\rankalph$ is a finite set of symbols and $\arity$ is a mapping $\arity: \rankalph \to \natnums$ ($\natnums$ denotes the set of non-negative integer numbers). $\arity(f)$, where $f \in \rankalph$, is the \emph{arity} of $f$. The set of symbols of arity $p$ is denoted by $\rankalph_p$. We assume that the set $\rankalph_0$ (the set of constants) is nonempty. Furthermore, we use parenthesis and commas for a short declaration of symbols with arity, such as $f(, )$ for a binary symbol $f$.

Let $\variables$ be a set of constants called \emph{variables} such that $\variables \cap \rankalph_0 = \emptyset$. We define a set of $n$ variables as $\variables_n$. The set $\terms$ of \emph{terms} over the ranked alphabet $\rankalph$ and the set of variables $\variables$ is the smallest set defined by:
\begin{itemize}
  \item  $\rankalph_0 \subseteq \terms$ and
  \item  $\variables \subseteq \terms$ and
  \item  if $p \ge 1$, $f \in \rankalph_p$ and $\seq[p]{t} \in \terms$, then $f(\seq[p]{t}) \in \terms$.
\end{itemize}
$T(\rankalph, \emptyset)$ is abbreviated as $\groundterms$. Terms in $\groundterms$ are called \emph{ground terms}. A term $t \in \terms$ is \emph{linear} if each variable occurs at most once in $t$.

\horbar
\begin{example}
\label{example:ground_term}

Let $\rankalph = \{a, g(), f(, )\}$ be a ranked alphabet. The ground term $\examplegt$ can be represented in a graphical way as:

$$
\begin{tikzpicture}
  [level distance=10mm,
   level 1/.style={sibling distance=15mm},
   level 2/.style={sibling distance=7mm},
   edge from parent path={(\tikzparentnode.south) -- (\tikzchildnode.north)}]
  \node{$f$}
    child{node{$f$}     
      child{node{$a$}}
      child{node{$a$}}
    }
    child{node{$g$}     
      child{node{$a$}}
    }
    ;
\end{tikzpicture}
$$

\end{example}

\horbar

\subsection{Trees}

Let $E$ be a set of labels and $\position{t} \subseteq \natnums^{*}$ be a prefix-closed set. Mapping $f : \position{t} \to E$ is called a finite ordered \emph{tree} $t$. A term $t \in \terms$ can then be viewed as a finite ordered ranked tree with its leaves labeled with variables or constants and its internal nodes labeled with symbols of positive arity, with out-degree equal to the arity of the label. Term $t \in \terms$ can then be defined as a partial function $t : \natnums^{*} \to \rankalph \cup \variables$ (with domain $\position{t}$) that satisfies the following properties:
\begin{enumerate}[{(\it i)}]
  \item  $\position{t}$ is nonempty and prefix-closed,
  \item  $\forall p \in \position{t}$, if $t(p) \in \rankalph_n, n \ge 1$, then $\left\{j \suchthat pj \in \position{t}\right\} = \{1, \dots, n\}$,
  \item  $\forall p \in \position{t}$, if $t(p) \in \variables \cup \rankalph_0$, then $\left\{j \suchthat pj \in \position{t}\right\} = \emptyset$.
\end{enumerate}
In the following we confuse trees and terms.

\subsection{Substitutions}

A \emph{substitution} $\sigma$ is a mapping $\sigma : \variables \to \terms$ (and a \emph{ground substitution} is a mapping $\sigma : \variables \to \groundterms$) where there are only finitely many variables which are not mapped to themselves. The \emph{domain} of a substitution $\sigma$ is the subset of variables $x \in \variables$ such that $\sigma(x) \neq x$. The substitution $\{x_1 \leftarrow t_1, \dots, x_n \leftarrow t_n\}$ is the identity of $\variables \setdiff \{\seq{x}\}$ and maps $x_i \in \variables$ on $t_i \in \terms$, for every index $1 \le i \le n$. The following extends substitutions to $\terms$:
\begin{equation}
  \forall f \in \rankalph_n, \forall \seq{t} \in \terms \qquad \sigma\left(f(\seq{t})\right) = f\left(\sigma(t_1), \dots, \sigma(t_n)\right).
\end{equation}
We confuse a substitution and its extension to $\terms$. Postfix notation is often used for substitutions: $t\sigma$ is the result of applying $\sigma$ to the term $t$.

\subsection{Contexts}

A linear term $C \in T(\rankalph, \variables_n)$ is called a \emph{context} and the expression $C[\seq{t}]$ for $\seq{t} \in \groundterms$ denotes the term in $\groundterms$ obtained from $C$ by replacing variable $x_i$ by $t_i$ for each $1 \le i \le n$, i.e.\ $C[\seq{t}] = C\{x_1 \leftarrow t_1, \dots, x_n \leftarrow t_n\}$. $\contexts^n(\rankalph)$ denotes the set of contexts over $(\seq{x})$.

Contexts with a single variable are denoted as $\contexts(\rankalph)$. A context is trivial if it is reduced to a variable. Given a context $C \in \contexts(\rankalph)$, we denote the trivial context by $C^0$, $C^1$ is equal to $C$ and, for $n > 1$, $C^n = C^{n-1}[C]$ is a context in $\contexts(\rankalph)$.

\section{Regular Tree Languages and Finite Tree Automata}

This section introduces various kinds of finite tree automata.

\subsection{Nondeterministic Finite Tree Automata}

A (bottom-up) \emph{nondeterministic finite tree automaton} (NFTA) over $\rankalph$ is a 4-tuple $\automaton = (Q, \rankalph, Q_f, \Delta)$, where $Q$ is a finite set of states ($Q \cap \rankalph = \emptyset$), $Q_f \subseteq Q$ is a set of final states and $\Delta$ is a set of transition rules 
\begin{equation}
  f(q_1(x_1), \dots, q_n(x_n)) \to q(f(\seq{x})) ,
  \label{eq_transition_function}
\end{equation}
where $n \in \natnums$, $f \in \rankalph_n$, $q, \seq{q} \in Q$ and $\seq{x} \in \variables$. The \emph{move relation} $\move{\automaton}$ is defined by: let $t, t' \in T(\rankalph \cup Q)$,
\begin{equation}
  t \move{\automaton} t' \ifandonlyif
  \left\{
    \begin{array}{l}
    \exists C \in \contexts(\rankalph \cup Q), \exists \seq{u} \in \groundterms, \\
    \exists f(q_1(x_1), \dots, q_n(x_n)) \to q(f(\seq{x})) \in \Delta, \\
    t = C[f(q_1(u_1), \dots, q_n(u_n))], \\
    t' = C[q(f(\seq{u}))]. \\
    \end{array}
  \right.
\end{equation}
$\move[*]{\automaton}$ is the reflexive and transitive closure of $\move{\automaton}$. A ground term $t \in \groundterms$ is \emph{accepted} by an NFTA $\automaton = (Q, \rankalph, Q_f, \Delta)$ if there exists $q \in Q_f$ such that
\begin{equation}
  t \move[*]{\automaton} q(t) .
\end{equation}
The set of all ground terms accepted by NFTA $\automaton$ (the \emph{language} of $\automaton$) is denoted as $\lang(\automaton)$. A set $\lang$ of ground terms is \emph{regular} if there exists such NFTA $\automaton$ that $\lang = \lang(\automaton)$. If two (or more) NFTA accept the same tree language, they are \emph{equivalent}.

\horbar
\begin{example}
Consider the ground term $\examplegt$ from Example~\ref{example:ground_term}. Let $\automaton = (Q, \rankalph, Q_f, \Delta)$ be an NFTA and $t_1 \in T(\rankalph \cup Q)$, $t_1 = f(q_1(f(a, a)), q_2(g(a)))$ be a partially processed term $t$ by $\automaton$, $t \move[*]{\automaton} t_1$. Assume that $f(q_1(x_1), q_2(x_2)) \to q_1(f(x_1, x_2)) \in \Delta$; then the following sequence of transitions is possible:
$$
t =
\parbox{2.5cm}{
\begin{tikzpicture}
  [level distance=10mm,
   level 1/.style={sibling distance=15mm},
   level 2/.style={sibling distance=7mm},
   edge from parent path={(\tikzparentnode.south) -- (\tikzchildnode.north)}]
  \node{$f$}
    child{node{$f$}     
      child{node{$a$}}
      child{node{$a$}}
    }
    child{node{$g$}
      child{node{$a$}}
    }
    ;
\end{tikzpicture}
}
\move[*]{\automaton}
t_1 = 
\parbox{2.5cm}{
\begin{tikzpicture}
  [level distance=10mm,
   level 1/.style={sibling distance=15mm},
   level 2/.style={sibling distance=7mm},
   edge from parent path={(\tikzparentnode.south) -- (\tikzchildnode.north)}]
  \node{$f$}
    child{node{$q_1$}       
      child{node{$f$}
        child{node{$a$}}
        child{node{$a$}}
      }
    }
    child{node{$q_2$}         
      child{node{$g$}
        child{node{$a$}}
      }
    }
    ;
\end{tikzpicture}
}
\move{\automaton}
\parbox{2.3cm}{
\begin{tikzpicture}
  [level distance=10mm,
   level 2/.style={sibling distance=15mm},
   level 3/.style={sibling distance=7mm},
   edge from parent path={(\tikzparentnode.south) -- (\tikzchildnode.north)}]
  \node{$q_1$}
    child{node{$f$}
      child{node{$f$}         
        child{node{$a$}}
        child{node{$a$}}
      }
      child{node{$g$}         
        child{node{$a$}}
      }
    }
    ;
\end{tikzpicture}
}
,
\qquad
\mathrm{i.e.}
\qquad
t \move[*]{\automaton} q_1(t) .
$$
If $q_1 \in Q_f$, then $t \in \lang(\automaton)$. 
\end{example}
\horbar

\noindent An NFTA $\automaton$ is \emph{complete} if there is at least one rule
\begin{equation}
  f(q_1(x_1), \dots, q_n(x_n)) \to q(f(\seq{x})) \in \Delta
\end{equation}
\noindent for all $n \ge 0$, $f \in \rankalph_n$, and $\seq{q} \in Q$. A state $q \in Q$ is \emph{accessible} if there exists a ground term $t$ such that $t \move[*]{\automaton} q(t)$. An NFTA is \emph{reduced} when all of its states are accessible.

The set of transition rules can also be defined as the set of rules of an alternative form: $f(\seq{q}) \to q$. A move relation can be defined as before, except that instead of preserving the structure of the term, the NFTA $\automaton$ replaces subtrees with its states. A term $t$ is then accepted by an NFTA $\automaton$ if
\begin{equation}
  t \move[*]{\automaton} q
\end{equation}
where $q \in Q_f$.

\subsection{Nondeterministic Finite Tree Automata with $\varepsilon$-rules}

The definition of \emph{NFTA with $\varepsilon$-rules} is similar to the definition of NFTA, except for the set of transition rules which may now also contain $\varepsilon$-rules of the form $q \to q'$, i.e.\ the state is changed without processing an input symbol.

\begin{theorem}
If $\lang$ is accepted by an NFTA with $\varepsilon$-rules, then $\lang$ is accepted by an NFTA without $\varepsilon$-rules.
\end{theorem}

\subsection{Deterministic Finite Tree Automata}

A \emph{deterministic finite tree automaton} (DFTA) is an NFTA where there are no two rules with the same left-hand side (and no $\varepsilon$-rules) in $\Delta$. It is \emph{unambiguous}, i.e.\ there is at most one run for every ground term, which means that there is at most one state $q \in Q$ such that $t \move[*]{\automaton} q$.

\begin{theorem}
Let $\lang$ be a regular set of ground terms. Then there exists a DFTA that accepts $\lang$.
\end{theorem}

\section{Closure properties}

\subsection{Union} \label{subsec_closure_union}

\begin{theorem}
The class of regular tree languages is closed under union.
\end{theorem}

\noindent Let us have the following two complete NFTAs (an NFTA can always be made complete by adding missing transitions that all point to a \emph{sink} nonaccepting state): $\automaton_1 = (Q_1, \rankalph, Q_{f1}, \Delta_1)$ and $\automaton_2 = (Q_2, \rankalph, Q_{f2}, \Delta_2)$. Now let us construct NFTA $\automaton = (Q, \rankalph, Q_f, \Delta)$ that accepts $\lang(\automaton) = \lang(\automaton_1) \cup \lang(\automaton_2)$, where $Q = Q_1 \times Q_2$, $Q_f = Q_{f1} \times Q_2 \cup Q_1 \times Q_{f2}$, and $\Delta = \Delta_1 \times \Delta_2$ where
\begin{eqnarray}
  \Delta_1 \times \Delta_2 &=& \{f((q_1, q_1'), \dots, (q_n, q_n')) \to (q, q')\suchthat \nonumber \\
   & & f(\seq{q}) \to q \in \Delta_1, f(\seq{q'}) \to q' \in \Delta_2\} .
\end{eqnarray}
This construction preserves determinism, i.e.\ if $\automaton_1$ and $\automaton_2$ are deterministic, then $\automaton$ is deterministic too.

\subsection{Complementation}

\begin{theorem}
The class of regular tree languages is closed under complementation.
\end{theorem}

\noindent Let $\lang(\automaton)$ be a regular tree language and $\automaton = (Q, \rankalph, Q_f, \Delta)$ be a complete DFTA. Then an NFTA $\automaton' = (Q, \rankalph, Q \setdiff Q_f, \Delta)$ accepts the complement of set $\lang$ in $\groundterms$.

\subsection{Intersection}

\begin{theorem}
The class of regular tree languages is closed under intersection.
\end{theorem}

\noindent Closure under intersection follows directly from closure under union and complementation using De~Morgan's law:
\begin{equation}
  \lang_1 \cap \lang_2 = \overline{\overline{\lang_1} \cup \overline{\lang_2}}
\end{equation}
\noindent where $\overline{\lang}$ denotes the complement of set $\lang$ in $\groundterms$. The construction that preserves determinism follows: Let $\automaton_1 = (Q_1, \rankalph, Q_{f1}, \Delta_1)$ and $\automaton_2 = (Q_2, \rankalph, Q_{f2}, \Delta_2)$ be NFTA. Consider NFTA $\automaton = (Q, \rankalph, Q_f, \Delta)$ such that $Q = Q_1 \times Q_2$, $Q_f = Q_{f1} \times Q_{f2}$ and $\Delta = \Delta_1 \times \Delta_2$. It holds that $\lang(\automaton) = \lang(\automaton_1) \cap \lang(\automaton_2)$.

\section{Minimisation of Tree Automata} \label{sec_myhill_nerode}

An equivalence relation $\congruence$ on $\groundterms$ is a \emph{congruence} on $\groundterms$ if for every $f \in \rankalph_n$
\begin{equation}
  u_i \congruence v_i, 1 \le i \le n \Rightarrow f(\seq{u}) \congruence f(\seq{v}) .
\end{equation}
It is of \emph{finite index} when there are only finitely many $\congruence$-classes. An equivalent definition is that a congruence is an equivalence relation closed under context, i.e.\ for all contexts $C \in \contexts(\rankalph)$, if $u \congruence v$, then $C[u] \congruence C[v]$. Assume $\lang$ is a regular tree language, then $\congruence_\lang$ on $\groundterms$ is defined by: $u \congruence_\lang v$ if for all contexts $C \in \contexts(\rankalph)$,
\begin{equation}
C[u] \in \lang \ifandonlyif C[v] \in \lang .
\end{equation}

\begin{myhillnerodetheorem}
The following three statements are equivalent:

\begin{enumerate}[(i)]
  \item  $\lang$ is a regular tree language, \label{mnt:isregular}
  \item  $\lang$ is the union of some equivalence classes of a congruence of finite index,
  \item  the relation $\congruence_\lang$ is a congruence of finite index. \label{mnt:relationiscongruence}
\end{enumerate}
\end{myhillnerodetheorem}

\noindent An interesting point of the proof of the theorem above is the proof of \textit{(\ref{mnt:relationiscongruence})} $\Rightarrow$ \textit{(\ref{mnt:isregular})}:

\begin{proof}
Let $Q_{min}$ be the finite set of equivalence classes of $\congruence_\lang$. Let us define the transition relation $\Delta_{min}$ as the smallest set such that
\begin{equation}
  f([u_1], \dots, [u_n]) \to [f(\seq{u})] \in \Delta_{min}
\end{equation}
for all $f \in \rankalph$, $\seq{u} \in \groundterms$, where $[u]$ denotes the equivalence class of term $u$. The definition of $\Delta_{min}$ is consistent because $\congruence_\lang$ is a congruence. Let $Q_{min_f} = \{[u] \suchthat u \in \lang\}$. The DFTA $\automaton_{min} = (Q_{min}, \rankalph, Q_{min_f}, \Delta_{min})$ accepts the tree language $\lang$.
\end{proof}

It can be proved that $\automaton_{min}$ is minimum (in the number of states) and unique up to a renaming of states.

\section{Top-down Tree Automata}

A nondeterministic \emph{top-down} finite tree automaton (top-down NFTA) over $\rankalph$ is a 4-tuple $\automaton = (Q, \rankalph, I, \Delta)$, where $Q$ is a finite set of states, $I \subseteq Q$ is a set of initial states and $\Delta$ is a set of transition rules
\begin{equation}
  q(f(\seq{x})) \to f(q_1(x_1), \dots, q_n(x_n)),
\end{equation}
where $n \ge 0$, $f \in \rankalph_n$, $q, \seq{q} \in Q$, $\seq{x} \in \variables$. The move relation is easily deduced from the move relation for bottom-up NFTA.

The tree language $\lang(\automaton)$ accepted by $\automaton$ is the set of ground terms $t$ for which there is an initial state $q \in I$ such that
\begin{equation}
  q(t) \move[*]{\automaton} t .
\end{equation}
Note that the expressive power of bottom-up and top-down nondeterministic finite tree automata is the same.
However, top-down DFTA are strictly less powerful than top-down NFTA.

\section{Decision Problems and their Complexity}

This section summarises some decision problems of regular tree languages and their complexity in the context of RAM machines.
Note the increased complexity with respect to regular word languages, which implies an even stronger need for a very careful design and various heuristic optimizations of working with finite tree automata.

\begin{itemize}
  \item  The \emph{fixed membership} problem (determining whether a certain ground term is accepted by a fixed finite tree automaton, i.e.\ the automaton \emph{is not} the input of the decision procedure) is \complexclass{ALOGTIME}-complete.
  \item  The \emph{uniform membership} problem (determining whether a certain ground term is accepted by a given finite tree automaton, i.e.\ the automaton \emph{is} also the input of the decision procedure) can be decided in linear time for DFTA and in polynomial time for NFTA.
  \item  The \emph{emptiness} problem (determining whether the language accepted by given finite tree automaton is empty) is decidable in linear time.
  \item  The \emph{intersection non-emptiness} problem (determining whether there is at least one ground term accepted by each finite tree automaton from a given finite sequence of tree automata) is \complexclass{EXPTIME}-complete.
  \item  The \emph{finiteness} problem (determining if the language of a given finite tree automaton is finite) is decidable in polynomial time.
  \item  The \emph{complement emptiness} problem (determining whether a given finite tree automaton accepts every ground term) can be decided in polynomial time for DFTA and it is \complexclass{EXPTIME}-complete for NFTA.
  \item  The \emph{equivalence} problem (determining whether two given finite tree automata accept the same language) is decidable.
  \item  The \emph{singleton set} problem (determining whether a given finite tree automaton accepts only a single ground term) is decidable in polynomial time.
\end{itemize}

\section{Tree Transducers}

\subsection{Bottom-up Tree Transducers}

A \emph{nondeterministic bottom-up tree transducer} (NBUTT) is a 5-tuple $U = (Q, \rankalph, \rankalph', Q_f, \Delta)$, where Q is a set of states ($Q \cap \rankalph = \emptyset$, $Q \cap \rankalph' = \emptyset$), $\rankalph$ and $\rankalph'$ are finite nonempty sets of input symbols and output symbols, $Q_f \subseteq Q$ is a set of final states and $\Delta$ is a set of transduction rules of the following two types:
\begin{equation}
  f(q_1(x_1), \dots, q_n(x_n)) \to q(u) ,
\end{equation}
where $f \in \rankalph_n$, $u \in T(\rankalph', \variables_n)$, $q, \seq{q} \in Q$, and $\seq{x} \in \variables_n$, and
\begin{equation}
  q(x_1) \to q'(u)\qquad\mbox{($\varepsilon$-rule)},
\end{equation}
where $u \in T(\rankalph', \variables_1)$, $q, q' \in Q$, and $x_1 \in \variables_1$.

Let $t, t' \in T(\rankalph \cup \rankalph' \cup Q)$. The move relation $\move{U}$ is defined as:
\begin{equation}
  t \move{U} t' \ifandonlyif
  \left\{
  \begin{array}{l}
    \exists f(q_1(x_1), \dots, q_n(x_n)) \to q(u) \in \Delta, \\
    \exists C \in \contexts(\rankalph \cup \rankalph' \cup Q), \\
    \exists \seq{u} \in T(\rankalph'), \\
    t = C[f(q_1(u_1), \dots, q_n(u_n))], \\
    t' = C[q(u\{x_1 \leftarrow u_1, \dots, x_n \leftarrow u_n\})]. \\
  \end{array}
  \right.
\end{equation}
The reflexive and transitive closure of $\move{U}$ is $\move[*]{U}$. The relation induced by $U$ (also denoted as $U$) is:
\begin{equation}
  U = \left\{(t, t') \suchthat t \move[*]{U} q(t'), t \in \groundterms, t' \in T(\rankalph'), q \in Q_f \right\} .
\end{equation}

A transducer is \emph{$\varepsilon$-free} if there is no $\varepsilon$-rule in $\Delta$. If all transduction rules are linear (no variable occurs twice in the right-hand side), then the transducer is \emph{linear}. It is \emph{non-erasing} if, for each rule, at least one symbol from $\rankalph'$ occurs in the right-hand side. In a \emph{complete} (or \emph{non-deleting}) transducer, for every rule $f(q_1(x_1), \dots, q_n(x_n)) \to q(u)$, for every $x_i, (1 \le i \le n)$, $x_i$ occurs at least once in $u$. An $\varepsilon$-free transducer where there are no two rules with the same left-hand side is called \emph{deterministic} (DBUTT).

\subsection{Top-down Tree Transducers}

A \emph{nondeterministic top-down tree transducer} (NTDTT) is a 5-tuple $D = (Q, \rankalph, \rankalph', Q_i, \Delta)$, where $Q$ is a set of states ($Q \cap \rankalph = \emptyset$, $Q \cap \rankalph' = \emptyset$), $\rankalph$ and $\rankalph'$ are finite nonempty sets of input and output symbols, $Q_i \subseteq Q$ is a set of initial states and $\Delta$ is a set of transduction rules of the following two types:
\begin{equation}
  q(f(\seq{x})) \to u[q_1(x_{i_1}), \dots, q_p(x_{i_p})] ,
\end{equation}
where $f \in \rankalph_n$, $u \in \contexts^p(\rankalph')$, $q, \seq[p]{q} \in Q$, $x_{i_1}, \dots, x_{i_p} \in \variables_n$, and
\begin{equation}
  q(x) \to u[q_1(x), \dots, q_p(x)]\qquad\mbox{($\varepsilon$-rule)},
\end{equation}
where $u \in \contexts^p(\rankalph')$, $q, \seq[p]{q} \in Q$, $x \in \variables$.

Let $t, t' \in T(\rankalph \cup \rankalph' \cup Q)$. The move relation $\move{D}$ is defined as:
\begin{equation}
  t \move{D} t' \ifandonlyif
  \left\{
  \begin{array}{l}
    \exists q(f(\seq{x})) \to u[q_1(x_{i_1}), \dots, q_p(x_{i_p})] \in \Delta, \\
    \exists C \in \contexts(\rankalph \cup \rankalph' \cup Q), \\
    \exists \seq{u} \in \groundterms, \\
    t = C[q(f(\seq{u}))], \\
    t' = C[u[q_1(v_1), \dots, q_p(v_p)]]\qquad\mbox{where $v_j = u_k$ if $x_{i_j} = x_k$}. \\
  \end{array}
  \right.
\end{equation}
The reflexive and transitive closure of $\move{D}$ is $\move[*]{D}$. The relation induced by $D$ (also denoted as $D$) is:
\begin{equation}
  D = \left\{(t, t') \suchthat q(t) \move[*]{D} t', t \in \groundterms, t' \in T(\rankalph'), q \in Q_i\right\} .
\end{equation}
$\varepsilon$-free, linear, non-erasing, complete, deterministic (DTDTT) top-down tree transducers are the same as in the bottom-up case.


\chapter{Existing Tree Automata Libraries} \label{chap_related_work}

This chapter describes several implementations of finite tree automata libraries focusing on a couple of the most interesting from our point of view: Timbuk and MONA.

\section{Timbuk} \label{sec_timbuk}

Timbuk~\cite{timbuk} is a collection of tools for achieving proofs of reachability over \emph{term rewriting systems} and for manipulating tree automata. This system is written in OCaml, a popular functional programming language. Version 2.2 of Timbuk was surveyed; although newer version 3.0 is currently available, this version has abandoned the tree automata library present in earlier versions as the tool now focuses on reachability analysis and equational approximations of term rewriting systems. This library is a free software (available under the GNU LGPLv2~\cite{lgpl2} licence) distributed for free, therefore it was possible to study the implementation.

The tree automaton is implemented as a tuple of lists: a list of symbols (an alphabet), a list of state operators, a list of states, a list of final states, a list of transitions and a list of prioritary transitions. The supported operations on tree automata are the standard ones: intersection, union, language emptiness, deletion of inaccessible states, determinisation and others. Since states and transitions are represented as lists, the aforementioned operations are implemented in a straightforward way. The library is able to construct a tree automaton directly from a given term rewriting system.

\section{MONA} \label{sec_mona}

MONA~\cite{mona} is a tool (released free of charge under the GNU GPLv2 licence~\cite{gpl2}) that implements decision procedures for the \emph{weak second-order theory of one or two successors} (WS1S/WS2S). These types of logic are notable for the following reasons:
\begin{description}
  \item[WS1S]  B\"uchi claims in~\cite{buchi59} that WS1S has an expressive power equivalent to regular expressions, i.e.\ it can be used to denote the class of regular languages.
  \item[WS2S]  According to~\cite{treinen00} (who further refers to Thatcher and Wright~\cite{thatcher68}), WS2S is equivalent to the class of regular tree languages.
\end{description}
Indeed, MONA uses finite automata and finite tree automata for determining the truth status of formulae in WS1S and WS2S, respectively. Independently of MONA, Glenn and Gasarch~\cite{glenn97} also implemented an automaton-based decision procedure for WS1S.

MONA was actively developed for six years, but since 2002 no further progress of the tool has appeared and only bugfixes have been applied.
However, Klarlund \latin{et al} \cite{monasecrets}~boldly claim that the developers of MONA tried many approaches to deal with common problems and tuned the tool to give the best performance.
The most important feature mentioned is symbolic representation of transition functions of automata by \emph{multi-terminal binary decision diagrams} (MTBDD), which are a generalisation of \emph{reduced ordered binary decision diagrams} (ROBDD, often abbreviated just as BDD, see~\cite{knuth_bdds} for further details).
The idea of generalising BDDs to MTBDDs is by assigning multiple values to the sink nodes of the diagram (i.e.\ generalising function $f$ represented by BDD, $f: \binaryset^n \to \binaryset$, to function $g$ represented by MTBDD, $g: \binaryset^n \to \mathbb{D}$, where $\mathbb{D}$ is an arbitrary domain such that it contains \emph{bottom} element $\bottom \in \mathbb{D}$).

Due to the fact that BDDs are only a compact representation of formulae in propositional logic with Boolean variables $\seq{x}$, Boolean formulae can be used for their description.
A BDD $f: \binaryset^n \to \binaryset$ maps to the Boolean formula
\begin{equation}
  \sum_{(\seq{a}) \in \binaryset^n} \left(\prod_{a_i = 0} \lnot x_i \cdot \prod_{a_i = 1} x_i \cdot f(\seq{a}) \right) .
  \label{equation_bdd_formula}
\end{equation}
The mapping for MTBDDs is analogous, however a few preconditions need to be imposed on domain $\mathbb{D}$:
\begin{enumerate}[{(\it i)}]
  \item  the product of $x \in \binaryset$ and $d \in \mathbb{D}$ is defined as
          \begin{equation}
            x \cdot d \stackrel{\mathrm{def}}{=} \left\{
              \begin{array}{r@{\quad\mathrm{if}\quad}l}
                \bottom & x = 0 \\
                d       & x = 1 \\
              \end{array}
            \right. ,
          \end{equation}

  \item  the addition operation on $\mathbb{D}$ needs to ensure that for $d \in \mathbb{D}$ it holds that
          \begin{equation}
            d + \bottom = \bottom + d = d ~.
          \end{equation}

\end{enumerate}
Then we define the mapping from MTBDD $g: \binaryset^n \to \mathbb{D}$ to the Boolean formula
\begin{equation}
  \sum_{(\seq{a}) \in \binaryset^n} \left(\prod_{a_i = 0} \lnot x_i \cdot \prod_{a_i = 1} x_i \cdot g(\seq{a}) \right) .
\end{equation}

\begin{figure}[ht]
  \centering
  \begin{minipage}{7cm}
    \centering
    \includegraphics[height=6cm]{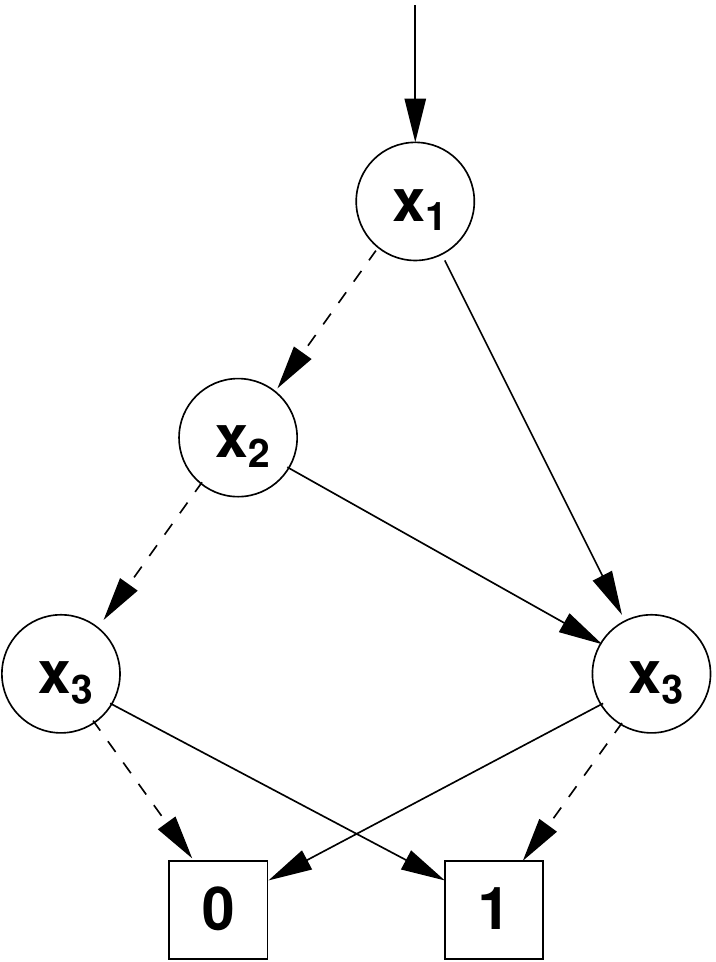}\\
    a) BDD
  \end{minipage}
  ~~~~~
  \begin{minipage}{7cm}
    \centering
    \includegraphics[height=6cm]{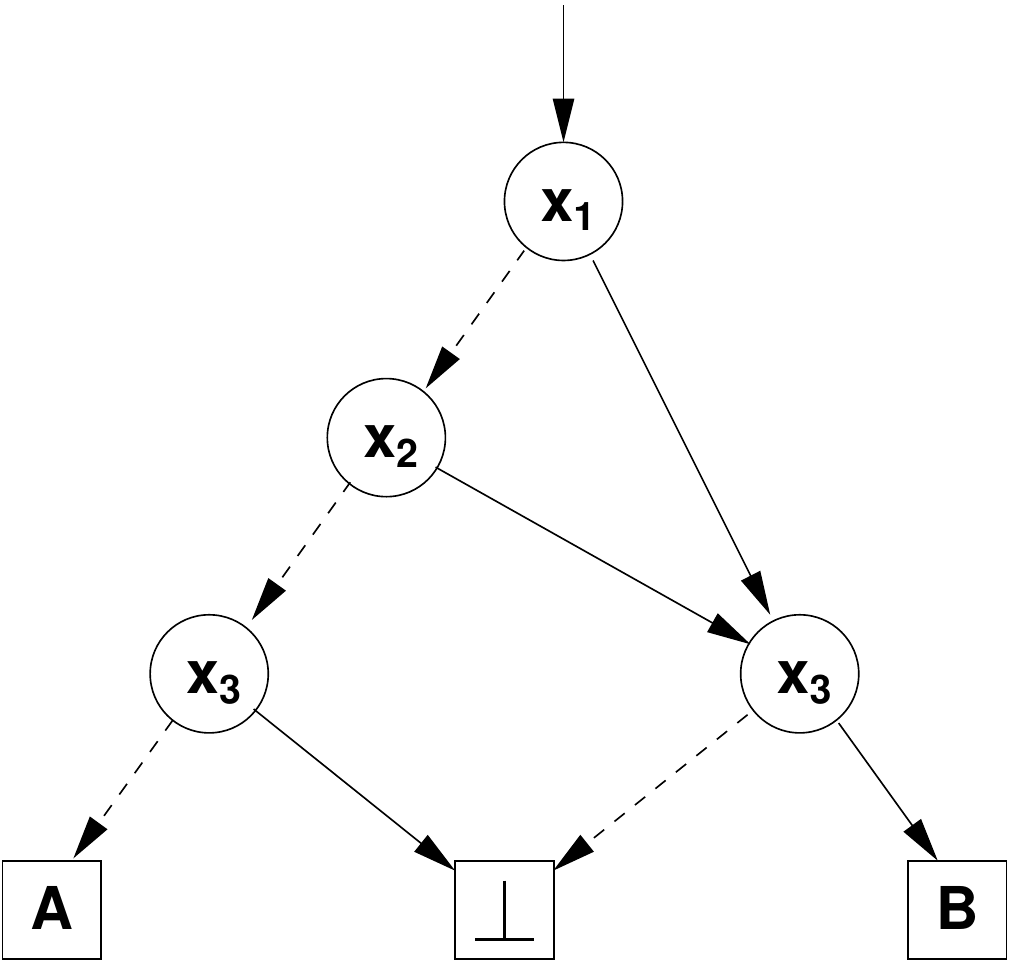}\\
    b) MTBDD
  \end{minipage}
  \caption{Examples of the structure of BDD and MTBDD.}
  \label{fig_bdd_examples}
\end{figure}

\horbar
\begin{example}

This example shows in Figure~\ref{fig_bdd_examples} the structure of the following decision diagrams:

\begin{enumerate}[]
\item
\begin{enumerate}[]
\item
\begin{enumerate}[a)]
  \item  a BDD representing formula:
          \begin{equation}
            x_1 \lnot x_3 + \lnot x_1 x_2 \lnot x_3 + \lnot x_1 \lnot x_2 x_3 ~,
          \end{equation}

  \item  an MTBDD representing formula:
          \begin{equation}
            \lnot x_1 \lnot x_2 \lnot x_3 A + x_1 x_3 B + \lnot x_1 x_2 x_3 B ~.
          \end{equation}
\end{enumerate}
\end{enumerate}
\end{enumerate}

\noindent  Note that, e.g., the expression $x_1 x_3$ represents the expression $x_1 (x_2 + \lnot x_2) x_3$ which fully expands to $x_1 x_2 x_3 + x_1 \lnot x_2 x_3$.
\end{example}
\horbar

\pagebreak

When MTBDDs are used for transition table representation, every symbol of the input alphabet $\Sigma$ is assigned a binary string, i.e.\ there exists an encoding function $\mathit{enc}: \Sigma~\to~\binaryset^n$, where $n = \left\lceil \lg \left|\Sigma\right| \right\rceil$. The values of sink nodes (set $\mathbb{D}$ from previously mentioned function $g$) is the state set of the automaton. Such an MTBDD can be used to describe the transition relation of the automaton for a single state: the input symbol is encoded by function $\mathit{enc}$ into a sequence of binary digits $(\seq{x})$, where $\seq{x}$ correspond to the Boolean variables of the MTBDD. The assignment to the variables denotes the path that is to be taken in the diagram and determines the sink node (i.e.\ the next state of the automaton). Such an MTBDD may either exist for every state of the automaton, or preferably \emph{shared} MTBDD is used. This is another generalisation which merges all diagrams into a single one with multiple root nodes (each corresponding to a different state) and changes the tree-like structure of an MTBDD into a \emph{directed acyclic graph} (DAG). This solution yields a compact representation of the transition function even for large input alphabets.

Another concept introduced by MONA developers is \emph{guided tree automaton}~\cite{biehl96}, which is supposed to tackle state space blow-up. Bottom-up tree automata often suffer from the problems of the way they work: while the automaton traverses the tree from its leaves to the root, it does not have any information about the position in the tree. The guided tree automaton provides a \emph{guide}, an additional top-down tree automaton that labels tree nodes by assigning state spaces to them, making them aware of their position in the tree. This assignment is done before the actual automaton starts working. When it does, it operates faster, since every state space has its own state set and transition table. The guide needs to be either provided by the programmer, or it can be synthesized automatically for certain domains (e.g.\ WSRT logic used for description of recursive data types, which is implemented in MONA).

Another noteworthy optimization applied in MONA is so-called \emph{eager minimisation}: whenever the structure of an automaton is modified, a Myhill-Nerode minimisation is performed. Although originally not expected, this strategy yields very good results. Despite the fact that formulae are often represented in the form of trees (at least during syntactic analysis of the formula), MONA uses DAGs for their representation. Common subexpressions are identified and collapsed, thus saving both space and time.

Although MONA only supports work with deterministic finite (tree) automata, there are formal verification techniques (such as~\cite{bouajjani08}) that can efficiently work directly with nondeterministic finite (tree) automata, thus avoiding possible time and space exponential blow-up caused by automata determinisation. After the construction of a finite tree automaton, MONA tries to find both a satisfying example and a counterexample. Therefore there is no efficient support for sophisticated manipulation with automata which may be required by some verification methods.

\section{Other Libraries}

Java library Lethal~\cite{lethal} supports numerous operations on tree automata, like checking whether some properties (determinism, completeness, \dots) hold for a given automaton, or standard operations on languages (such as union, intersection, complement or difference). The implementation appears to be quite na\"\i{}ve, with a primary focus on education. However, as the only studied library, Lethal also implements tree transducers and hedge automata (a modification of tree automata for unranked trees).

Binary Tree Automata Library~\cite{grappa} is a Caml library for tree automata. The implementation provides only basic functions and is close to Timbuk (see section~\ref{sec_timbuk}), although it uses hash tables for a transition table representation and language-provided sets for state sets.

A simple implementation of a tree automata library in ELAN can be found in~\cite{elan}. Also this library provides only basic functionality.


\chapter{Analysis} \label{chap_analysis}

This chapter starts with an introduction to abstract regular tree model checking and an analysis of potential use of the library and collects requirements for the library.

\section{Abstract Regular Tree Model Checking}

The basic idea of \emph{regular tree model checking} is to decide the emptiness of the language
\begin{equation}
  \tau^{*}\left(\lang(\mathit{Init})\right) \cap \lang(\mathit{Bad}) ,
\end{equation}
where $\mathit{Init}$ is a tree automaton denoting the set of initial states of the system, $\mathit{Bad}$ is a tree automaton expressing the set of states violating the safety properties of the system, and $\tau$ is a linear tree transducer representing the transition relation of the system. Because an iterative computation of $\tau^{*}\left(\lang(\mathrm{Init})\right)$ may not terminate, several acceleration methods have been proposed. One of them is \emph{abstract regular tree model checking}~\cite{bouajjani06}, which is an acceleration technique based on the \emph{abstract-check-refine} paradigm. \emph{Abstraction} $\alpha$ is a function from the set of all tree automata $\mathcal{M}_{\rankalph}$ over ranked alphabet $\rankalph$ to its subset $\mathcal{A}_{\rankalph}$, $\mathcal{A}_{\rankalph} \subseteq \mathcal{M}_{\rankalph}$, such that $\forall M \in \mathcal{M}_{\rankalph} : \lang(M) \subseteq \lang(\alpha(M))$.

\subsection{Abstraction Based on Languages of Finite Height}

\emph{Abstraction based on languages of finite height}, which was introduced in~\cite{bouajjani06}, defines two states of a tree automaton as equivalent if their languages up to a given height $n$ are identical. The implementation can be done similar to the Myhill-Nerode minimisation, except that the procedure stops after $n$ iterations.

\subsection{Abstraction Based on Predicate Languages}

Given a set of \emph{predicate} tree automata $\mathcal{P} = (\seq{P})$, \emph{abstraction based on predicate languages} (introduced also in~\cite{bouajjani06}) defines two states of a tree automaton equivalent if their languages have a nonempty intersection with exactly the same subset of languages represented by tree automata from $\mathcal{P}$. This can be done by labeling every state with predicates that have a nonempty intersection with the language of the automaton and collapsing states with identical labeling.

\section{Requirements}

An important requirement for the library is to enable a direct work with nondeterministic tree automata without determinising the automaton first.
This is convenient for avoiding state explosion connected with automaton determinisation in some verification techniques (see~\cite{bouajjani08}).
The following standard operations are necessary to be implemented in the library:
\begin{itemize}
  \item  creating a finite tree automaton denoting the union of languages of given finite tree automata,
  \item  creating a finite tree automaton denoting the intersection of languages of given finite tree automata,
  \item  creating a finite tree automaton denoting the complement of the language of a given finite tree automaton,
  \item  determinisation of a finite tree automaton,
  \item  minimisation of a finite tree automaton,
  \item  determining emptiness of language of a finite tree automaton,
  \item  reducing the size of a given nondeterministic finite tree automaton without determinisation, and
  \item  determining inclusion of languages of given finite tree automata while avoiding determinisation of any automaton.
\end{itemize}
%


The library also needs to implement tree transducers at least in their \emph{structure-preserving} form.
Certain techniques~\cite{bouajjani06, abdulla10, abdulla07, abdulla08, holik09} need to efficiently traverse all states of the automaton in order to, for instance, compute the abstraction of the automaton. Support for this is also necessary.


\chapter{Design} \label{chap_design}

This chapter starts with a description of the representation that we propose for the transition function of nondeterministic finite tree automata.
This is followed by a description of algorithms for operations on finite tree automata that use this representation.
Relabelling tree transducers and operations with them are described at the end of the chapter.

\section{Representation of a Finite Tree Automaton} \label{sec_representation}

The performance of operations on a nondeterministic finite tree automaton $\automaton = (Q, \rankalph, Q_f, \transfunc)$ is mostly affected by the choice of the data structure for representing the transition function $\transfunc$. Two major approaches are possible:

\begin{description}
  \item[Explicit representation]  This approach represents the transition function of a tree automaton by enumerating all transitions in a data structure used for a representation of the set.

  \item[Symbolic representation]  This method is a popular approach in model checking that is based on a representation of the transition function using Boolean formulae. The exact form of the representation varies depending on the application, however the most popular data structure used for representing Boolean formulae is the BDD.
\end{description}

The analysis in Chapter~\ref{chap_related_work} showed that \emph{symbolic representation} using MTBDDs is a very promising approach. Therefore we chose MTBDDs for representation of transition function. To recap, MTBDD is a data structure that stores mapping $g: \binaryset^n \to \mathbb{D}$, where $\mathbb{D}$ is an arbitrary set. 

Our design attempts to tackle the problem of large alphabets by using a shared MTBDD such that the domain of the MTBDD, i.e.\ the sequence of Boolean variables $\binaryset^n$, represents binary encodings of symbols from $\rankalph$ according to some encoding function $\mathit{enc}: \rankalph \to \binaryset^n$, where $n \le \left\lceil \lg \left|\rankalph\right| \right\rceil$ (note that $n$ may be smaller than $\left\lceil \lg \left|\rankalph\right| \right\rceil$ because when arity of symbols is implicit, more symbols with different arity may map to one assignment of Boolean variables; in conflicting cases, we will denote symbol $f \in \rankalph_p$ as $f_p$).
Using function $\mathit{enc}$ to encode symbols from $\rankalph$, MTBDD may represent function $g: \rankalph \to \mathbb{D}$. 
Before we proceed, let us first define the set of \emph{\superstate{}s} $S(\transfunc)$ of the transition function $\transfunc$ as 
\begin{eqnarray}
  S(\transfunc) &=& \left\{(\seq[p]{q})\suchthat p \ge 0, \right. \nonumber \\
            & & \left. f(\seq[p]{q}) \to D \in \transfunc, f \in \rankalph_p, D \subseteq Q, D \neq \emptyset\right\}
\end{eqnarray}
or in case of a complete automaton with a sink state $q_{\mathit{sink}}$:
\begin{eqnarray}
  S(\transfunc) &=& \left\{(\seq[p]{q})\suchthat p \ge 0, \right. \nonumber \\
            & & \left. f(\seq[p]{q}) \to D \in \transfunc, f \in \rankalph_p, D \subseteq Q, D \neq \left\{q_{\mathit{sink}}\right\}\right\} .
\end{eqnarray}
Let $S_n(\transfunc)$ be the set of \superstate{}s of $\transfunc$ of arity $n$.
Note that an empty sequence $()$ represents \emph{initial \superstate{}}, i.e.\ the \superstate{} from which transitions over leaf nodes are possible.
We also extend the definition of membership relation $\in$ to \superstate{}s in the following way:
\begin{equation}
  q \in (\seq{q}) \stackrel{\mathrm{def}}{\ifandonlyif} \exists 1 \le i \le n : q = q_i
\end{equation}

Using the previous definition of \superstate{}s and definition of $\transfunc$ (see Equation~\ref{eq_transition_function}), we may alternatively define the transition function of an automaton $\automaton$ as a mapping $\transfuncmtbdd$ in the following way:
\begin{eqnarray}
  \transfuncmtbdd &:& S \to (\rankalph \to 2^Q) \nonumber \\
  & & (\seq[p]{q}) \mapsto \left\{(f, D) \suchthat f(\seq[p]{q}) \to D \in \transfunc \right\}
\end{eqnarray}
This means that we may represent the transition function $\transfunc$ of a tree automaton $\automaton$ as a data structure that associates each \superstate{} with an MTBDD that is indexed using a binary encoding of symbols from $\rankalph$ and has subsets of $Q$ in its sink nodes.
When shared MTBDD is used, each \superstate{} is mapped to a root of a given MTBDD.
In case $\transfuncmtbdd$ is not \emph{total}, we make it total by assigning MTBDD where all symbols map to a sink state $\{q_{\mathit{sink}}\}$ to each \superstate{} that has no image in $\transfuncmtbdd$, which yields a complete automaton.
We further confuse $\transfunc$ and $\transfuncmtbdd$.

\horbar
\begin{example}

Consider the nondeterministic finite tree automaton $\automaton = (Q, \rankalph, Q_f, \transfunc)$, $Q = \{q_1, q_2, q_3\}$, $\rankalph = \{a, b_0, b_2, c_0, c_1, d_1\}$, and $\transfunc = \{b_0 \to \{q_1, q_2\}, c_0 \to \{q_2\}, d_1(q_2) \to \{q_3\},\linebreak
b_2(q_1, q_3) \to \{q_1, q_2\}, c_1(q_3) \to \{q_1, q_2\} \}$ ($Q_f$ is not important at this point). A shared MTBDD corresponding to $\transfunc$ is in Figure~\ref{fig_smtbdd}.
\begin{figure}[h]
  \vspace{-0.6cm}
  \begin{center}
    \includegraphics[width=10cm]{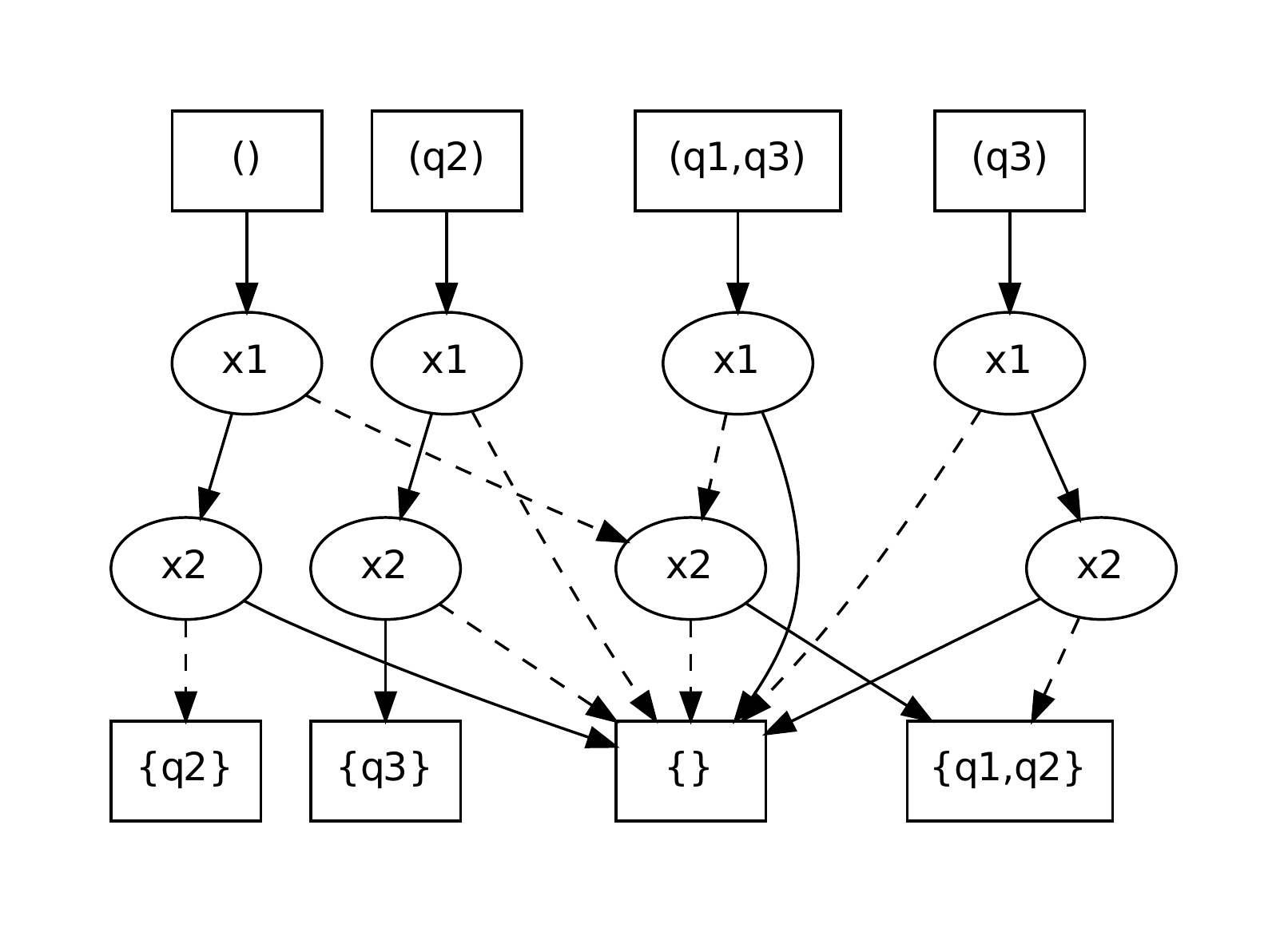}
  \end{center}
  \vspace{-1.0cm}
  \caption{A representation of $\transfunc$ by a shared MTBDD. Encoding of symbols from $\rankalph$: $a$: \texttt{00}, $b$: \texttt{01}, $c$: \texttt{10}, $d$: \texttt{11}. Dashed lines represent \texttt{0} value of given variable, solid lines represent value \texttt{1}.}
  \label{fig_smtbdd}
  \vspace{-0.6cm}
\end{figure}
\end{example}
\horbar

\pagebreak

\section{Operations on MTBDDs} \label{sec_operations}

This section describes algorithms that perform operations on finite tree automata with a transition function represented by an MTBDD.
These algorithms manipulate MTBDDs using the following two sufficient functions:

\begin{description}

  \item[Apply] The standard $\apply$ function that performs a given binary operation \texttt{op} on all respective sink nodes (this means sink nodes accessible over the same symbol) of two input MTBDDs (\texttt{lhs} and \texttt{rhs} for left-hand side MTBDD and right-hand side MTBDD respectively) and returns the resulting MTBDD.
    \begin{eqnarray}
      \apply &:& (\rankalph \to 2^Q) \to (\rankalph \to 2^Q) \to (2^Q \to 2^Q \to 2^Q) \to (\rankalph \to 2^Q) \nonumber \\
      & & \apply~\mathtt{lhs~rhs~op} = \lambda x~.~\mathtt{op}~(\mathtt{lhs}~x)~(\mathtt{rhs}~x)
    \end{eqnarray}

  \item[MonadicApply]  The monadic version of $\apply$ function that performs a given unary operation \texttt{op} on all sink nodes of input MTBDD \texttt{tf} and returns the resulting MTBDD.
    \begin{eqnarray}
      \monadicapply &:& (\rankalph \to 2^Q) \to (2^Q \to 2^Q) \to (\rankalph \to 2^Q) \nonumber \\
      & & \monadicapply~\mathtt{tf~op} = \lambda x~.~\mathtt{op}~(\mathtt{tf}~x)
    \end{eqnarray}

\end{description}

Note that $\lambda$-calculus is used for definitions and applications of functions that work with MTBDDs in order to make them more comprehensible.
In case the result of the $\apply$ operation is not stored (the operation is performed solely for the side effect of \texttt{op}), \texttt{op} does not need to return a value.
Further, we assume that transition functions for all automata are stored in a single shared MTBDD.

\subsection{Insertion of a Transition} \label{subsec_insert}

\emph{Inserting a transition} into an MTBDD-represented transition function of finite tree automaton $\automaton = (Q, \rankalph, Q_f, \transfunc)$ is done by creating a new MTBDD with only given transition and merging it with the original MTBDD representing $\transfunc$ by substituting the sink node at position given by the symbol of the transition with the new value as described in Algorithm~\ref{alg_insertion}.
In order to create a new MTBDD with a given transition, an additional function is necessary:

\begin{description}
  \item[CreateMTBDD]  This function creates an MTBDD which maps a single symbol to a single set of states.
    \begin{eqnarray}
      \mathit{CreateMTBDD} &:& \rankalph \to 2^Q \to (\rankalph \to 2^Q) \nonumber \\
                           & & \mathit{CreateMTBDD}~\mathtt{k~D} = \lambda x~.~\mathbf{if}~x = \mathtt{k}~\mathbf{then}~\mathtt{D}~\mathbf{else}~\{q_{sink}\}~~~~~~~~
    \end{eqnarray}
\end{description}

\enlargethispage{5mm}

\IncMargin{1em}
\begin{algorithm}[H]

  \SetKwData{Tmp}{tmp}
  \SetKwData{Superstate}{sp}

  \caption{Transition insertion}
  \KwIn{Transition function $\transfunc_{IN}$ \linebreak
        Transition $f(\seq{q}) \to D$ to be inserted}
  \KwOut{$\transfunc_{OUT} = \left(\transfunc_{IN} \setdiff \{ f(\seq{q}) \to E \suchthat E \subseteq Q \} \right) \cup \{f(\seq{q}) \to D\}$}

  \Begin
  {
    $\Tmp \asgn \createmtbdd~f~D$\;
    $\Superstate \asgn (\seq{q})$\;
    $\transfunc_{OUT} \asgn \transfunc_{IN}$\;
    $\transfunc_{OUT}~\Superstate \asgn \apply~\left(\transfunc_{IN}~\Superstate\right)~\Tmp~(\lambda X~Y~.~$ \lIf{$Y = \{q_{sink}\}$}{$X$} \lElse{$Y$}$)$\;

    \Return{$\transfunc_{OUT}$}\;
  }

  \label{alg_insertion}
\end{algorithm}
\DecMargin{1em}

\subsection{Retrieval of a Transition} \label{subsec_retrieval}

The algorithm that \emph{retrieves a transition} (i.e.\ for a given \superstate{} $(\seq{q})$ and a symbol $f$ returns $D$ such that $f(\seq{q}) \to D \in \transfunc$) from an MTBDD-represented transition function $\transfunc$ first creates a \emph{projection BDD} and makes a projection of the MTBDD representing the transition function for given \superstate{} according to given symbol of the input alphabet. A projection BDD $p : \rankalph \to \binaryset$ is a BDD over the same set of Boolean variables as the transition function MTBDD which identifies the nodes that are to be excluded from the MTBDD with value $0$ and the others with value $1$. After the projection is done, $\monadicapply$ collects the sink nodes of the resulting MTBDD. The algorithm, which is described in Algorithm~\ref{alg_retrieval}, needs the following two additional functions for working with projection BDDs:

\begin{description}

  \item[CreateProjection]  This function creates a projection BDD for symbol $k$.
    \begin{eqnarray}
      \mathit{CreateProjection} &:& \rankalph \to (\rankalph \to \binaryset) \nonumber \\
                                & & \mathit{CreateProjection}~\mathtt{k} = \lambda x~.~\mathbf{if}~x = \mathtt{k}~\mathbf{then}~1~\mathbf{else}~0
    \end{eqnarray}

  \item[Project]  Makes a projection of MTBDD \texttt{lhs} using a projection BDD \texttt{rhs} and returns the resulting MTBDD.
    \begin{eqnarray}
      \project &:& (\rankalph \to 2^Q) \to (\rankalph \to \binaryset) \to (\rankalph \to 2^Q) \nonumber \\
      & & \project~\mathtt{lhs~rhs} = \lambda x~.~\mathbf{if}~(\mathtt{rhs}~x) = 1~\mathbf{then}~(\mathtt{lhs}~x)~\mathbf{else}~\{q_{sink}\}~~~~~~~~
    \end{eqnarray}

\end{description}

\IncMargin{1em}
\begin{algorithm}[H]

  \SetKwData{Tmp}{tmp}
  \SetKwData{Superstate}{sp}
  \SetKwData{Proj}{proj}
  \SetKwData{States}{states}
  \SetKwFunction{Collect}{collect}

  \caption{Transition retrieval}
  \KwIn{Transition function $\transfunc$ \linebreak
        Symbol $f$ and \superstate{} $(\seq{q})$}
  \KwOut{$D = \left\{E \suchthat f(\seq{q}) \to E \in \transfunc \right\}$}

  \Begin
  {
    $\States \asgn \emptyset$\;
    $\Tmp \asgn \createprojection~f$\;
    $\Proj \asgn \project~\left(\transfunc~(\seq{q})\right)~\Tmp$\;
    $\monadicapply~\Proj~(\Collect~\States)$\;

    \Return{$\States$}\;
  }

  \label{alg_retrieval}
\end{algorithm}
\DecMargin{1em}

\IncMargin{1em}
\begin{function}[H]
  \addtocounter{algocf}{-1}

  \caption{collect(states, leaf)}

  \Begin
  {
    $\mathit{states} \asgn \mathit{states} \cup \mathit{leaf}$\;
  }
\end{function}
\DecMargin{1em}

\pagebreak

\subsection{Language Union} \label{subsec_union}

The task of the operation of \emph{language union} is, for two input tree automata $\automaton_1 = (Q_1, \rankalph, Q_{f1}, \transfunc_1)$ and $\automaton_2 = (Q_2, \rankalph, Q_{f2}, \transfunc_2)$, to create a tree automaton $\automaton_{\cup} = (Q_{\cup}, \rankalph, Q_{f\cup}, \transfunc_{\cup})$ such that $\lang(\automaton_{\cup}) = \lang(\automaton_1) \cup \lang(\automaton_2)$.
Although the algorithm presented in Section~\ref{subsec_closure_union} preserves determinism, we chose to use a more simple approach that does not create a product automaton but rather reuses transition functions of input automata as much as possible (and may introduce nondeterminism when input automata are deterministic).

The idea of this construction is to create such an automaton that makes nondeterministic transitions over leaf symbols to either $\automaton_1$ or $\automaton_2$ and then continues its run in the target automaton.
Assume without loss of generality that $Q_1 \cap Q_2 = \emptyset$, then $Q_{\cup} = Q_1 \cup Q_2$, $Q_{f\cup} = Q_{f1} \cup Q_{f2}$, and
\begin{eqnarray}
  \transfunc_{\cup} &=& \left( \transfunc_1 \setdiff \left\{f \to D_1 \suchthat f \in \rankalph_0, D_1 \subseteq Q_1 \right\} \right) \cup \nonumber \\
  & & \left( \transfunc_2 \setdiff \left\{f \to D_2 \suchthat f \in \rankalph_0, D_2 \subseteq Q_2 \right\} \right) \cup \nonumber \\
  & & \left\{ f \to D \suchthat f \in \rankalph_0, D =  D_1 \cup D_2, f \to D_1 \in \transfunc_1, f \to D_2 \in \transfunc_2 \right\} .
\end{eqnarray}
Computations of $Q_{\cup}$ and $Q_{f\cup}$ are trivial. Since all transitions are stored in a single MTBDD the computation of $\transfunc_{\cup}$ needs only one $\apply$ operation:
\begin{equation}
  \transfunc_{\cup}~\left(\right) \asgn \apply~\left(\transfunc_1~\left(\right)\right)~\left(\transfunc_2~\left(\right)\right)~\left( \lambda X~Y~.~X \cup Y \right) .
\end{equation}

\begin{figure}[h]
  \begin{center}
    \includegraphics[width=13cm]{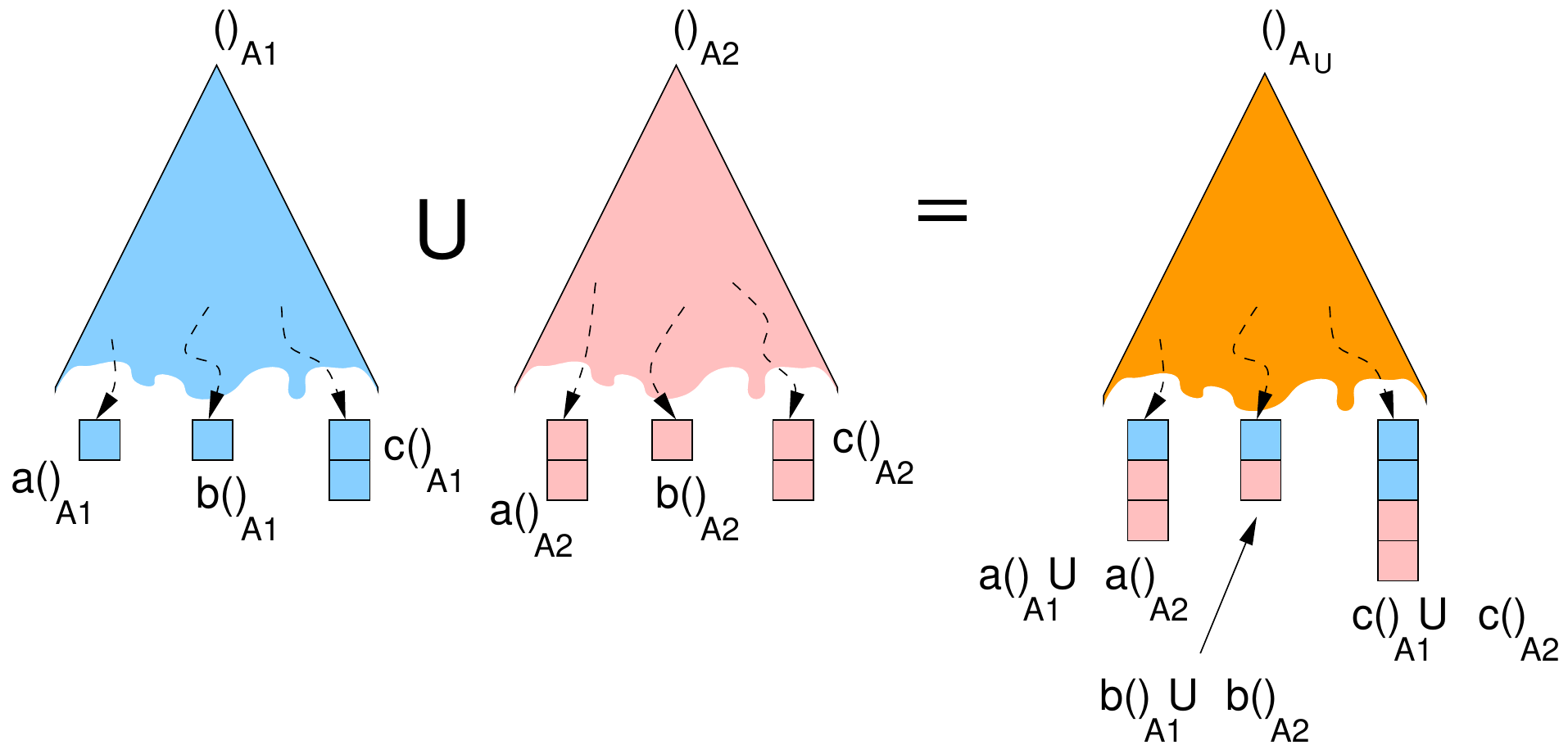}
  \end{center}
  \caption{Construction of automaton $\automaton_{\cup}$ such that $\lang(\automaton_{\cup}) = \lang(\automaton_1) \cup \lang(\automaton_2)$.}
  \label{fig_union}
\end{figure}

Figure~\ref{fig_union} shows the process of construction of the transition function for the union automaton. The procedure is described in Algorithm~\ref{alg_union}.

\IncMargin{1em}
\begin{algorithm}[ht]

  \caption{Union automaton construction}
  \KwIn{Input automata $\automaton_1 = (Q_1, \rankalph, Q_{f1}, \transfunc_1)$ and $\automaton_2 = (Q_2, \rankalph, Q_{f2}, \transfunc_2)$}
  \KwOut{$\automaton_{\cup} = (Q_{\cup}, \rankalph, Q_{f\cup}, \transfunc_{\cup})$ such that $\lang(\automaton_{\cup}) = \lang(\automaton_1) \cup \lang(\automaton_2)$}

  \Begin
  {
    $Q_{\cup} \asgn Q_1 \cup Q_2$\;
    $Q_{f\cup} \asgn Q_{f1} \cup Q_{f2}$\;
    $\transfunc_{\cup} \asgn \transfunc_1 \cup \transfunc_2$\;
    $\transfunc_{\cup}~\left(\right) \asgn \apply~\left(\transfunc_1~\left(\right)\right)~\left(\transfunc_2~\left(\right)\right)~\left( \lambda X~Y~.~X \cup Y \right)$\;

    \Return{$\automaton_{\cup} = (Q_{\cup}, \rankalph, Q_{f\cup}, \transfunc_{\cup})$}\;
  }

  \label{alg_union}
\end{algorithm}
\DecMargin{1em}

\subsection{Language Intersection} \label{subsec_intersection}

The requirements on the \emph{language intersection} operation are very similar to language union: given two finite tree automata $\automaton_1 = (Q_1, \rankalph, Q_{f1}, \transfunc_1)$ and $\automaton_2 = (Q_2, \rankalph, Q_{f2}, \transfunc_2)$ construct a finite tree automaton $\automaton_{\cap} = (Q_{\cap}, \rankalph, Q_{f\cap}, \transfunc_{\cap})$ such that $\lang(\automaton_{\cap}) = \lang(\automaton_1) \cap \lang(\automaton_2)$.

\begin{figure}[ht]
  \begin{center}
    \includegraphics[width=13cm]{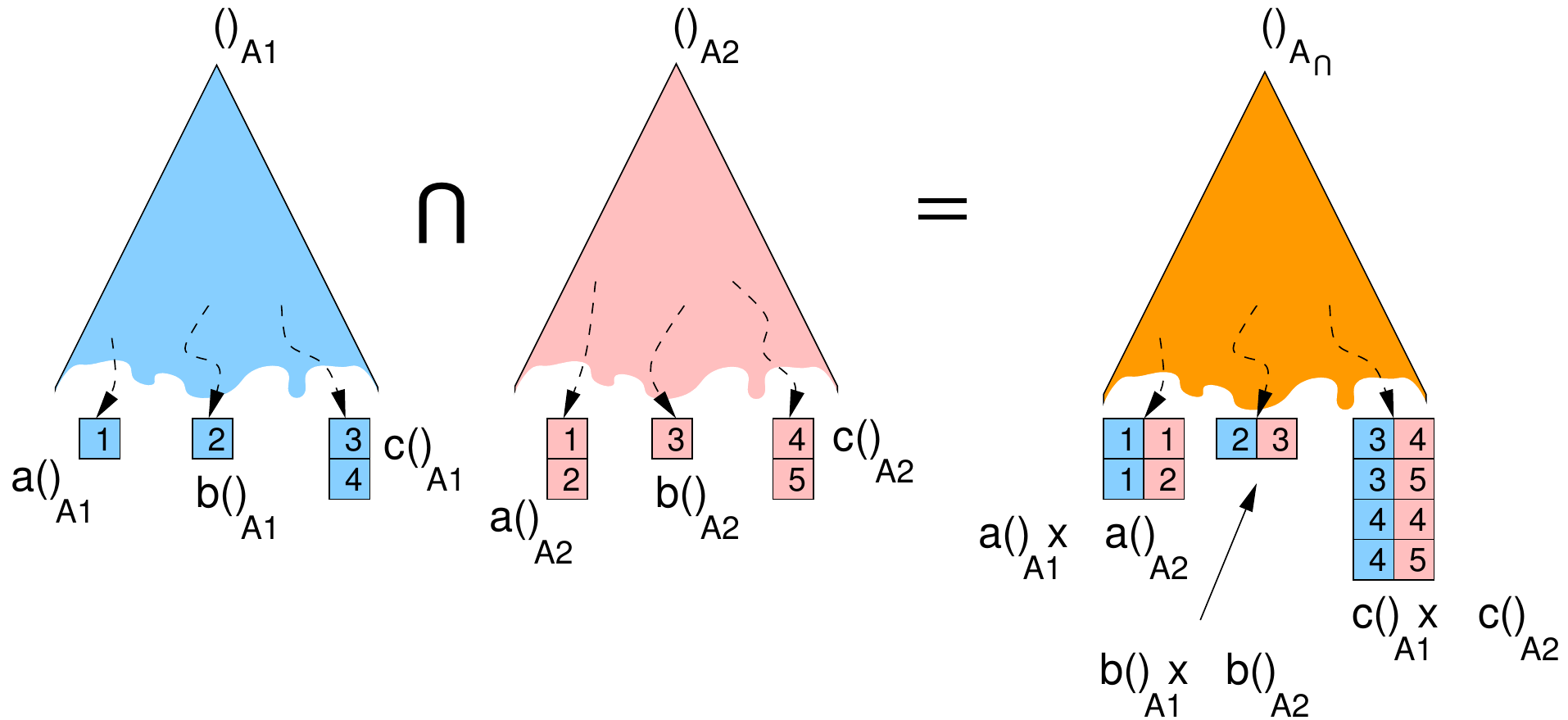}
  \end{center}
  \caption{Construction of automaton $\automaton_{\cap}$ such that $\lang(\automaton_{\cap}) = \lang(\automaton_1) \cap \lang(\automaton_2)$.}
  \label{fig_intersection}
\end{figure}

The construction is done by creating a \emph{product automaton} (a tree automaton with state set that is the Cartesian product of state sets of input automata) $\automaton_{\cap}$ which simulates parallel run of both input automata:
\begin{equation}
  \automaton_{\cap} = (Q_1 \times Q_2, \rankalph, Q_{f1} \times Q_{f2}, \transfunc_{\cap})
\end{equation}
where
\begin{eqnarray}
  \transfunc_{\cap} &=& \left\{f\left((q_{11}, q_{21}), \dots, (q_{1n}, q_{2n}) \right) \to (q_1, q_2) \suchthat f \in \rankalph_n, \right. \nonumber \\
  & & \left. f(q_{11}, \dots, q_{1n}) \to q_1 \in \transfunc_1, f(q_{21}, \dots, q_{2n}) \to q_2 \in \transfunc_2 \right\}
\end{eqnarray}
such that $\automaton_{\cap}$ contains only reachable states and transitions.
Detection of reachable states is done by starting from initial \superstate{}s of automata, analysing all transitions from reachable \superstate{}s and collecting states that may be reached in this way until the algorithm has no unanalysed state.
A \superstate{} $(\seq{q})$ is reachable if $\forall 1 \le i \le n : q_i$ is reachable.
Due to the fact that we work with complete automata (with sink state $q_{\mathit{sink}} \notin Q_{f}$), whenever we reach a product state $(q_1, q_{\mathit{sink}})$ or $(q_{\mathit{sink}}, q_2)$, we may stop generating further states (this is because $q_{sink}$ has only transitions to $q_{sink}$ so no accepting state can be reached from such state).
Figure~\ref{fig_intersection} shows the first step of construction of the product automaton.
The construction process is described in Algorithm~\ref{alg_intersection}.

\IncMargin{1em}
\begin{algorithm}[H]

  \SetKwFunction{Intersect}{intersect}
  \SetKwData{NewStates}{newStates}
  \SetKwData{ProductState}{productState}
  \SetKwData{ProcessedStates}{processed}
  \SetKwData{Superstatea}{sp1}
  \SetKwData{Superstateb}{sp2}

  \caption{Intersection automaton construction}
  \KwIn{Input automata $\automaton_1 = (Q_1, \rankalph, Q_{f1}, \transfunc_1)$ and $\automaton_2 = (Q_2, \rankalph, Q_{f2}, \transfunc_2)$}
  \KwOut{$\automaton_{\cap} = (Q_{\cap}, \rankalph, Q_{f\cap}, \transfunc_{\cap})$ such that $\lang(\automaton_{\cap}) = \lang(\automaton_1) \cap \lang(\automaton_2)$}

  \Begin
  {
    $Q_{\cap} \asgn Q_{f\cap} \asgn \transfunc_{\cap} \asgn \emptyset$\;
    $\NewStates \asgn$ \desc{empty queue}\;

    $\transfunc_{\cap}~\left(\right) \asgn \apply~\left(\transfunc_1~\left(\right)\right)~\left(\transfunc_2~\left(\right)\right)~(\Intersect~\NewStates)$\;

    \While{\NewStates \desc{is not empty}}
    {
      $(q_a, q_b) \asgn \NewStates.\Dequeue()$\;
      \If{$(q_a, q_b) \notin Q_{\cap}$}
      {
        $Q_{\cap} \asgn Q_{\cap} \cup \{ (q_a, q_b) \}$\;

        \lIf{$q_a = q_{sink} \vee q_b = q_{sink}$}
        {
          \Continue\;
        }

        \lIf{$q_a \in Q_{f1} \wedge q_b \in Q_{f2}$}
        {
          $Q_{f\cap} \asgn Q_{f\cap} \cup \{ (q_a, q_b) \} $\;
        }

        \ForEach{$n \in \natnums$ \desc{such that} $S_n(\transfunc_1) \neq \emptyset \wedge S_n(\transfunc_2) \neq \emptyset$}
        {
          \ForEach{$(q_{11}, \dots, q_{1n}) \in S_n(\transfunc_1)$ \desc{such that} $q_a \in (q_{11}, \dots, q_{1n})$}
          {
            \ForEach{$(q_{21}, \dots, q_{2n}) \in S_n(\transfunc_2)$ \desc{such that} $q_b \in (q_{21}, \dots, q_{2n})$}
            {
              \If{$\forall 1 \le i \le n : (q_{1i}, q_{2i}) \in Q_{\cap}$}
              {
                $\Superstatea \asgn \left(q_{11}, \dots, q_{1n}\right)$\;
                $\Superstateb \asgn \left(q_{21}, \dots, q_{2n}\right)$\;
                $\transfunc_{\cap}~\left((q_{11}, q_{21}), \dots, (q_{1n}, q_{2n})\right) \asgn \apply~\left(\transfunc_1~\Superstatea\right)~\left(\transfunc_2~\Superstateb\right)~(\Intersect~\NewStates)$\;
              }
            }
          }
        }
      }
    }

    \Return{$\automaton_{\cap} = (Q_{\cap}, \rankalph, Q_{f\cap}, \transfunc_{\cap})$}\;
  }

  \label{alg_intersection}
\end{algorithm}
\DecMargin{1em}

\IncMargin{1em}
\begin{function}[H]
  \addtocounter{algocf}{-1}

  \SetKwData{ProductSet}{productSet}

  \caption{intersect(newStates, lhs, rhs)}

  \Begin
  {
    $\ProductSet \asgn lhs \times rhs$\;
    \ForEach{$(q_a, q_b) \in \ProductSet$}
    {
      $\mathit{newStates}.\Enqueue((q_a, q_b))$\;
    }

    \Return{\ProductSet}\;
  }
\end{function}
\DecMargin{1em}

\subsection{Determinisation} \label{subsec_determinisation}

The \emph{determinisation} operation takes an input finite tree automaton $\automaton = (Q, \rankalph, Q_f, \transfunc)$ and transforms it into a deterministic finite tree automaton $\automaton_d = (Q_d, \rankalph, Q_{fd}, \transfunc_d)$ such that $\lang(\automaton_d) = \lang(\automaton)$.

The determinisation algorithm that is described in Algorithm~\ref{alg_determinisation} works with \emph{\macrostate{}s}.
A \macrostate{} $M \subseteq Q$ is a state in the deterministic automaton that represents all states which might have been accessed during the run of the nondeterministic automaton over the same sequence of symbols.
The algorithm starts from the initial \superstate{}, creates a new \macrostate{} for each sink node of the MTBDD for the initial \superstate{} and proceeds with finding all \superstate{}s $(\seq{q})$ such that there exist \macrostate{}s $\seq{M} : \forall 1 \le i \le n : q_i \in M_i$.
For each such \superstate{} an MTBDD with union of sets at sink nodes of all MTBDDs that can be accessed by combinations of states in given \macrostate{}s is created; new \macrostate{}s are retrieved as sets of states from sink node of this MTBDD.
This guarantees that only reachable states are present in the result automaton.

\IncMargin{1em}
\begin{algorithm}[H]

  \SetKwData{NewStates}{newStates}
  \SetKwData{Tmp}{tmp}
  \SetKwFunction{CollectSets}{collectSets}

  \caption{Automaton determinisation}
  \KwIn{Input automaton $\automaton = (Q, \rankalph, Q_{f}, \transfunc)$}
  \KwOut{Deterministic automaton $\automaton_d = (Q_d, \rankalph, Q_{fd}, \transfunc_d)$, $\lang(\automaton_d) = \lang(\automaton)$}

  \Begin
  {
    $Q_d \asgn Q_{fd} \asgn \transfunc_d \asgn \emptyset$\;
    $\NewStates \asgn $ empty queue\;

    $\transfunc_d~() \asgn \monadicapply~\left(\transfunc~()\right)~(\CollectSets~\NewStates)$\;

    \While{\NewStates \desc{is not empty}}
    {
      $s \asgn \NewStates.\Dequeue()$\;
      \If{$s \notin Q_d$}
      {
        $Q_d \asgn Q_d \cup \{ s \}$\;
        \lIf{$\exists q_f \in s$ \desc{such that} $q_f \in Q_f$}
        {
          $Q_{fd} \asgn Q_{fd} \cup \{s\} $\;
        }

        \ForEach{$n \in \natnums$ \desc{such that} $S_n(\transfunc) \neq \emptyset$}
        {
          \ForEach{$(\seq{q}) \in S_n(\transfunc)$ \desc{such that} $\exists 1 \le i \le n : q_i \in s$}
          {
            \ForEach{$\seq{s} \in Q_d$ \desc{such that} $q_1 \in s_1, \dots, q_n \in s_n, s_i = s$}
            {
              \tcc{Create empty MTBDD}

              $\Tmp \asgn \emptyset$\;
              \ForEach{$(\seq{p}) \in S_n(\transfunc)$ \desc{such that} $p_1 \in s_1, \dots, p_n \in s_n$}
              {
                $\Tmp \asgn \apply~\Tmp~\left(\transfunc~(\seq{p})\right)~\left(\lambda X~Y~.~X \cup Y \right)$\;
              }

              $\transfunc_d~(\seq{s}) \asgn \monadicapply~\Tmp~(\CollectSets~\NewStates)$\;
            }
          }
        }
      }
    }

    \Return{$\automaton_d = (Q_d, \rankalph, Q_{fd}, \transfunc_d)$}\;
  }

  \label{alg_determinisation}
\end{algorithm}
\DecMargin{1em}

\IncMargin{1em}
\begin{function}[H]
  \addtocounter{algocf}{-1}

  \SetKwData{SubsetState}{s}

  \caption{collectSets(newStates, tf)}

  \Begin
  {
    $\mathit{newStates}.\Enqueue(\mathit{tf})$\;

    \Return{$\{\mathit{tf}\}$}\;
  }
\end{function}
\DecMargin{1em}

\subsection{Language Complementation} \label{subsec_complementation}

Given a finite tree automaton $\automaton = (Q, \rankalph, Q_f, \transfunc)$, the task of \emph{language complementation} is to construct automaton $\automaton_c$ such that $\lang(\automaton_c) = \overline{\lang(\automaton)}$. This is done by first transforming $\automaton$ to a deterministic automaton $\automaton_d = (Q_d, \rankalph, Q_{fd}, \transfunc_d)$ by the procedure described in Section~\ref{subsec_determinisation} and then complementing the set of accepting states: $\automaton_c = (Q_d, \rankalph, Q_d \setdiff Q_{fd}, \transfunc_d)$.

\subsection{Automaton Reduction} \label{subsec_reduction}

\emph{Reduction of a finite tree automaton} is a generic operation that takes a finite tree automaton $\automaton = (Q, \rankalph, Q_f, \transfunc)$ and a quotient set $Q /\!\sim$ of some equivalence relation $\sim$ and returns a reduced finite tree automaton $\automaton_r = (Q_r, \rankalph, Q_{fr}, \transfunc_r)$ such that $Q_r = Q /\!\sim$, $Q_{fr} = \{D \in Q_r \suchthat \exists q \in D : q \in Q_f\}$, and
\begin{eqnarray}
  \transfunc_r &=& \left\{ f(\seq{B}) \to B \suchthat \right. \nonumber \\
               & & \left. f(\seq{q}) \to q \in \Delta, f \in \rankalph, q_1 \in B_1, \dots, q_n \in B_n, q \in B\right\} .
\end{eqnarray}

Various methods can be used for obtaining the equivalence relation $\sim$, e.g.\ Myhill-Nerode minimisation (see Section~\ref{subsec_minimisation}) or downward simulation (see Section~\ref{subsec_simulation}).
Note that while the former approach can be used over deterministic finite tree automata only, the latter may be used to reduce the size of nondeterministic finite tree automata as well (in their case, however, the result is not a minimal nondeterministic finite tree automaton but reduced nondeterministic finite tree automaton only).
The algorithm for reduction of a finite tree automaton is given in Algorithm~\ref{alg_reduction}.

\IncMargin{1em}
\begin{algorithm}[H]

  \SetKwData{Superstate}{sp}

  \caption{Automaton reduction}
  \KwIn{Input automaton $\automaton = (Q, \rankalph, Q_{f}, \transfunc)$ \linebreak
        Quotient set $Q /\!\sim$}
  \KwOut{Reduced automaton $\automaton_r = (Q_r, \rankalph, Q_{fr}, \transfunc_r)$}

  \Begin
  {
    $Q_r \asgn Q /\!\sim$\;
    $Q_{fr} \asgn \left\{[q]_{\sim} \suchthat q \in Q_f\right\}$\;
    $\transfunc_r \asgn \emptyset$\;
    \ForEach{$n \in \natnums$ \desc{such that} $S_n(\transfunc) \neq \emptyset$}
    {
      \ForEach{$(\seq{q}) \in S_n(\transfunc)$}
      {
        $\Superstate \asgn \left([q_1]_{\sim}, \dots, [q_n]_{\sim}\right)$\;
        $\transfunc_r~\Superstate \asgn \apply~\left(\transfunc_r~\Superstate\right)~\left(\transfunc~(\seq{q})\right)~\left(\lambda X~Y~.~X \cup \{[y]_{\sim} \suchthat y \in Y\}\right)$\;
      }
    }

    \Return{$\automaton_r = (Q_r, \rankalph, Q_{fr}, \transfunc_r)$}\;
  }

  \label{alg_reduction}
\end{algorithm}
\DecMargin{1em}

\subsection{Pruning Unreachable States} \label{subsec_pruning}

The task of \emph{pruning unreachable states} of a finite tree automaton $\automaton = (Q, \rankalph, Q_f, \transfunc)$ is removal of states $q$ (and corresponding transitions, which means removing MTBDDs for all \superstate{}s that contain $q$) for which there does not exist a tree $t \in T(\rankalph)$ such that $t \move[*]{\automaton} q$.
The algorithm attempts to simulate the run of the automaton for all possible trees and collect states that can be reached.
The description of the algorithm is in Algorithm~\ref{alg_pruning}.

\IncMargin{1em}
\begin{algorithm}[h]

  \SetKwData{ReachStates}{reachStates}
  \SetKwData{Superstate}{sp}
  \SetKwFunction{CollectReachable}{collectReachable}

  \caption{Unreachable states pruning}
  \KwIn{Input automaton $\automaton = (Q, \rankalph, Q_f, \transfunc)$}
  \KwOut{Automaton $\automaton_p = (Q_p, \rankalph, Q_{fp}, \transfunc_p)$ without unreachable states, such that $\lang(\automaton_p) = \lang(\automaton)$}

  \Begin
  {
    $Q_p \asgn Q_{fp} \asgn \transfunc_p \asgn \emptyset$\;
    $\ReachStates \asgn$ \desc{empty queue}\;

    $\transfunc_p~\left(\right) \asgn \monadicapply~\left(\transfunc~\left(\right)\right)~(\CollectReachable~\ReachStates)$\;

    \While{\ReachStates \desc{is not empty}}
    {
      $q \asgn \ReachStates.\Dequeue()$\;
      \If{$q \notin Q_p$}
      {
        $Q_p \asgn Q_p \cup \{ q \}$\;

        \lIf{$q \in Q_f$}
        {
          $Q_{fp} \asgn Q_{fp} \cup \{ q \} $\;
        }

        \ForEach{$n \in \natnums$ \desc{such that} $S_n(\transfunc) \neq \emptyset$}
        {
          \ForEach{$(\seq{q}) \in S_n(\transfunc)$ \desc{such that} $q \in (\seq{q})$}
          {
            \If{$\forall 1 \le i \le n : q_i \in Q_p$}
            {
              $\Superstate \asgn (\seq{q})$\;
              $\transfunc_p~\Superstate \asgn \monadicapply~\left(\transfunc~\Superstate\right)~(\CollectReachable~\ReachStates)$\;
            }
          }
        }
      }
    }

    \Return{$\automaton_p = (Q_p, \rankalph, Q_{fp}, \transfunc_p)$}\;
  }

  \label{alg_pruning}
\end{algorithm}
\DecMargin{1em}

\IncMargin{1em}
\begin{function}[H]
  \addtocounter{algocf}{-1}

  \caption{collectReachable(reachStates, leaf)}

  \Begin
  {
    \ForEach{$q \in \mathit{leaf}$}
    {
      $\mathit{reachStates}.\Enqueue(q)$\;
    }

    \Return{$\mathit{leaf}$}\;
  }
\end{function}
\DecMargin{1em}

\subsection{Minimisation} \label{subsec_minimisation}

\emph{Automaton minimisation} is an operation on a finite tree automaton $\automaton = (Q, \rankalph, Q_f, \transfunc)$ which returns deterministic finite tree automaton $\automaton_m = (Q_m, \rankalph, Q_{fm}, \transfunc_m)$ such that $\lang(\automaton) = \lang(\automaton_m)$ and $\automaton_m$ is an automaton that has the least states from all deterministic finite tree automata that accept $\lang(\automaton)$. Existence of a minimum deterministic finite tree automaton is guaranteed by the proof of Myhill-Nerode Theorem (see Section~\ref{sec_myhill_nerode}).

The minimisation process starts with pruning unreachable states and determinising the input automaton. Once done, Algorithm~\ref{alg_minimisation} computes equivalence relation for congruence (used in Myhill-Nerode Theorem) $\sim$ on $Q$. This is done by refining the equivalence relation from the start point with two initial classes: the accepting states and the non-accepting states. All \superstate{}s $(\seq{q})$ are then searched and in case there exists an equivalence class $[q_i]_{\sim}$ such that when $q_i$ is substituted in $(\seq{q})$ for some other element from $[q_i]_{\sim}$ and the target class of transitions over respective symbols differs, then $\sim$ is refined.

In the following step, the quotient set of $\sim$ is passed to the reduction procedure (see Section~\ref{subsec_reduction}) and obtaining the minimum automaton is straightforward.

\IncMargin{1em}
\begin{algorithm}[H]

  \SetKwData{Eq}{eq}
  \SetKwData{PrevEq}{prevEq}
  \SetKwData{Refined}{refined}
  \SetKwData{Superstate}{sp}
  \SetKwFunction{RefineEquivalence}{refineEq}

  \caption{Computation of $\sim$ equivalence over states}
  \KwIn{Deterministic automaton without unreachable states $\automaton = (Q, \rankalph, Q_f, \transfunc)$}
  \KwOut{Equivalence relation $\sim \subseteq Q \times Q$}

  \Begin
  {
    $\Eq \asgn \left\{(p, q) \suchthat p \in Q_f \ifandonlyif q \in Q_f\right\}$\;
    $\PrevEq \asgn \emptyset$\;

    \While{$\Eq \neq \PrevEq$}
    {
      $\PrevEq \asgn \Eq$\;

      \ForEach{$n \in \natnums$ \desc{such that} $S_n(\transfunc) \neq \emptyset$}
      {
        \ForEach{$(\seq{q}) \in S_n(\transfunc)$}
        {
          \ForEach{$1 \le i \le n$}
          {
            \ForEach{$q \in [q_i]_{\PrevEq}$}
            {
              $\Superstate_{qi} \asgn (q_1, \dots, q_{i-1}, q_i, q_{i+1}, \dots, q_n)$\;
              $\Superstate_{q} \asgn (q_1, \dots, q_{i-1}, q, q_{i+1}, \dots, q_n)$\;
              \uIf{$\Superstate_q \in S_n(\transfunc)$}
              {
                $\Refined \asgn \False$\;
                $\apply~\left(\transfunc~\Superstate_{qi}\right)~\left(\transfunc~\Superstate_q\right)~(\RefineEquivalence~\PrevEq~\Refined)$\;

                \lIf{\Refined}
                {
                  $\Eq \asgn \Eq \setdiff \{ (q, q_i), (q_i, q)\}$\;
                }
              }
              \Else
              {
                $\Eq \asgn \Eq \setdiff \{ (q, q_i), (q_i, q)\}$\;
              }
            }
          }
        }
      }
    }

    \Return{$\sim = \Eq$}\;
  }

  \label{alg_minimisation}
\end{algorithm}
\DecMargin{1em}

\IncMargin{1em}
\begin{function}[H]
  \addtocounter{algocf}{-1}

  \caption{refineEq(prevEq, refined, \{lhs\}, \{rhs\})}

  \Begin
  {
    \If{$[\mathit{lhs}]_{\mathit{prevEq}} \neq [\mathit{rhs}]_{\mathit{prevEq}}$}
    {
      $\mathit{refined} \asgn \True$\;
    }
  }
\end{function}
\DecMargin{1em}

\subsection{Checking Language Emptiness} \label{subsec_emptiness}

The problem of \emph{determining emptiness} of a language is defined as given a finite tree automaton $\automaton = (Q, \rankalph, Q_f, \transfunc)$, is $\lang(\automaton) = \emptyset$?
The algorithm for deciding language emptiness first removes unreachable states from automaton $\automaton$ using the method described in Section~\ref{subsec_pruning}.
This constructs a finite tree automaton $\automaton_p = (Q_p, \rankalph, Q_{fp}, \transfunc_p)$ without unreachable states.
It holds that language $\lang(\automaton_p)$ is empty if and only if $Q_{fp} = \emptyset$ (i.e.\ there is no reachable final state in $\automaton_p$).
Note that a slightly more efficient algorithm can determine that $\lang(\automaton) \neq \emptyset$ immediately when the analysis of reachable states reaches state $q$ such that $q \in Q_f$.

\subsection{Downward Simulation Reduction} \label{subsec_simulation}

\emph{Downward simulation}~\cite{abdulla07} $\issimulatedby$ for a finite tree automaton $\automaton = (Q, \rankalph, Q_f, \transfunc)$ is a binary relation on $Q$ such that if $q \issimulatedby r$ and $f(\seq{q}) \to q \in \transfunc$, then there are $\seq{r}$ such that $f(\seq{r}) \to r \in \transfunc$ and $q_i \issimulatedby r_i$ for each $1 \le i \le n$. Formally:
\begin{eqnarray}
  \forall f \in \rankalph &:& \left[q \issimulatedby r \wedge f(\seq{q}) \to q \in \transfunc\right] \Rightarrow \nonumber \\
                          & & \left[\exists \seq{r} \in Q : f(\seq{r}) \to r \in \transfunc \wedge \forall 1 \le i \le n : q_i \issimulatedby r_i\right]~~~~~~~~
  \label{eq_simulation}
\end{eqnarray}
From the previous equation, the following can be inferred using \emph{modus tollens}:
\begin{eqnarray}
  \forall f \in \rankalph &:& \neg \left[\exists \seq{r} \in Q : f(\seq{r}) \to r \in \Delta \wedge \forall 1 \le i \le n : q_i \issimulatedby r_i\right] \Rightarrow \nonumber \\
                          & & \left[\neg \left(q \issimulatedby r\right) \vee \neg \left(f(\seq{q}) \to q \in \Delta\right)\right]
  \label{eq_simulation_after_modus_tollens}
\end{eqnarray}
We further expand $\issimulatedby$ relation to \superstate{}s:
\begin{equation}
  (\seq{q}) \issimulatedby (\seq{r}) \stackrel{\mathrm{def}}{\ifandonlyif} \forall 1 \le i \le n : q_i \issimulatedby r_i
\end{equation}
It can be proved that $\issimulatedby$ is reflexive and transitive.
It is possible to use downward simulation for reduction of the size of an automaton by identifying states that simulate each other and collapsing those states together.
Even though an automaton obtained in this way is often not \emph{minimum}, the reduction can be significant and computation is faster than minimisation which needs to first convert the automaton to deterministic one.

The algorithm for computation of downward simulation, described in Algorithm~\ref{alg_simulation}, starts with declaring $\issimulatedby = Q \times Q$ and then for each \superstate{} $(\seq{q})$ finds all \superstate{}s $(\seq{r})$ such that $(\seq{q}) \issimulatedby (\seq{r})$ and makes a new MTBDD with uniting the sink nodes of those.
This union MTBDD represents all states $r$ that can be reached using \superstate{}s $(\seq{r})$ simulating \superstate{} $(\seq{q})$.
Now for each state $q$ accessible from $(\seq{q})$ over symbol $f \in \rankalph$ we check for each $r$ such that $q \issimulatedby r$ that $r$ is in the union MTBDD accessible over $f$.
In case it is not, according to Equation~\ref{eq_simulation_after_modus_tollens} the simulation relation $\issimulatedby$ can be is refined by removing $(q, r)$ from $\issimulatedby$.
This is repeated until $\issimulatedby$ reaches the fixpoint.

As downward simulation is reflexive and transitive but generally not symmetric, symmetric closure of the relation needs to be performed in order to obtain equivalence relation.
Reduction is then performed using the generic reduction procedure as described in Section~\ref{subsec_reduction}.

\IncMargin{1em}
\begin{algorithm}[h]

  \SetKwData{Simulation}{sim}
  \SetKwData{PreviousSimulation}{prevSim}
  \SetKwData{Tmp}{tmp}
  \SetKwFunction{SimulationRefinement}{simulationRefinement}

  \caption{Downward simulation computation}
  \KwIn{Input automaton $\automaton = (Q, \rankalph, Q_{f}, \transfunc)$}
  \KwOut{Simulation relation $\issimulatedby \subseteq Q \times Q$}

  \Begin
  {
    $\PreviousSimulation \asgn \emptyset$\;
    $\Simulation \asgn Q \times Q$\;

    \While{$\PreviousSimulation \neq \Simulation$}
    {
      $\PreviousSimulation \asgn \Simulation$\;

      \ForEach{$n \in \natnums$ \desc{such that} $S_n(\transfunc) \neq \emptyset$}
      {
        \ForEach{$(\seq{q}) \in S_n(\transfunc)$}
        {
          \tcc{Create empty MTBDD}
          $\Tmp \asgn \emptyset$\;

          \ForEach{$(\seq{r}) \in S_n(\transfunc)$ \desc{such that} $\forall 1 \le i \le n : (q_i, r_i) \in \Simulation$}
          {
            $\Tmp \asgn \apply~\Tmp~\left(\transfunc~(\seq{r})\right)~(\lambda X~Y~.~X \cup Y)$\;
          }

          $\apply~\left(\transfunc~(\seq{q})\right)~\Tmp~(\SimulationRefinement~\Simulation)$\;
        }
      }
    }

    \Return{$\issimulatedby = \Simulation$}\;
  }

  \label{alg_simulation}
\end{algorithm}
\DecMargin{1em}

\IncMargin{1em}
\begin{function}[h]
  \addtocounter{algocf}{-1}

  \caption{simulationRefinement(sim, lhs, rhs)}

  \Begin
  {
    \ForEach{$q \in \mathit{lhs}$}
    {
      \ForEach{$r$ \desc{such that} $(q, r) \in \mathit{sim}$}
      {
        \lIf{$r \notin \mathit{rhs}$}
        {
          $\mathit{sim} \asgn \mathit{sim} \setdiff \left\{(q, r)\right\}$\;
        }
      }
    }
  }
\end{function}
\DecMargin{1em}

\subsection{Checking Language Inclusion Using Antichains} \label{subsec_antichains}

The \emph{language inclusion} decision problem is to determine for two input finite tree automata $\automaton_1 = (Q_1, \rankalph, Q_{f1}, \transfunc_1)$ and $\automaton_2 = (Q_2, \rankalph, Q_{f2}, \transfunc_2)$ whether it holds that $\lang(\automaton_1) \subseteq \lang(\automaton_2)$.
The standard approach of checking language inclusion is by determinising $\automaton_2$, complementing it, and checking whether $\lang(\automaton_1) \cap \overline{\lang(\automaton_2)} = \emptyset$.
In case the intersection is not empty, it means that there are some trees which are in $\lang(\automaton_1)$ and not in $\lang(\automaton_2)$ and therefore the inclusion does not hold.
Nevertheless complementation needs determinisation of $\automaton_2$ which is often very expensive.
Therefore it is desirable to find approaches that do not need this operation.

One approach that avoids determinisation is based on \emph{antichains}~\cite{bouajjani08}.
An antichain over $Q_1 \times 2^{Q_2}$ is a set $S \subseteq Q_1 \times 2^{Q_2}$ such that for every $(p,s), (p^{\prime}, s^{\prime}) \in S$ if $p = p^{\prime}$ then $s \not\subset s^{\prime}$.
For $(p, s) \in S$, $p$ denotes a state from $\automaton_1$ that is reachable over some tree and $s$ denotes a set of states of automaton $\automaton_2$ that are reachable over the same tree.
If such a pair $(p, s)$ can be reached so that $p \in Q_{f1}$ and $\forall r \in s : r \notin Q_{f2}$, the inclusion $\lang(\automaton_1) \subseteq \lang(\automaton_2)$ does not hold.
The algorithm is given in Algorithm~\ref{alg_antichains}.

\IncMargin{1em}
\begin{algorithm}[h]

  \SetKwData{Antichain}{antichain}
  \SetKwData{PrevAntichain}{prevAntichain}
  \SetKwData{Tmp}{tmp}
  \SetKwFunction{CollectProducts}{collectProducts}

  \caption{Antichain-based inclusion}
  \KwIn{Input automata $\automaton_1 = (Q_1, \rankalph, Q_{f1}, \transfunc_1)$ and $\automaton_2 = (Q_2, \rankalph, Q_{f2}, \transfunc_2)$}
  \KwOut{$\True$ if $\lang(\automaton_1) \subseteq \lang(\automaton_2)$, $\False$ otherwise}

  \Begin
  {
    $\PrevAntichain \asgn \emptyset$\;
    $\Antichain \asgn \emptyset$\;
    $\apply~\left(\transfunc_1~()\right)~\left(\transfunc_2~()\right)~(\CollectProducts~\Antichain)$\;
    
    \While{$\Antichain \neq \PrevAntichain$}
    {
      $\PrevAntichain \asgn \Antichain$\;

      \ForEach{$(q, D) \in \PrevAntichain$}
      {
        \lIf{$q \in Q_{f1} \wedge \forall p \in D : p \notin Q_{f2}$}
        {
          \Return{$\False$}\;
        }
      }

      \ForEach{$n \in \natnums$ \desc{such that} $S_n(\transfunc_1) \neq \emptyset$}
      {
        \ForEach{$(\seq{q}) \in S_n(\transfunc_1)$ \desc{such that} $\forall 1 \le i \le n : \exists R_i \subseteq Q_2 : (q_i, R_i) \in \PrevAntichain$}
        {
          $\Tmp \asgn \emptyset$\;

          \ForEach{$(\seq{s}) \in S_n(\transfunc_2)$ \desc{such that} $\forall 1 \le i \le n : s_i \in R_i$}
          {
            $\Tmp \asgn \apply~\Tmp~\left(\transfunc_2~(\seq{s})\right)~(\lambda X~Y~.~X \cup Y)$\;
          }

          $\apply~\left(\transfunc_1~(\seq{q})\right)~\Tmp~(\CollectProducts~\Antichain)$\;
        }
      }
    }

    \Return{$\True$}\;
  }

  \label{alg_antichains}
\end{algorithm}
\DecMargin{1em}

\IncMargin{1em}
\begin{function}[h]
  \addtocounter{algocf}{-1}

  \SetKwData{Lhs}{lhs}
  \SetKwData{Rhs}{rhs}

  \caption{collectProducts(antichain,lhs, rhs)}

  \Begin
  {
    \ForEach{$q \in \mathit{lhs}$}
    {
      \If{$\nexists (q, E) \in \mathit{antichain}$ \desc{such that} $\mathit{rhs} \subseteq E$}
      {
        $\mathit{antichain} \asgn (\mathit{antichain} \setdiff \left\{ (q, F) \suchthat F \subset \mathit{rhs}\right\}) \cup \left\{(q, \mathit{rhs})\right\} $\;
      }
    }
  }
\end{function}
\DecMargin{1em}

\enlargethispage{2mm}

\section{Transducers} \label{sec_transducers}

This section starts with a description of the representation of \emph{relabelling} (or sometimes called \emph{structure-preserving}) tree transducers.
These are transducers that do not change the structure of input trees but only change symbols in their nodes.
The section continues by a definition of two operations that are necessary in regular tree model checking: performing a transduction step on a finite tree automaton and composition of transducers.

\subsection{Representation of a Relabelling Tree Transducer} \label{subsec_transducer_representation}

We represent only relabelling tree transducers that use the same alphabet $\rankalph$ for both input and output, we therefore refer to transducer $\tau = (Q, \rankalph, \rankalph^{\prime} = \rankalph, Q_f, \transfunc)$ by $\transducer = (Q, \rankalph, Q_f, \transfunc)$.
Relabelling tree transducers contain transduction rules of the following type:
\begin{equation}
  f(q_1(x_1), \dots, q_n(x_n)) \to q(g(\seq{x})) ,
\end{equation}
where $n \in \natnums$, $f,g \in \rankalph_n$, $q, \seq{q} \in Q$ and $\seq{x} \in \variables$, or using an alternative notation as
\begin{equation}
  f(\seq{q}) \to q(g) .
\end{equation}

The representation of a transduction function $\transfunc$ of a relabelling tree transducer is therefore very similar to the representation of a transition function of a finite tree automaton and can again be symbolic.
We naturally expand the definition of a \superstate{} $S(\transfunc)$ to the transduction function.
The transduction function $\transfunc$ of a relabelling tree transducer $\transducer$ may then be alternatively defined as a mapping $\transfuncmtbdd$ in the following way:
\begin{eqnarray}
  \transfuncmtbdd &:& S \to \left(\rankalph \to (\rankalph \to 2^Q)\right) \nonumber \\
  & & (\seq[p]{q}) \mapsto \left\{\left(f, (g, D)\right) \suchthat D = \left\{ q \suchthat f(\seq[p]{q}) \to q(g) \in \transfunc \right\} \right\} .
  \label{eq_transduction_func}
\end{eqnarray}
However, since the composition of functions is associative, the formula in Equation~\ref{eq_transduction_func} can be rewritten as
\begin{eqnarray}
  \transfuncmtbdd &:& S \to \left((\rankalph \to \rankalph) \to 2^Q\right) \nonumber \\
  & & (\seq[p]{q}) \mapsto \left\{\left((f, g), D\right) \suchthat D = \left\{ q \suchthat f(\seq[p]{q}) \to q(g) \in \transfunc \right\} \right\}
\end{eqnarray}
(we again confuse $\transfunc$ and $\transfuncmtbdd$).
This means that we can represent a transduction function of a relabelling tree transducer using MTBDDs in the same way as a transition function of a finite tree automaton, provided we expand the function $\mathit{enc} : \rankalph \to \binaryset^n$, which is defined in Section~\ref{sec_representation}, to $\mathit{enc}_T : (\rankalph \times \rankalph) \to \binaryset^{2n}$ in the following way:
\begin{eqnarray}
  \mathit{enc}_T &:& (\rankalph \times \rankalph) \to \binaryset^{2n} \nonumber \\
  & & (a, b) \mapsto (\seq{a}, \seq{b}) \\
  & & \qquad \mathrm{where} \qquad (\seq{a}) = \mathit{enc}(a) \quad \mathrm{and} \quad(\seq{b}) = \mathit{enc}(b) \nonumber .
\end{eqnarray}
Note that the actual ordering of $\seq{a}$ and $\seq{b}$ is not important provided that it remains consistent for $\mathit{enc}_T$.
Another ordering which may be useful for some cases is for instance $(a_1, b_1, \dots, a_n, b_n)$.

In case we denote the MTBDD for a \superstate{} $s_{\automaton}$ of the transition function $\transfunc_{\automaton}$ of a finite tree automaton $\automaton$ as
\begin{equation}
  \sum_{f \in \rankalph \atop \mathit{enc}(f) = (\seq{a})} \left(\prod_{a_i = 0} \lnot x_i \cdot \prod_{a_i = 1} x_i \cdot s_{\automaton}(f) \right)
\end{equation}
(see Section~\ref{sec_mona} for further details of this notation), we may represent MTBDD for a \superstate{} $s_{\transducer}$ of the transduction function $\transfunc_{\transducer}$ of a relabelling tree transducer $\transducer$ as
\begin{equation}
  \sum_{{(f, g) \in \rankalph \times \rankalph \atop \mathit{enc}_T(f, g) = } \atop (\seq{a}, \seq{b})} \left(\prod_{a_i = 0} \lnot x_i \cdot \prod_{a_i = 1} x_i \cdot \prod_{b_i = 0} \lnot y_i \cdot \prod_{b_i = 1} y_i \cdot s_{\transducer}(f, g) \right) .
\end{equation}

This representation works with an MTBDD extended by Boolean variables $\seq{y}$ (we assume that the representation of finite tree automata described in Section~\ref{sec_representation} uses variables $\seq{x}$).
In order to support operations that work with both finite tree automata and relabelling tree transducers, the following two functions that work directly with the structure of MTBDDs are necessary:

\begin{description}

  \item[TrimVariables]  This function receives an MTBDD $M$ and $x$, which is a base name of Boolean variables $\seq{x}$ such that $\seq{x}$ are in $M$, and returns MTBDD $M_{-x}$ that does not contain $\seq{x}$. Hence, $\trimvariables(M, x) = M_{-x}$.
  Since this operation may cause collisions (e.g.\ producing formula $y_1 y_2 A + y_1 y_2 B$ where $A \neq B, A \neq \bottom, B \neq \bottom$), they need to be properly handled by uniting colliding state sets (e.g.\ producing $y_1 y_2 (A \cup B)$ for the previous example).
  The following formula formally defines the function:
\begin{eqnarray}
  & & \trimvariables\left(\sum_{{(f, g) \in \rankalph \times \rankalph \atop \mathit{enc}_T(f, g) = } \atop (\seq{a}, \seq{b})} \left(\prod_{a_i = 0} \lnot x_i \cdot \prod_{a_i = 1} x_i \cdot \prod_{b_i = 0} \lnot y_i \cdot \prod_{b_i = 1} y_i \cdot J(f, g) \right), x\right) \nonumber \\
  & & = \sum_{g \in \rankalph \atop \mathit{enc}(g) = (\seq{b})} \left(\prod_{b_i = 0} \lnot y_i \cdot \prod_{b_i = 1} y_i \cdot \bigcup_{f \in \rankalph} J(f, g) \right)
\end{eqnarray}
  The implementation of $\trimvariables(M, x)$ can be done in the following way:

  \begin{enumerate}[]
    \item  Traverse $M$ from the root to sink nodes and for each node $k$ on the path such that $k$ represents some $x_i$ do the following: take both child nodes of $k$, $k_0$ and $k_1$, and set $k$ to $k \asgn \apply~k_0~k_1~(\lambda X~Y~.~X \cup Y)$.
   \end{enumerate}

  \item[RenameVariables]  A function that receives an MTBDD $M$ and names of two Boolean variables $x$ and $y$ such that $\seq{x}$ are in $M$.
  The function renames all occurrences of $x_i$ to $y_i$ for each $1 \le i \le n$.
  The function is formally defined by the following formula:
\begin{eqnarray}
  & & \renamevariables\left(\sum_{f \in \rankalph \atop \mathit{enc}(f) = (\seq{a})} \left(\prod_{a_i = 0} \lnot x_i \cdot \prod_{a_i = 1} x_i \cdot J(f) \right), x, y\right) \nonumber \\
  & & = \sum_{f \in \rankalph \atop \mathit{enc}(f) = (\seq{b})} \left(\prod_{b_i = 0} \lnot y_i \cdot \prod_{b_i = 1} y_i \cdot J(f) \right)
\end{eqnarray}
  The implementation of $\renamevariables(M, x, y)$ simply traverses $M$ and renames all occurrences of $x_i$ to $y_i$ (assuming that $\seq{y}$ are not in $M$).

\end{description}

\subsection{Performing a Transduction Step} \label{subsec_transduction_step}

The operation of \emph{performing a transduction step} of a finite tree automaton $\automaton = (Q_{\automaton}, \rankalph, Q_{f\automaton}, \transfunc_{\automaton})$ according to the transduction denoted by a relabelling tree transducer $\transducer = (Q_{\transducer}, \rankalph, Q_{f\transducer}, \transfunc_{\transducer})$ constructs a finite tree automaton $\automaton_w = (Q_w, \rankalph, Q_{fw}, \transfunc_w)$ such that $\lang(\automaton_w) = \transducer\left(\lang(\automaton)\right)$.
Informally, if $\automaton$ represents a set of configurations of a system and $\transducer$ represents transitions in the system, then $\transducer\left(\lang(\automaton)\right)$ is a finite tree automaton that represents the set of configurations of the system after one transition.

\IncMargin{1em}
\begin{algorithm}[H]

  \SetKwData{Tmp}{tmp}
  \SetKwData{NewStates}{newStates}
  \SetKwData{Superstatea}{sp1}
  \SetKwData{Superstateb}{sp2}
  \SetKwFunction{Intersect}{intersect}

  \caption{Performing transduction step}
  \KwIn{Input automaton $\automaton = (Q_{\automaton}, \rankalph, Q_{f\automaton}, \transfunc_{\automaton})$ \linebreak
        Relabelling tree transducer $\transducer = (Q_{\transducer}, \rankalph, Q_{f\transducer}, \transfunc_{\transducer})$}
  \KwOut{Automaton $\automaton_w = (Q_w, \rankalph, Q_{fw}, \transfunc_w)$ such that $\lang(\automaton_w) = \transducer\left(\lang(\automaton)\right)$}

  \Begin
  {
    $Q_w \asgn Q_{fw} \asgn \transfunc_w \asgn \emptyset$\;
    $\NewStates \asgn$ \desc{empty queue}\;

    $\Tmp \asgn \apply~\left(\transfunc_{\automaton}~\left(\right)\right)~\left(\transfunc_{\transducer}~\left(\right)\right)~(\Intersect~\NewStates)$\;
    $\Tmp \asgn \trimvariables(\Tmp, x)$\;
    $\transfunc_w~() \asgn \renamevariables(\Tmp, y, x)$\;

    \While{\NewStates \desc{is not empty}}
    {
      $(q_a, q_b) \asgn \NewStates.\Dequeue()$\;
      \If{$(q_a, q_b) \notin Q_w$}
      {
        $Q_w \asgn Q_w \cup \{ (q_a, q_b) \}$\;

        \lIf{$q_a = q_{sink} \vee q_b = q_{sink}$}
        {
          \Continue\;
        }

        \lIf{$q_a \in Q_{f\automaton} \wedge q_b \in Q_{f\transducer}$}
        {
          $Q_{fw} \asgn Q_{fw} \cup \{ (q_a, q_b) \} $\;
        }

        \ForEach{$n \in \natnums$ \desc{such that} $S_n(\transfunc_{\automaton}) \neq \emptyset \wedge S_n(\transfunc_{\transducer}) \neq \emptyset$}
        {
          \ForEach{$(q_{11}, \dots, q_{1n}) \in S_n(\transfunc_{\automaton})$ \desc{such that} $q_a \in (q_{11}, \dots, q_{1n})$}
          {
            \ForEach{$(q_{21}, \dots, q_{2n}) \in S_n(\transfunc_{\transducer})$ \desc{such that} $q_b \in (q_{21}, \dots, q_{2n})$}
            {
              \If{$\forall 1 \le i \le n : (q_{1i}, q_{2i}) \in Q_w$}
              {
                $\Superstatea \asgn \left(q_{11}, \dots, q_{1n}\right)$\;
                $\Superstateb \asgn \left(q_{21}, \dots, q_{2n}\right)$\;
                $\Tmp \asgn \apply~\left(\transfunc_{\automaton}~\Superstatea\right)~\left(\transfunc_{\transducer}~\Superstateb\right)~(\Intersect~\NewStates)$\;
                $\Tmp \asgn \trimvariables(\Tmp, x)$\;
                $\transfunc_w~\left((q_{11}, q_{21}), \dots, (q_{1n}, q_{2n})\right) \asgn \renamevariables(\Tmp, y, x)$\;
              }
            }
          }
        }
      }
    }

    \Return{$\automaton_w = (Q_w, \rankalph, Q_{fw}, \transfunc_w)$}\;
  }

  \label{alg_performing_transduction}
\end{algorithm}
\DecMargin{1em}

\begin{figure}[ht]
  \centering
  \begin{minipage}{4cm}
    \centering
    \includegraphics[height=5cm]{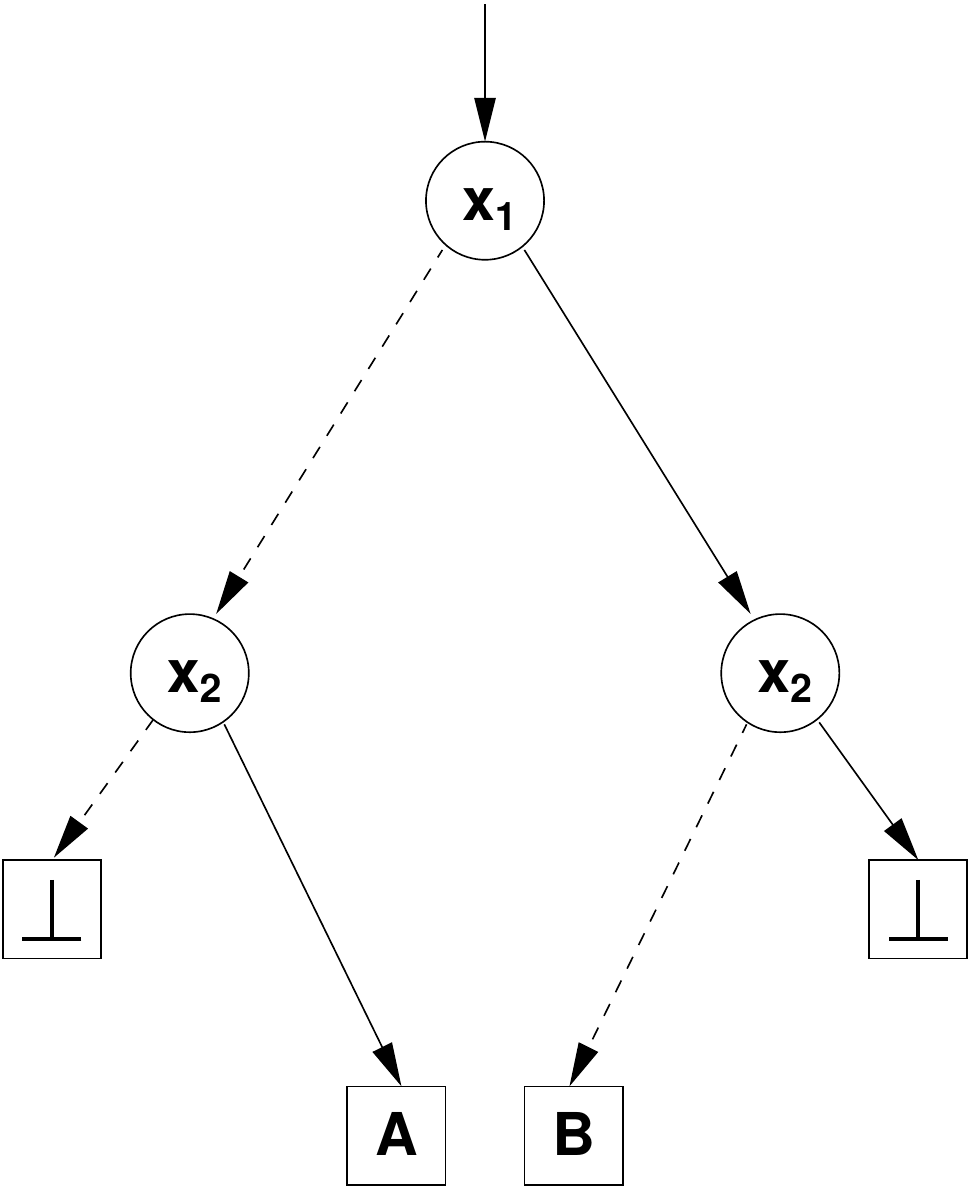}\\
    a) MTBDD $s_{\automaton}$
  \end{minipage}
  \hfill
  \begin{minipage}{4cm}
    \centering
    \includegraphics[height=5cm]{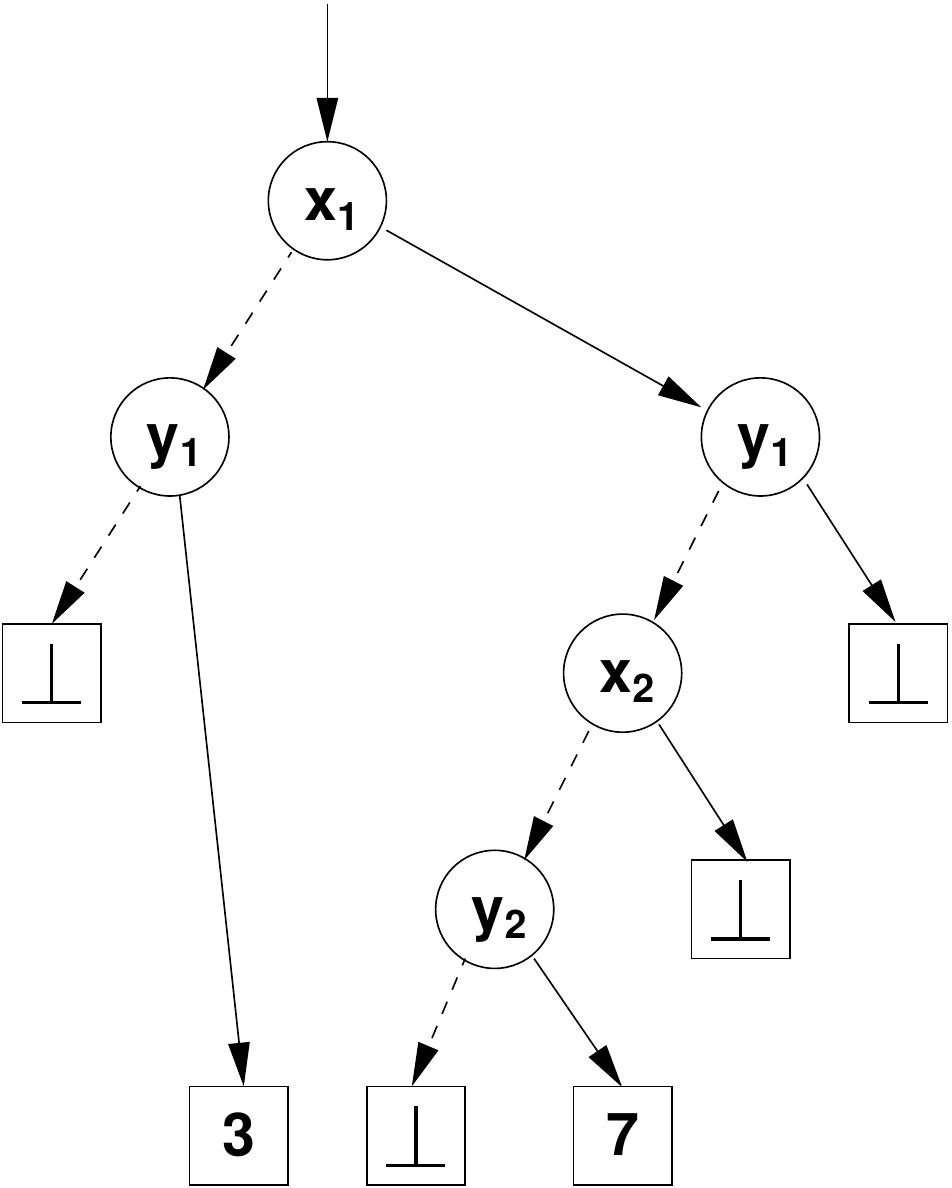}\\
    b) MTBDD $s_{\transducer}$
  \end{minipage}
  \hfill
  \begin{minipage}{6cm}
    \centering
    \includegraphics[height=5cm]{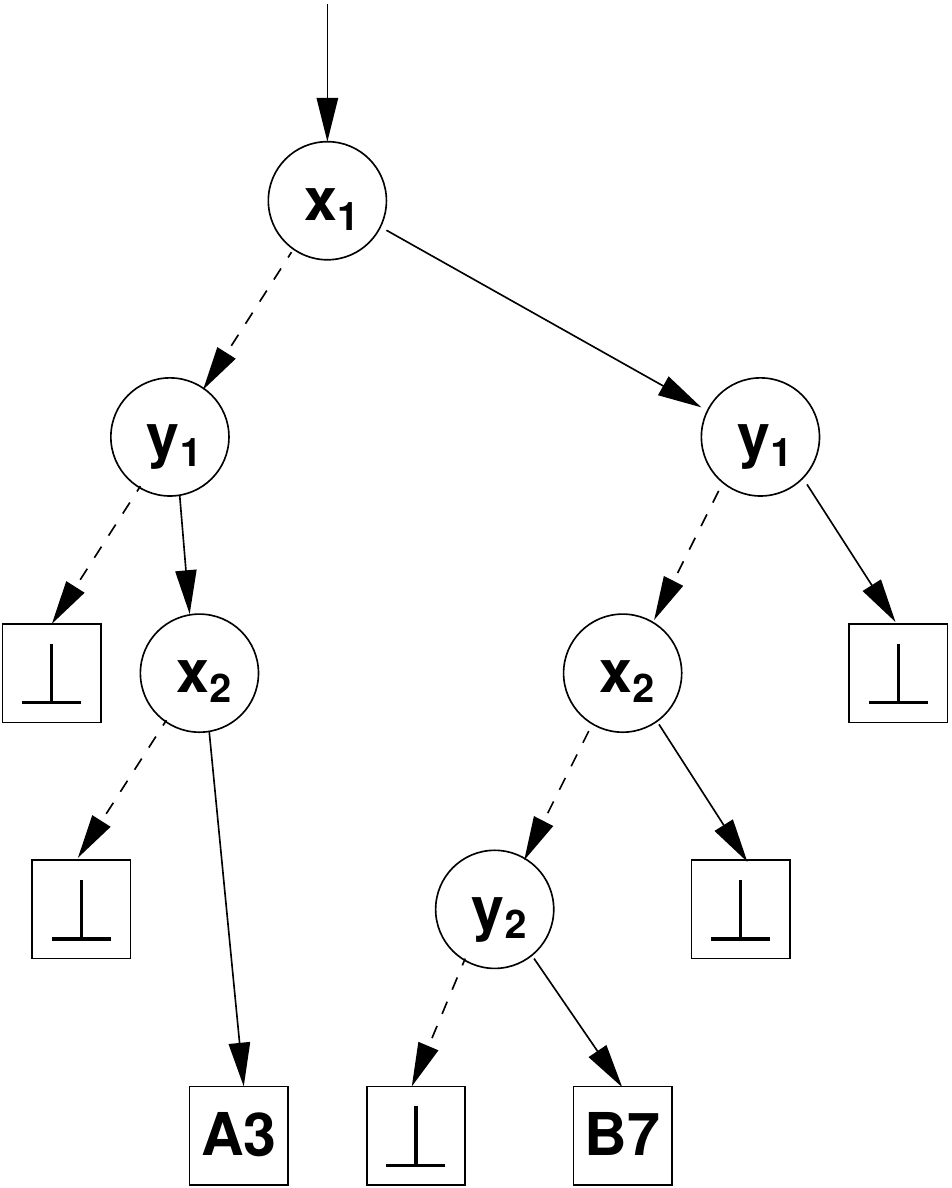}\\
    c) $\mathtt{res} \asgn \apply~s_{\automaton}~s{\transducer}~\mathtt{intersect}$
  \end{minipage}\\
  \vspace{3mm}
  \begin{minipage}{7cm}
    \centering
    \includegraphics[height=5cm]{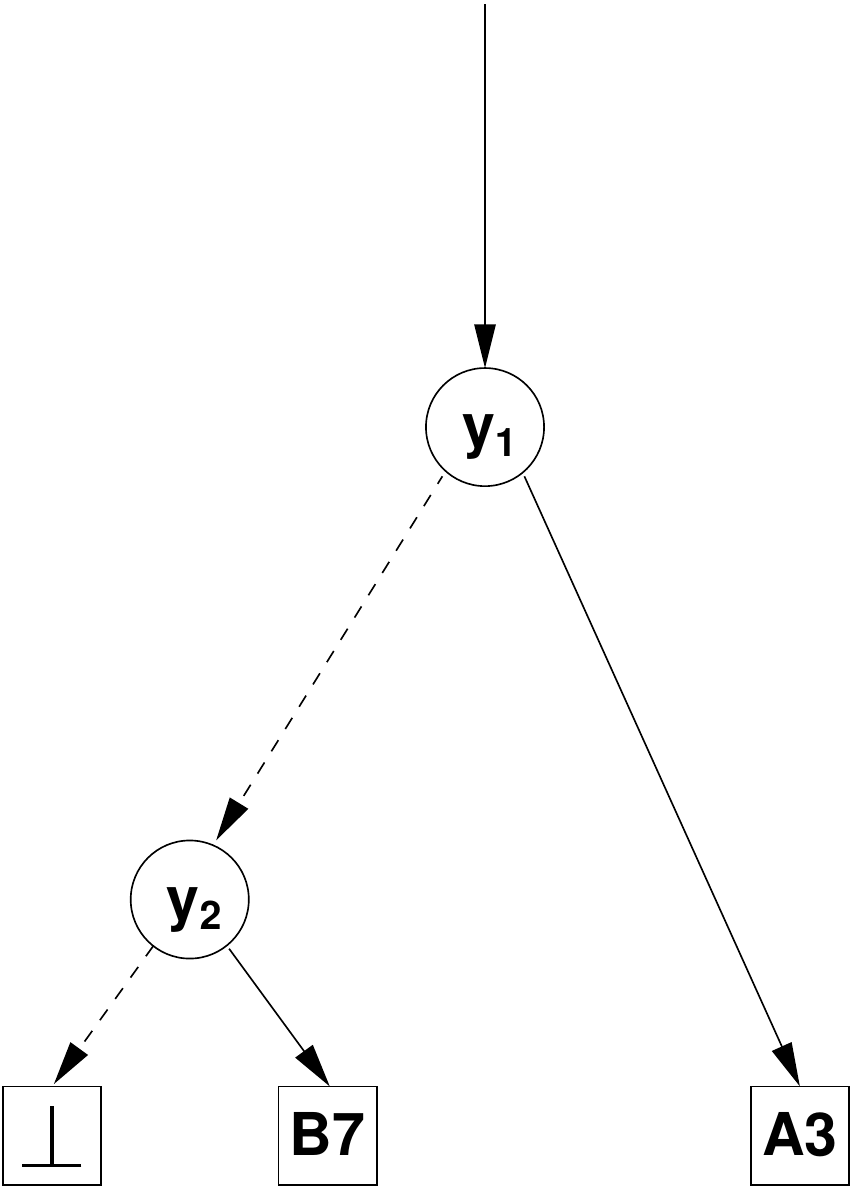}\\
    d) $\mathtt{trimmed} \asgn \trimvariables(\mathtt{res}, x)$
  \end{minipage}
  ~~~~~~~~
  \begin{minipage}{7cm}
    \centering
    \includegraphics[height=5cm]{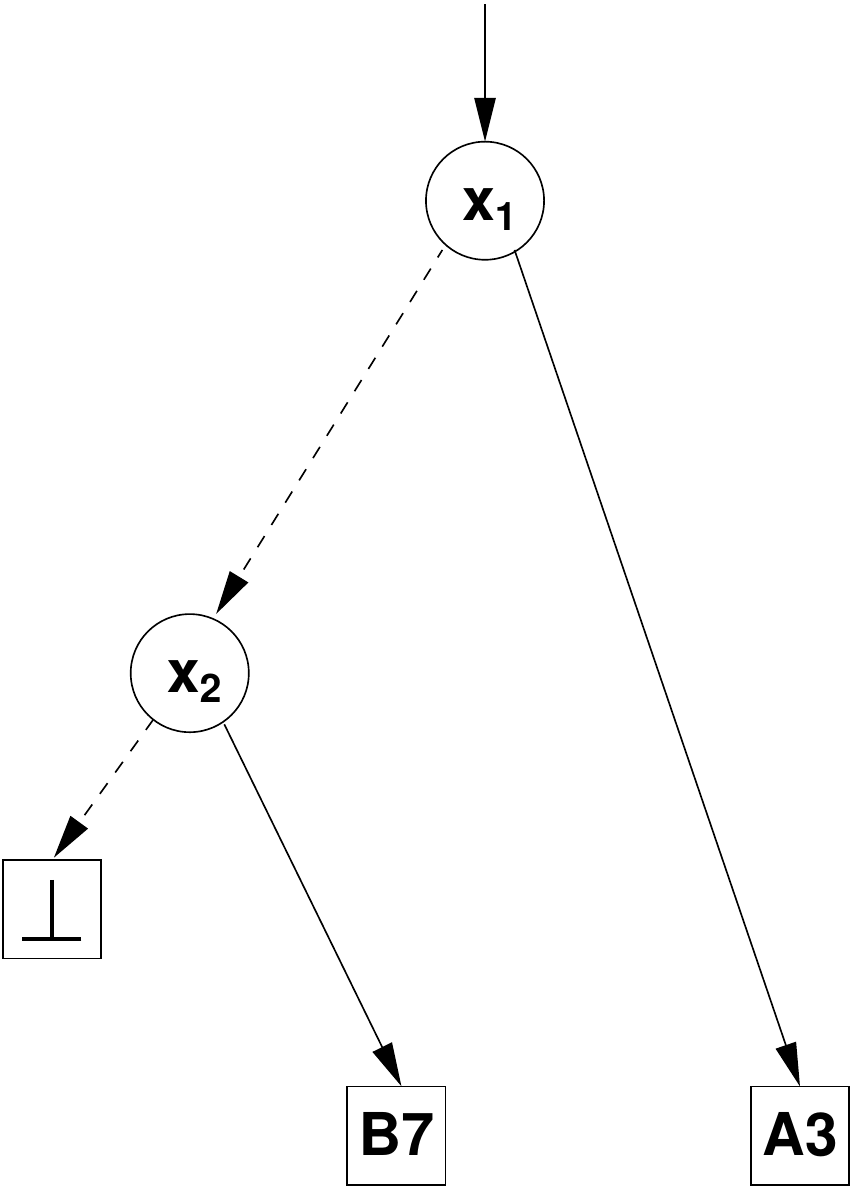}\\
    e) $\renamevariables(\mathtt{trimmed}, y, x)$
  \end{minipage}
  \caption{An example of performing transduction step of transducer $\transducer$ on automaton $\automaton$ for one pair of \superstate{}s $s_{\transducer}$ and $s_{\automaton}$, such that $\mathtt{01}(s_{\automaton}) \to A, \mathtt{10}(s_{\automaton}) \to B$, and $\mathtt{0X}(s_{\transducer}) \to 3(\mathtt{1X}), \mathtt{10}(s_{\transducer}) \to 7(\mathtt{01})$. Ordering $(a_1, b_1, \dots, a_n, b_n)$ is used.}
  \label{fig_transduction_step_example}
\end{figure}

The algorithm for this operation is described in Algorithm~\ref{alg_performing_transduction}.
The algorithm assumes that MTBDDs for transition function $\transfunc_{\automaton}$ are defined over Boolean variables $\seq{x}$ and that MTBDDs for transduction function $\transfunc_{\transducer}$ are defined over Boolean variables $\seq{x}$ and $\seq{y}$, where $\seq{x}$ are used for input symbols of the transducer and $\seq{y}$ are used for output symbols.
MTBDDs for the output automaton are again over Boolean variables $\seq{x}$.
The algorithm works by traversing both the automaton and the transducer in parallel and performing relabelling of transitions which are in both (the algorithm may resemble the computation of intersection, it actually uses function \texttt{intersect()} which is defined in Section~\ref{subsec_intersection}).
Figure~\ref{fig_transduction_step_example} attempts to give the idea about how the algorithm works for a pair of \superstate{}s.

\subsection{Transducer Composition} \label{subsec_trans_composition}

\emph{Transducer composition} is an operation that, when given two relabelling tree transducers $\transducer_1 = (Q_1, \rankalph, Q_{f1}, \transfunc_1)$ and $\transducer_2 = (Q_2, \rankalph, Q_{f2}, \transfunc_2)$, creates a relabelling tree transducer $\transducer = (Q, \rankalph, Q_f, \transfunc)$ such that for all finite tree automata $\automaton$, it holds that $\transducer\left(\lang(\automaton)\right) = \transducer_2\left(\transducer_1\left(\lang(\automaton)\right)\right)$ (or $\transducer = \transducer_2 \circ \transducer_1$).

The algorithm described as Algorithm~\ref{alg_transducer_composition} assumes that MTBDDs for transduction functions $\transfunc_1$ and $\transfunc_2$ are over Boolean variables $\seq{x}$ (which encode the input symbol) and $\seq{y}$ (which encode the output symbol).
The MTBDDs for the transduction function of the constructed transducer are again over Boolean variables $\seq{x}$ for the input and $\seq{y}$ for the output.
However, the MTBDDs need to be able to also work with Boolean variables $\seq{z}$ as they are used inside the algorithm.

\begin{figure}[ht]
  \centering
  \begin{minipage}{4cm}
    \centering
    \includegraphics[height=7cm]{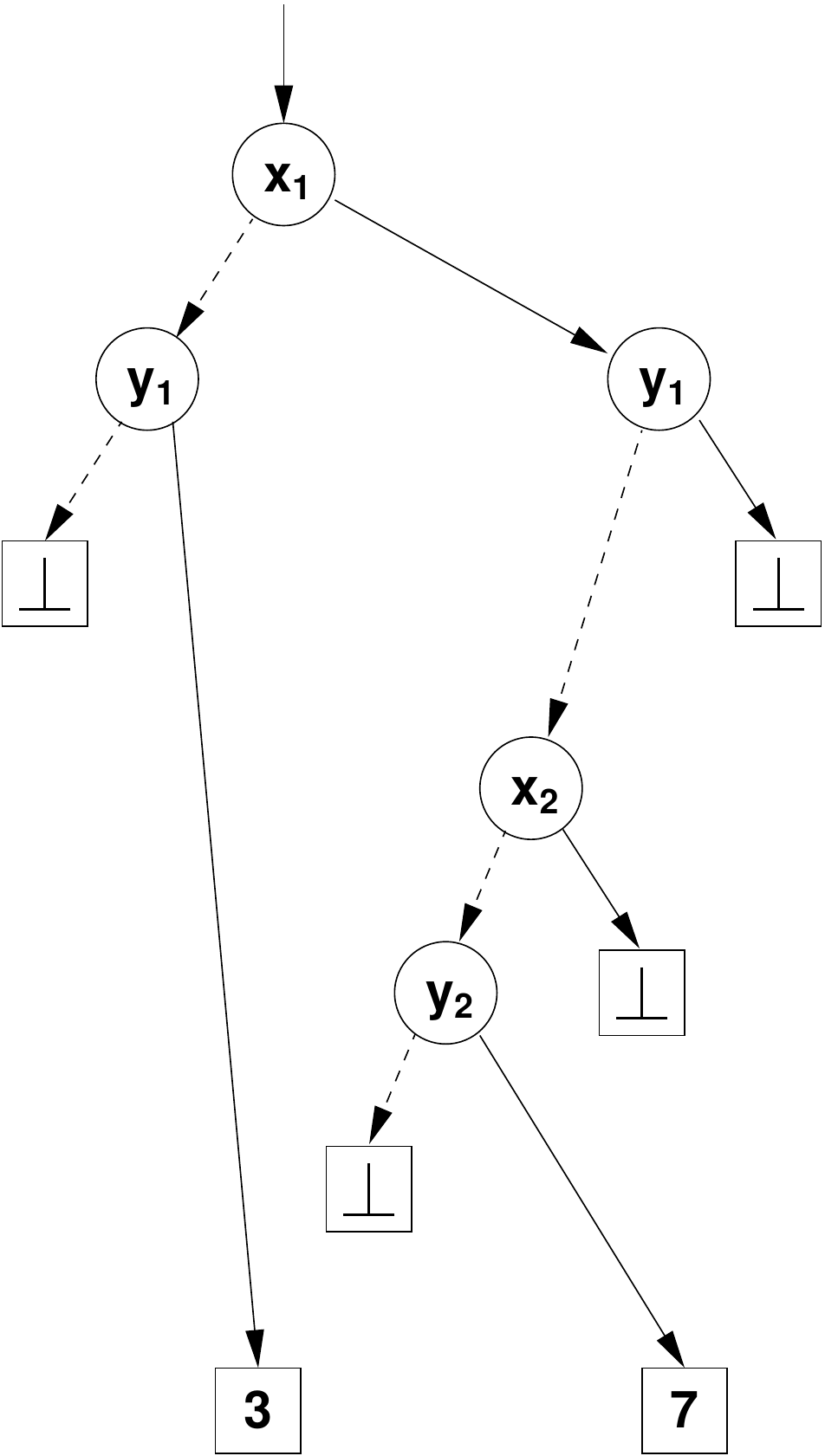}\\
    a) MTBDD $s_1$
  \end{minipage}
  \hfill
  \begin{minipage}{4cm}
    \centering
    \includegraphics[height=7cm]{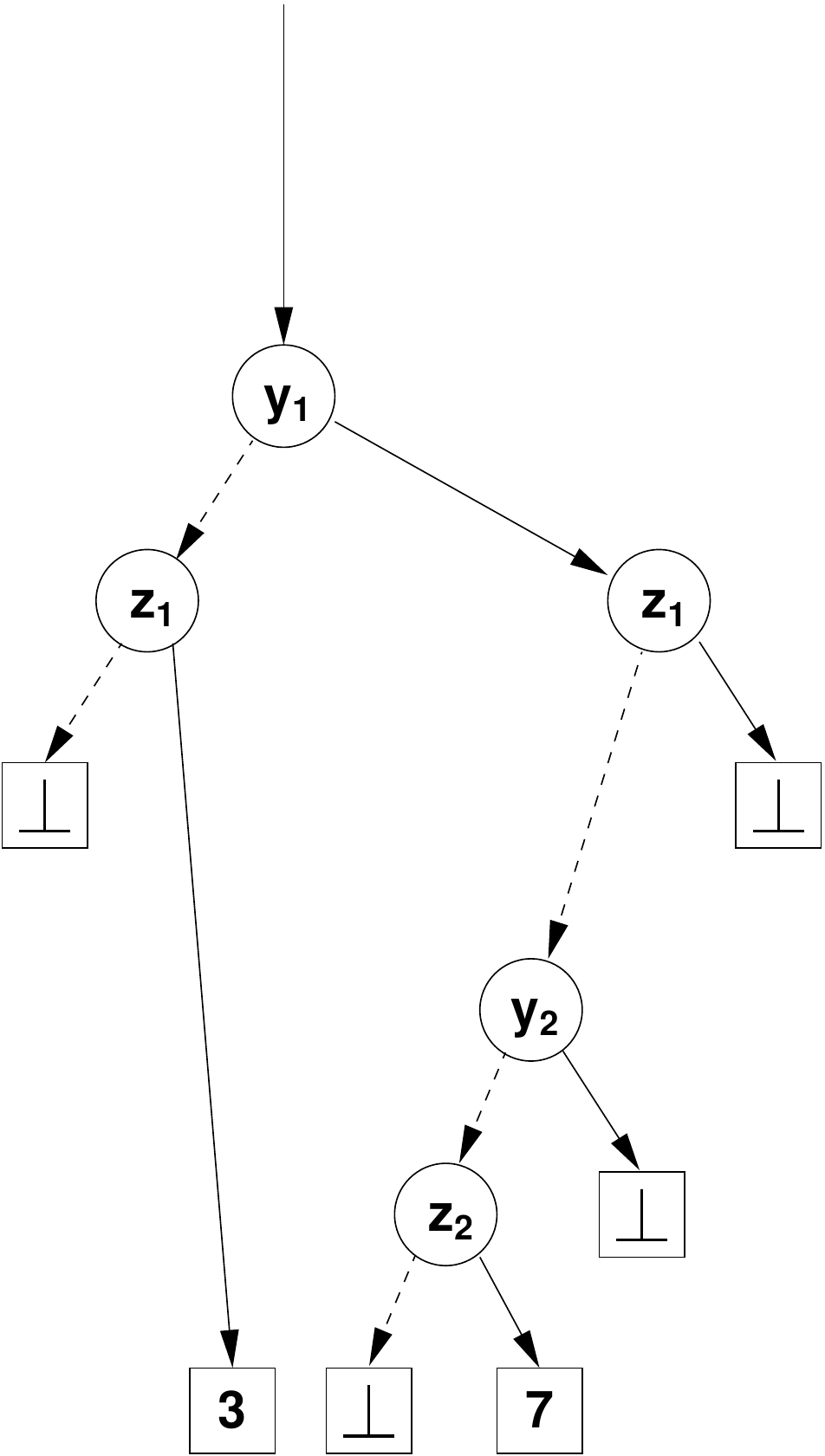}\\
    b) MTBDD $s_2$
  \end{minipage}
  \hfill
  \begin{minipage}{6.2cm}
    \centering
    \includegraphics[height=7cm]{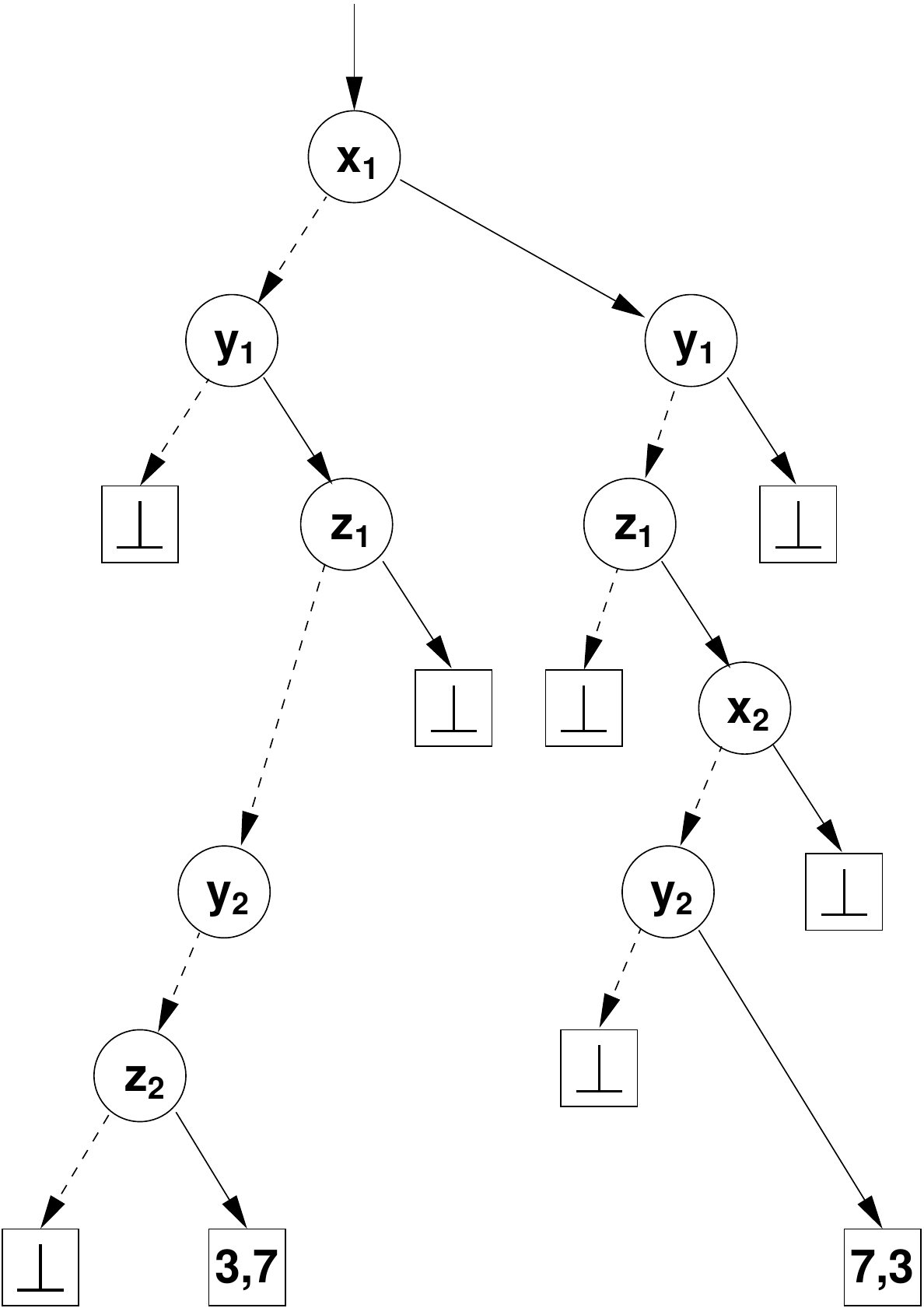}\\
    c) $\mathtt{res} \asgn \apply~s_1~s_2~\mathtt{intersect}$
  \end{minipage}\\
  \vspace{3mm}
  \begin{minipage}{7cm}
    \centering
    \includegraphics[height=7cm]{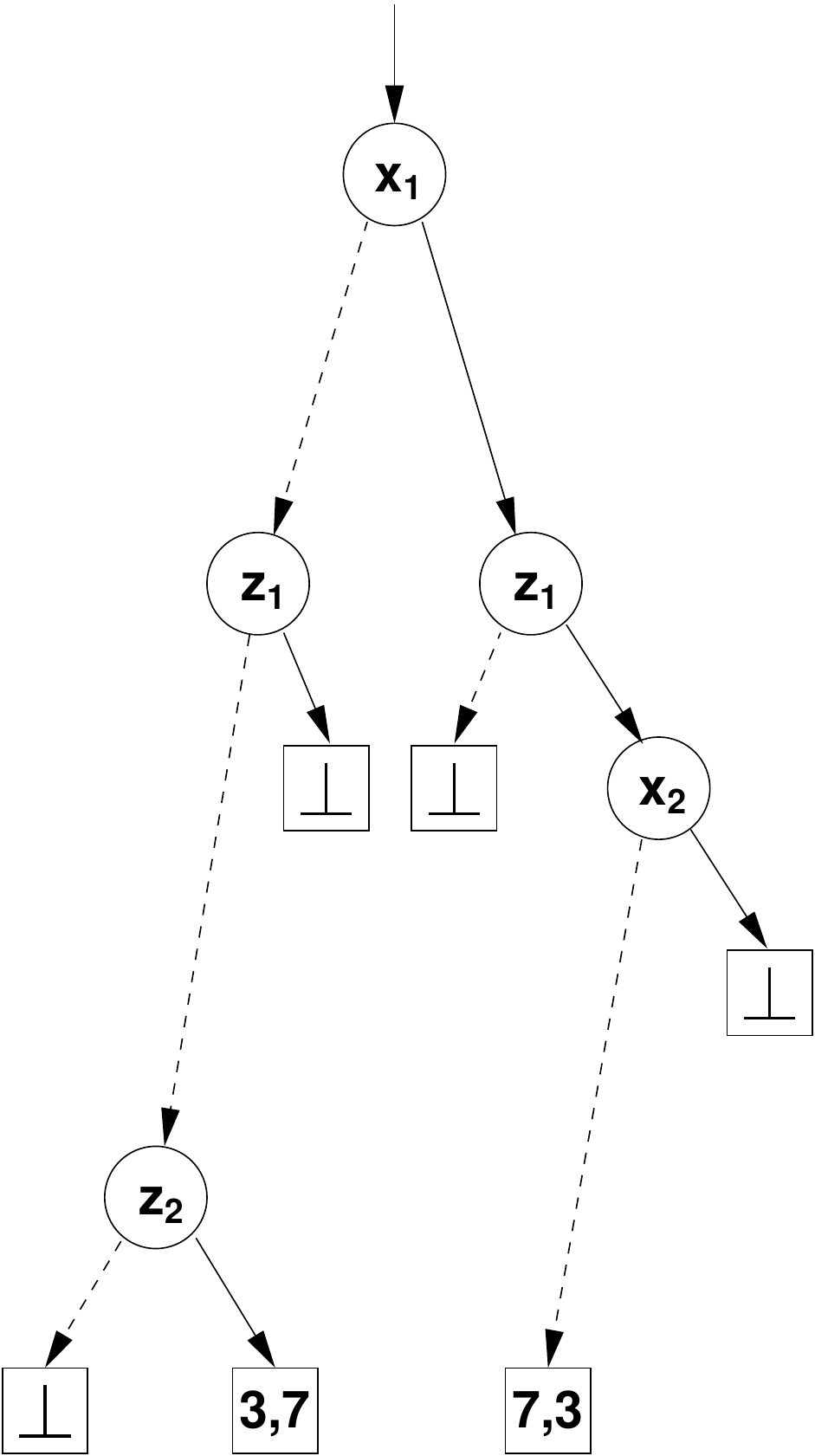}\\
    d) $\mathtt{trimmed} \asgn \trimvariables(\mathtt{res}, y)$
  \end{minipage}
  ~~~~~~~~
  \begin{minipage}{7cm}
    \centering
    \includegraphics[height=7cm]{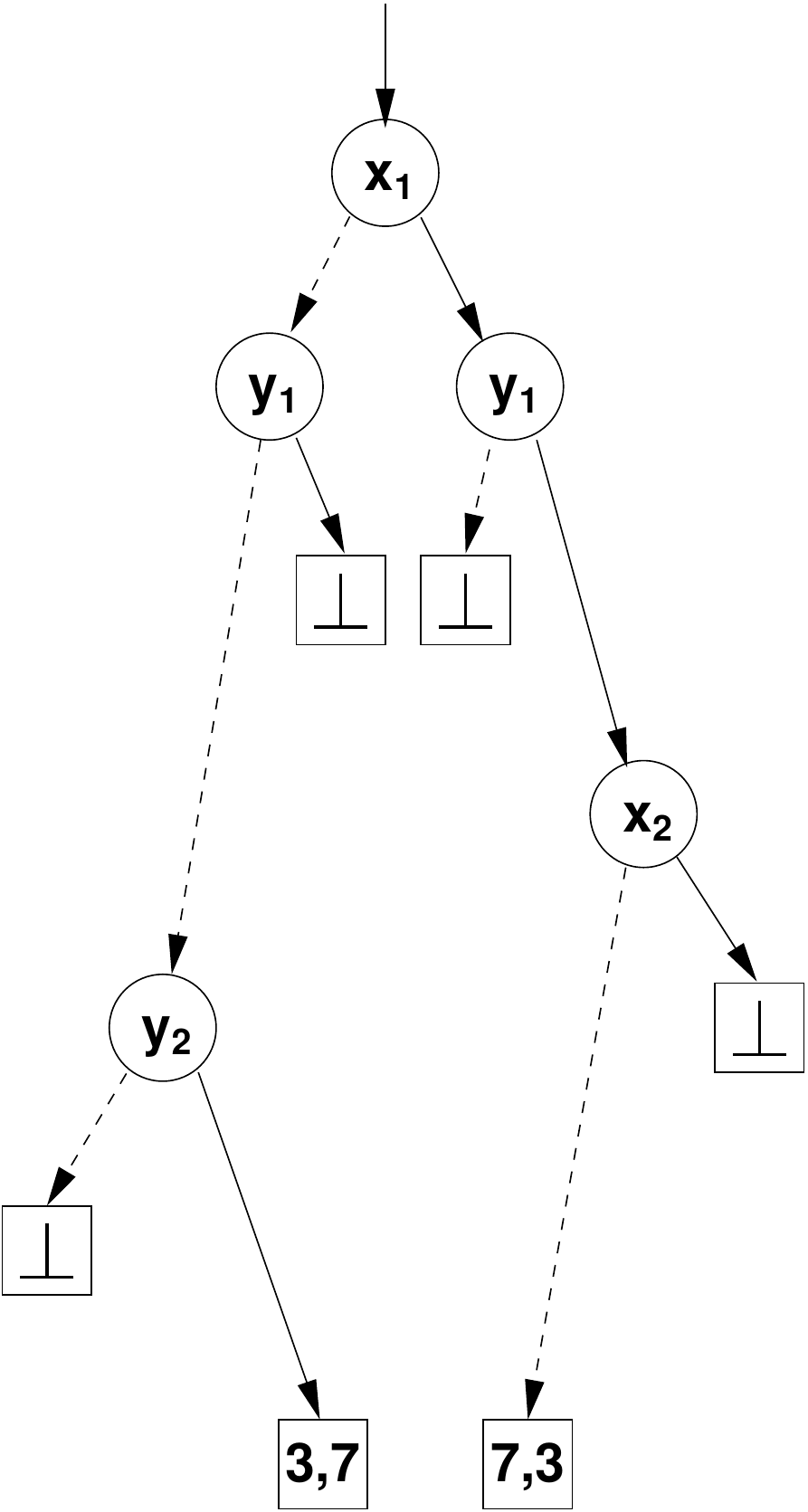}\\
    e) $\renamevariables(\mathtt{trimmed}, z, y)$
  \end{minipage}
  \caption{An example of performing transducer composition of transducer $\transducer$ on itself: $\transducer \circ \transducer$, for one \superstate{} $s_{\transducer}$, such that $\mathtt{0X}(s_{\transducer}) \to 3(\mathtt{1X}), \mathtt{10}(s_{\transducer}) \to 7(\mathtt{01})$. We assume that $s_1 = s_{\transducer}$ and $s_2 = \renamevariables(\renamevariables(s_{\transducer},y,z),x,y )$. Ordering $(a_1, b_1, c_1, \dots, a_n, b_n, c_n)$ is used.}
  \label{fig_transducer_composition_example}
\end{figure}

The algorithm is very similar to the algorithm that performs a transduction step on a finite tree automaton (see Section~\ref{subsec_transduction_step}).
Figure~\ref{fig_transducer_composition_example} gives an example of the operations carried out by the algorithm for one pair of \superstate{}s.

\IncMargin{1em}
\begin{algorithm}[h]

  \SetKwData{Tmp}{tmp}
  \SetKwData{NewStates}{newStates}
  \SetKwData{Superstate}{sp}
  \SetKwFunction{Intersect}{intersect}

  \caption{Transducer composition}
  \KwIn{Input transducers  $\transducer_1 = (Q_1, \rankalph, Q_{f1}, \transfunc_1)$ and $\transducer_2 = (Q_2, \rankalph, Q_{f2}, \transfunc_2)$}
  \KwOut{Transducer $\transducer = (Q, \rankalph, Q_f, \transfunc)$ such that $\transducer = \transducer_2 \circ \transducer_1$}

  \Begin
  {
    $Q \asgn Q_f \asgn \transfunc \asgn \emptyset$\;
    $\NewStates \asgn$ \desc{empty queue}\;

    $\Tmp \asgn \renamevariables(\transfunc_2~(), y, z)$\;
    $\Tmp \asgn \renamevariables(\Tmp, x, y)$\;
    $\Tmp \asgn \apply~\left(\transfunc_1~\left(\right)\right)~\Tmp~(\Intersect~\NewStates)$\;
    $\Tmp \asgn \trimvariables(\Tmp, y)$\;
    $\transfunc~() \asgn \renamevariables(\Tmp, z, y)$\;

    \While{\NewStates \desc{is not empty}}
    {
      $(q_a, q_b) \asgn \NewStates.\Dequeue()$\;
      \If{$(q_a, q_b) \notin Q$}
      {
        $Q \asgn Q \cup \{ (q_a, q_b) \}$\;

        \lIf{$q_a = q_{sink} \vee q_b = q_{sink}$}
        {
          \Continue\;
        }

        \lIf{$q_a \in Q_{f1} \wedge q_b \in Q_{f2}$}
        {
          $Q_f \asgn Q_f \cup \{ (q_a, q_b) \} $\;
        }

        \ForEach{$n \in \natnums$ \desc{such that} $S_n(\transfunc_1) \neq \emptyset \wedge S_n(\transfunc_2) \neq \emptyset$}
        {
          \ForEach{$(q_{11}, \dots, q_{1n}) \in S_n(\transfunc_1)$ \desc{such that} $q_a \in (q_{11}, \dots, q_{1n})$}
          {
            \ForEach{$(q_{21}, \dots, q_{2n}) \in S_n(\transfunc_2)$ \desc{such that} $q_b \in (q_{21}, \dots, q_{2n})$}
            {
              \If{$\forall 1 \le i \le n : (q_{1i}, q_{2i}) \in Q$}
              {
                $\Tmp \asgn \renamevariables(\transfunc_2~(q_{21}, \dots, q_{2n}), y, z)$\;
                $\Tmp \asgn \renamevariables(\Tmp, x, y)$\;
                $\Superstate \asgn \left(q_{11}, \dots, q_{1n}\right)$\;
                $\Tmp \asgn \apply~\left(\transfunc_1~\Superstate\right)~\Tmp~(\Intersect~\NewStates)$\;
                $\Tmp \asgn \trimvariables(\Tmp, y)$\;
                $\transfunc~\left((q_{11}, q_{21}), \dots, (q_{1n}, q_{2n})\right) \asgn \renamevariables(\Tmp, z, y)$\;
              }
            }
          }
        }
      }
    }

    \Return{$\transducer = (Q, \rankalph, Q_f, \transfunc)$}\;
  }

  \label{alg_transducer_composition}
\end{algorithm}
\DecMargin{1em}

\chapter{Implementation} \label{chap_implementation}

This chapter describes design and implementation of a prototype of the library.
It starts with description of the implementation of the type of MTBDDs as defined in Section~\ref{sec_representation}.
This is followed by description of the object-oriented design of the implementation.

\section{MTBDD Package} \label{sec_mtbdd_package}

Since a smart and efficient implementation of MTBDDs is not trivial, it was decided that an existing library should be used instead of implementing an own BDD package.
For this purpose, CUDD~\cite{cudd} (distributed free of charge under the new and simplified BSD licence~\cite{bsdlic}), which is a C~library implementing shared BDDs, ADDs (algebraic decision diagrams) and ZDDs (zero-suppressed decision diagrams), has been chosen.

Using this library, we represent an MTBDD by an ADD~\cite{bahar93}, which is in fact an MTBDD that puts emphasis on performing algebraic operations (such as addition, multiplication, or computation of logarithm) on sets of floating point numbers represented by the diagram. Despite such broad range of operations we use the data structure only for storage and retrieval of data and performing $\apply$ operation. Because CUDD only allows to store floating point numbers into the sink nodes of MTBDDs, we had to deal with the problem to substitute those floating point numbers for sets of states of an automaton (as described in Section~\ref{sec_representation}). We solved this problem by patching the library so that sink nodes would contain pointers to sets stored in another data structure, which serves as a pool of sets of states. In order to make use of MTBDD's space reduction, it must hold that there are never two equal sets of states in the pool. This means that two pointers point to the same set of states if and only if they are equal.

\begin{figure}[ht]
  \begin{center}
    \includegraphics[width=9cm]{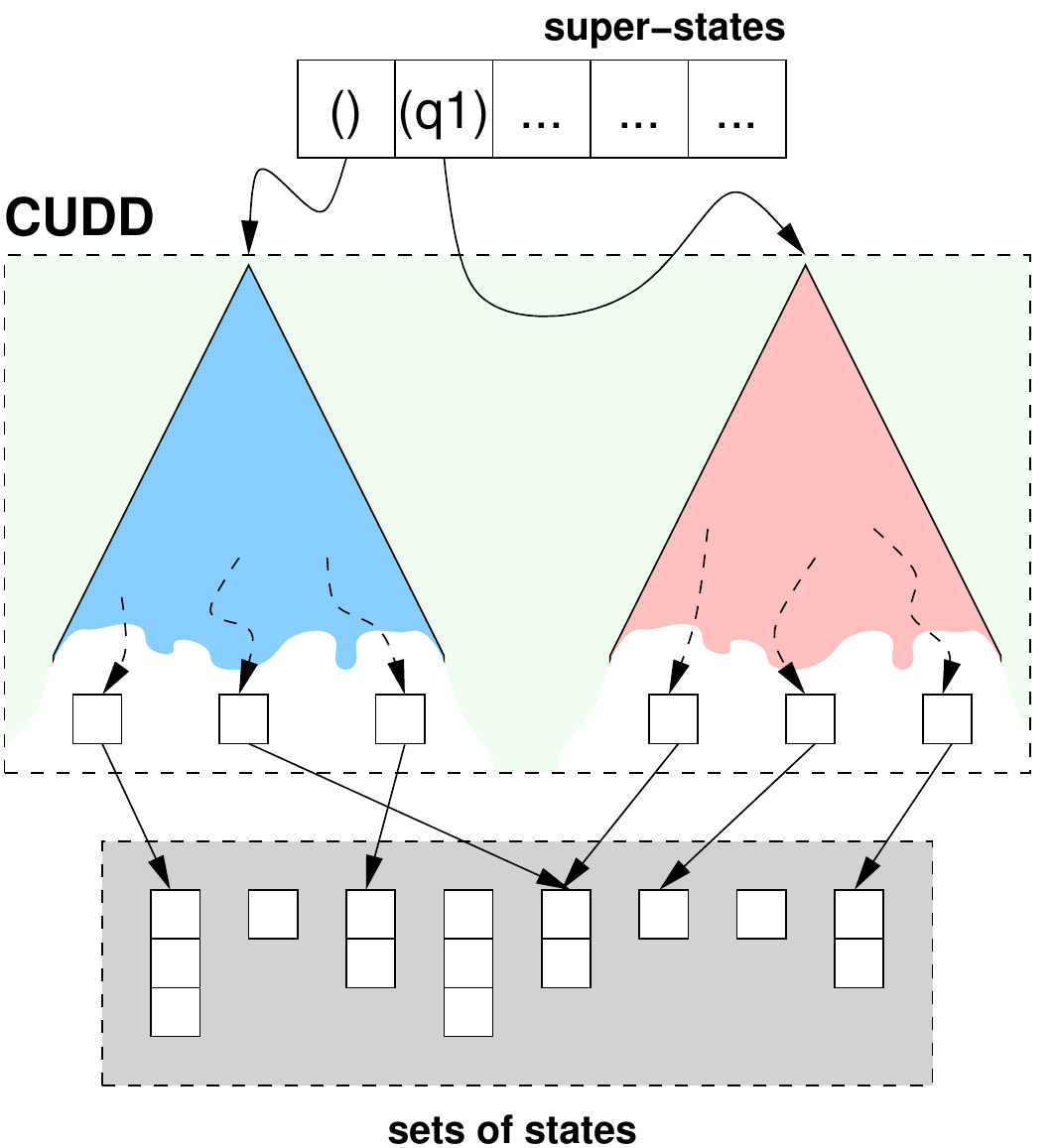}
  \end{center}
  \vspace{-0.4cm}
  \caption{Wrapper over CUDD-provided MTBDDs.}
  \label{fig_mtbdd_wrapper}
\end{figure}

As shared variation of MTBDD is used, a way to distinguish among individual MTBDDs in such structure needs to be defined. We use another data structure that provides mapping from the set of \superstate{}s of the transition function to roots of the shared MTBDD. The resulting wrapper over MTBDDs provided by CUDD is shown in Figure~\ref{fig_mtbdd_wrapper}.

Due to the fact that many algorithms in Chapter~\ref{chap_design} need to alter some data outside of MTBDD during an $\apply$ operation, we also patched CUDD and support the following $\apply$ operation: $\apply$\texttt{(lhs, rhs, op)}, where \texttt{lhs} and \texttt{rhs} are sets of states of the left and right MTBDD respectively, and \texttt{op} is a function object: an object that can be called like an ordinary function. Using \texttt{op} to pass pointers to data structures in main subroutines, we can avoid using global variables and thus making the code re-entrant and less prone to programming errors.

\section{Object-Oriented Design} \label{sec_oo_design}

C++ has been chosen as the implementation language because of its efficiency, good support, means for modular design and an extensive standard library.
We employ C++'s support of object-oriented programming paradigm to create a generic and modular design, which is further described in this section.

\begin{figure}[b]
  \begin{center}
    \includegraphics[width=9cm]{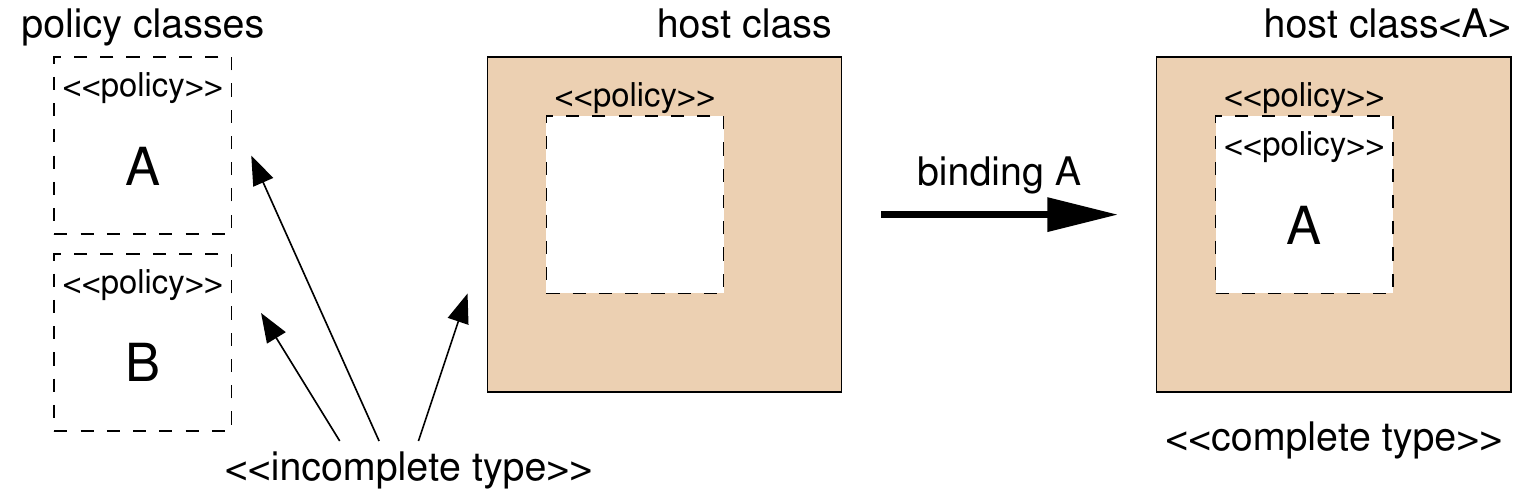}
  \end{center}
  \caption{Binding of policy classes to host class.}
  \label{fig_policy_classes}
\end{figure}

In order to provide both modularity and good performance, policy-based design~\cite{alexandrescu_modern_cpp_design} is exploited in the object-oriented design of the library.
This approach uses \emph{policy classes}, which are classes that are not supposed to be instantiated (which can be enforced by making their constructor protected) or to only provide interface, but rather to provide a certain functionality when inherited by some class called the \emph{host class}.
Each policy class implements a particular interface called a \emph{policy}.
The host class is a class template, i.e.\ an incomplete class that does not name a type by itself but needs to be have its template arguments bound in order to do so, as shown in Figure~\ref{fig_policy_classes}.
In addition to standard template arguments the host class also defines its policies.
Using multiple inheritance, several orthogonal policy classes may be inherited by host class.
Due to the fact that policy classes of a host class are resolved statically during compile time, the compiler may perform certain optimizations, such as inlining of code, which is an advantage over using e.g.\ virtual methods.

\subsection{MTBDD Wrapper}

\cuddfacade{} is a class that was designed according to the fa\c{c}ade design pattern~\cite{gof_design_patterns}.
It is used as the access point to CUDD library that provides very extensive and confusing API.
\cuddfacade{} provides a clean and type-safe interface with only those operations which are necessary for the implementation of the library, while hiding the others.
The class is compiled with all CUDD's object files into a single static library which is then further used by the tree automata library.

\begin{figure}[ht]
  \begin{center}
    \includegraphics[width=14cm]{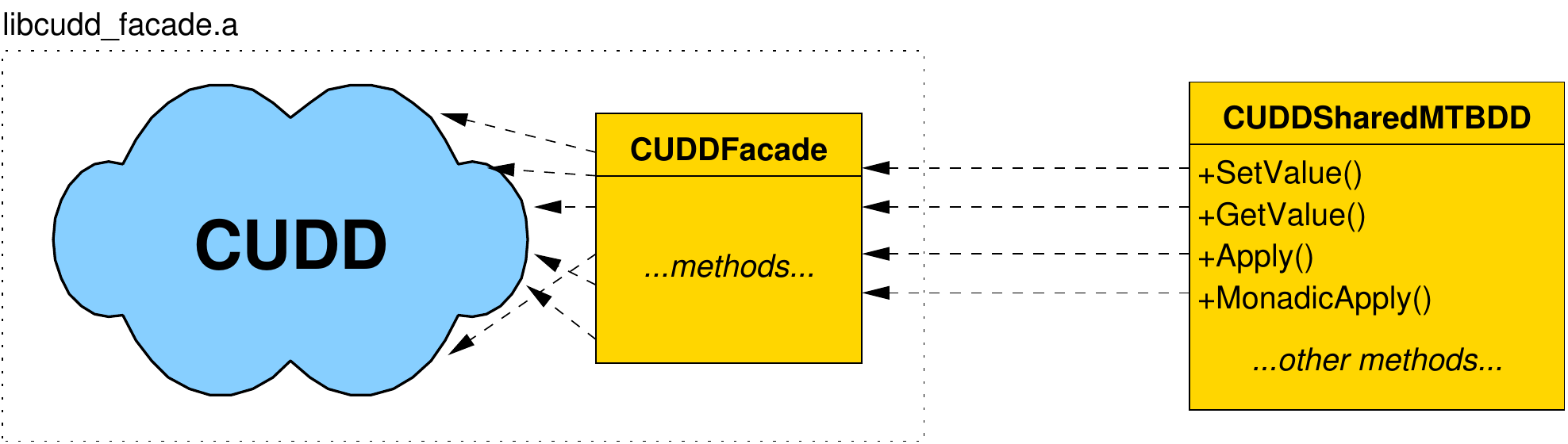}
  \end{center}
  \caption{Interface to CUDD package.}
  \label{fig_cudd_facade}
\end{figure}

\cuddsharedmtbdd{} is an object-oriented representation of a shared MTBDD as described in Section~\ref{sec_mtbdd_package}.
The class uses \cuddfacade{} to access CUDD as depicted in Figure~\ref{fig_cudd_facade}.
\cuddsharedmtbdd{} is a class template with the following template parameters:
\begin{description}

  \item[\RootType]  Defines the type for accessing MTBDDs for individual \superstate{}s. This may be an arbitrary type, the prototype implementation uses \unsigned{}.

  \item[\LeafType]  The type for the sink node of the MTBDD. This may again be an arbitrary type, the prototype implementation uses a set of states. However, deterministic automata may store the target state directly in the sink node thus making the access time shorter.

  \item[\VariableAssignmentType]  This template parameter determines the data type for representation of assignment to Boolean variables of the MTBDD, i.e.\ the data type for the symbol. To fully utilise the potential of MTBDDs, our representation of assignment to Boolean variables of MTBDD can assign 3 different values to a variable: \texttt{true} (\texttt{1}), \texttt{false} (\texttt{0}) and \texttt{don't care} (\texttt{X}). By using the \texttt{don't care} value, we may work with whole sets of transitions over various symbols as with a single transition (for example to work with four transitions over symbols encoded as \texttt{100}, \texttt{101}, \texttt{110}, and \texttt{111} at once, it is enough to use only one encoding \texttt{1XX}).

  \item[\RootAllocator]  The implementation of this policy defines the mapping of roots of \RootType{} to CUDD-related pointers to root nodes of corresponding MTBDDs.

  \item[\LeafAllocator]  This policy determines the exact implementation of the mapping between the \unsigned{} sink nodes stored in the patched CUDD data structures and those of type \LeafType. 

\end{description}
These template parameters allow high configurability of \cuddsharedmtbdd{}, for instance to be used for deterministic finite tree automata or for finite (word) automata transition functions.

\cuddsharedmtbdd{} also defines \applyfunctor{} which is an abstract class of a function object with a single pure virtual method, which is overloaded operator \texttt{()}:
\begin{center}
  \texttt{virtual LeafType operator()(const LeafType\& lhs, const LeafType\& rhs) = 0;}
\end{center}
Classes that inherit this abstract function object need to implement the only method by defining a function for $\apply$ operation.
The $\apply$ operation takes root nodes of two MTBDDs and an object of a class that implements the \applyfunctor{} interface:
\begin{center}
  \begin{minipage}{12cm}
    \verb#RootType Apply(const RootType& lhs, const RootType& rhs,# \\
    \verb#               AbstractApplyFunctorType* op);#
  \end{minipage}
\end{center}

\subsection{Transition Function}

The \mtbddtransitionfunction{} class represents transition functions of several automata using single MTBDD.
This is because CUDD only allows executing $\apply$ operation on MTBDD roots from the same shared MTBDD.
When an automaton is being created, it \emph{registers} to some \mtbddtransitionfunction{} and inserts all its transitions into this object.

A challenging issue that needs to be faced is the choice of data structure for storage of \superstate{}s, i.e.\ mapping of \superstate{}s to their respective MTBDDs.
Storage of nullary and unary \superstate{}s is obvious.
Since there is only one nullary \superstate{} for each automaton, these \superstate{}s are stored in a single array indexed by automaton number.
Unary \superstate{}s of each automaton are stored in a separate array indexed by the only state's number.
The prototype implementation also deals with storage of binary \superstate{}s by using a 2-dimensional matrix indexed by the two states of the \superstate{}.
Due to the fact that space requirements for $n$-dimensional matrix grow exponentially and the utilisation drops with almost the same speed for real-world problems, more sophisticated data structures need to be found.
Our prototype implementation uses for storage of \superstate{}s with arity greater than 2 a hash table with an arbitrarily long vector of states as the key and pointer to MTBDD as the value.

\subsection{Tree Automaton}

The \treeautomatonclass{} class represents a finite tree automaton with a high-level interface.
The interface allows the use of human-readable names of states and symbols and provides mapping to their inner representation.
\treeautomatonclass{} enables adding a state, transition or marking a state as final.
It also supports importing and exporting a finite tree automaton to or from a file.

\begin{figure}[ht]
  \label{fig_timbuk_builder}
  \begin{center}
    \includegraphics[width=10cm]{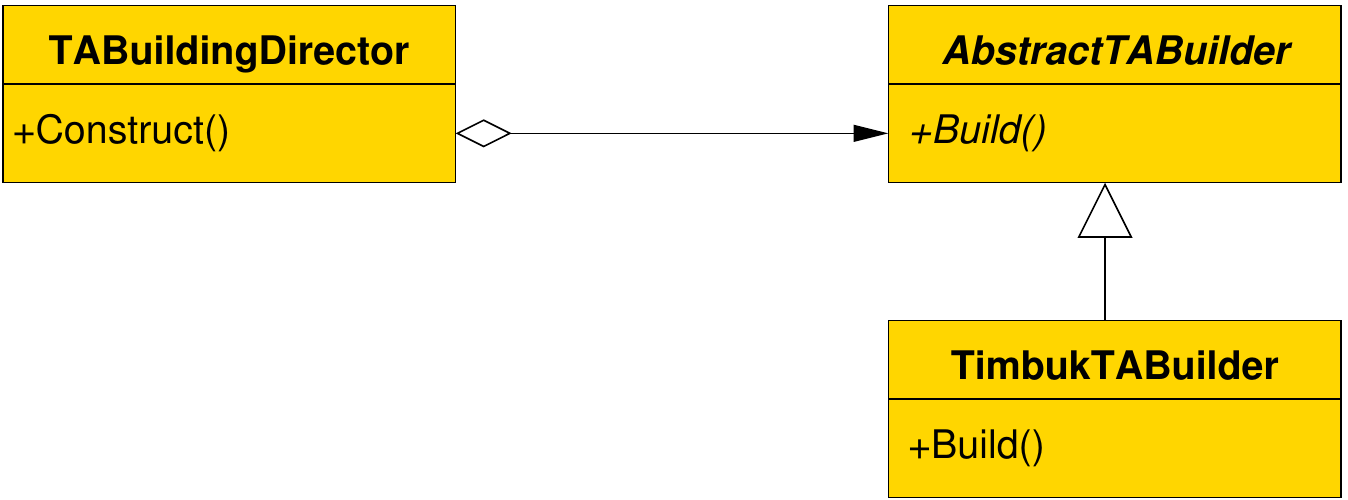}
  \end{center}
  \caption{\tabuildingdirector{} structure.}
\end{figure}

\subsection{Automaton Import}

In order to support direct user interface to the library, the library supports reading a finite tree automaton from a file.
The reading interface is designed according to the builder design pattern~\cite{gof_design_patterns}.
Building a new finite tree automaton from a file is done by creating an object of class \tabuildingdirector{} and assigning an instance of class implementing the \abstracttabuilder{} interface to it.
Then calling the \texttt{Construct()} method of \tabuildingdirector{} (which further calls the \texttt{Build()} method of \abstracttabuilder{}) with a data stream that has a format recognized by the concrete builder constructs proper tree automaton.
For testing purposes one concrete builder was implemented: \linebreak \timbuktabuilder{} which accepts input data stream with description of automata in Timbuk-like format (see Section~\ref{sec_timbuk}).

\subsection{Automaton Export}

This section deals with exporting description of a tree automaton into a human-readable format.
In order to do so, the most difficult task is to extract transitions from symbolic representation into explicit.
This operation needs to know the structure of the MTBDD used for representation of the transition function.
This is achieved by using \emph{test symbols} $(\seq{x})$ that start with value \texttt{XXX\dots{}X} and for each Boolean variable $\seq{x}$ attempt to bind its value to both \texttt{0} and \texttt{1}.
If it holds that the resulting MTBDDs for both bindings are the same, then the value of the variable is not important, it is left in \texttt{X} (\texttt{don't care}) and the procedure carries on to the following variables.
In case the MTBDDs are not the same, the procedure splits into two branches and continues for both bindings.
This continues until all variables have been either bound or left in \texttt{X}.

A simple script that converts a file in the output format into a graphical representation of the automaton for \texttt{dot} tool~\cite{graphviz} has been created.

\begin{figure}[ht]
  \begin{center}
    \includegraphics[width=8cm]{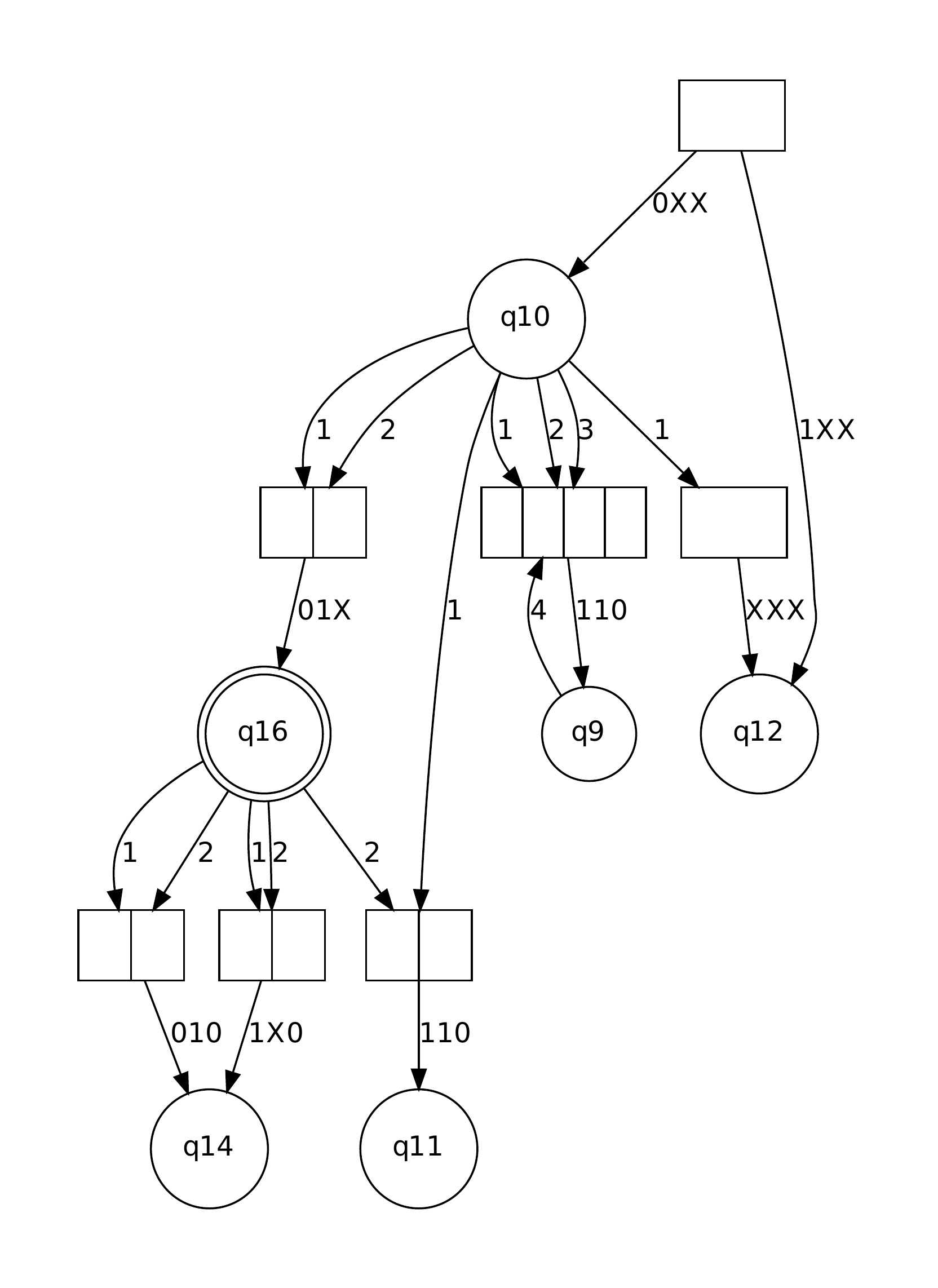}
  \end{center}
  \vspace{-1cm}
  \caption{Example of graphical representation of a tree automaton.}
  \label{fig_dot_example}
\end{figure}

Unlike finite word automata, transition function and run of a finite tree automaton cannot be expressed simply as a labelled graph and \emph{walk} (i.e.\ sequence of vertices and edges such that each vertex or edge may occur several times in the sequence) in the graph.
Up to our knowledge there is no widely accepted standard graphical representation of finite tree automata, we therefore attempted to choose a simple and understandable representation resembling finite (word) automata as much as possible.
As in finite (word) automata, states are represented by circles, final states by double circles.
We introduce new type of vertices representing \superstate{}s which make the graph bipartite: an edge is either from a state to a \superstate{} or from a \superstate{} to a state (this may resemble a Petri net).
Representation of \superstate{}s is by rectangles with \emph{boxes}.
The number of boxes determines the arity of the \superstate{}.
The labelling of edges from states to \superstate{}s denotes the position of the state in the \superstate{} vector; the labelling of edges from \superstate{}s to states denotes the symbol over which the transition is to be made.
An example of graphical representation of a finite tree automaton is in Figure~\ref{fig_dot_example}.

\subsection{Operations}

Operations on finite tree automata with transition function represented using an MTBDD are provided by \mtbddoperation{} class.
The prototype implementation implements the following operations: language union, language intersection, reduction of an automaton according to some equivalence class, and computation of downward simulation.
All these operations are performed only using the interface provided by \cuddsharedmtbdd{}.


\chapter{Evaluation} \label{chap_evaluation}

This chapter provides an evaluation of the prototype implementation (called libSFTA) of the library described in the previous chapters.
The tests were run on a laptop with a dual-core Intel Core 2 Duo CPU at 1.80\,GHz and 2\,GiB of available memory with Debian Squeeze GNU/Linux installed.
We measured performance of the following three finite tree automata operations: language union, language intersection, and automaton reduction according to the downward simulation relation.

\section{Language Union} \label{sec_eval_union}

\enlargethispage{5mm}

The performance of the language union operation on two input finite tree automata was measured and compared to Timbuk, a tree automata library (described in Section~\ref{sec_timbuk}) that performs operations on nondeterministic finite tree automata using an \emph{explicit} representation of the transition function (note that the implemented library uses a \emph{symbolic} representation).
We made this choice because MONA immediately determinises input automata so the comparison would not be fair.

\begin{figure}[h]
  \hspace{-1.3cm}
  \begin{minipage}{8.5cm}
    \centering
    \includegraphics[width=8.5cm]{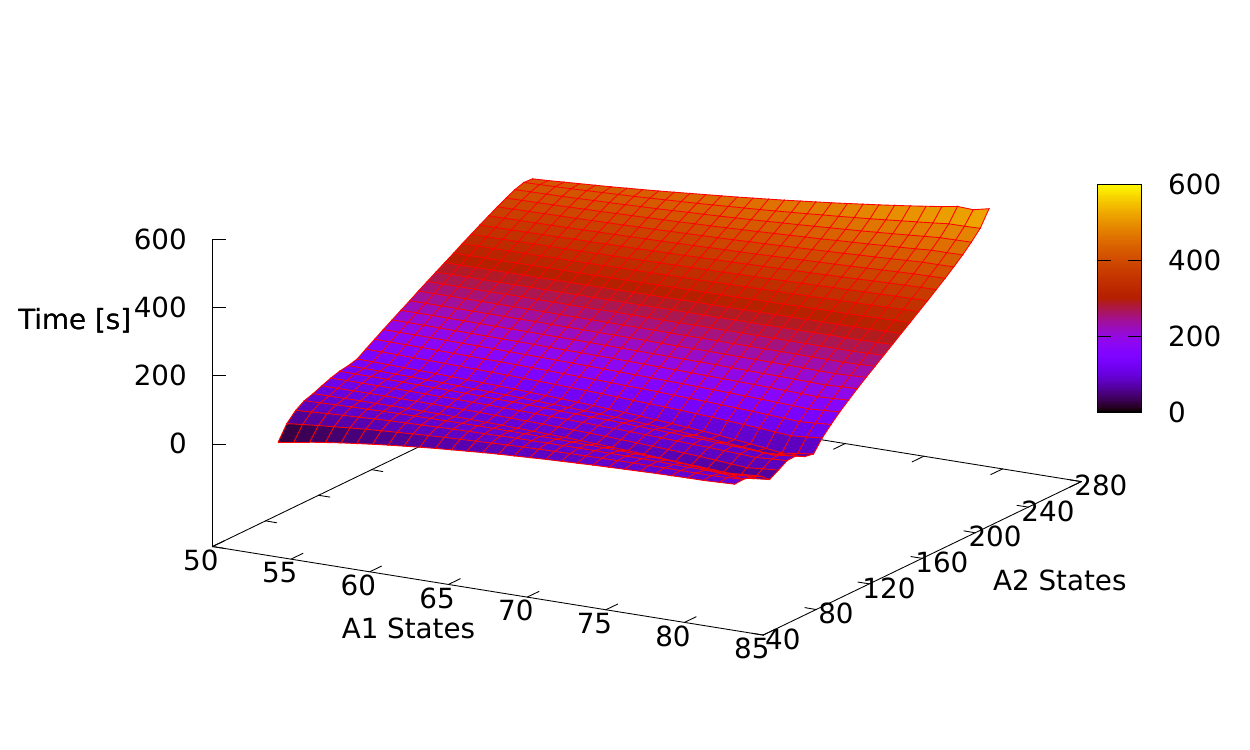}
    a) Timbuk
  \end{minipage}
  \hfill
  \begin{minipage}{8.5cm}
    \centering
    \includegraphics[width=8.5cm]{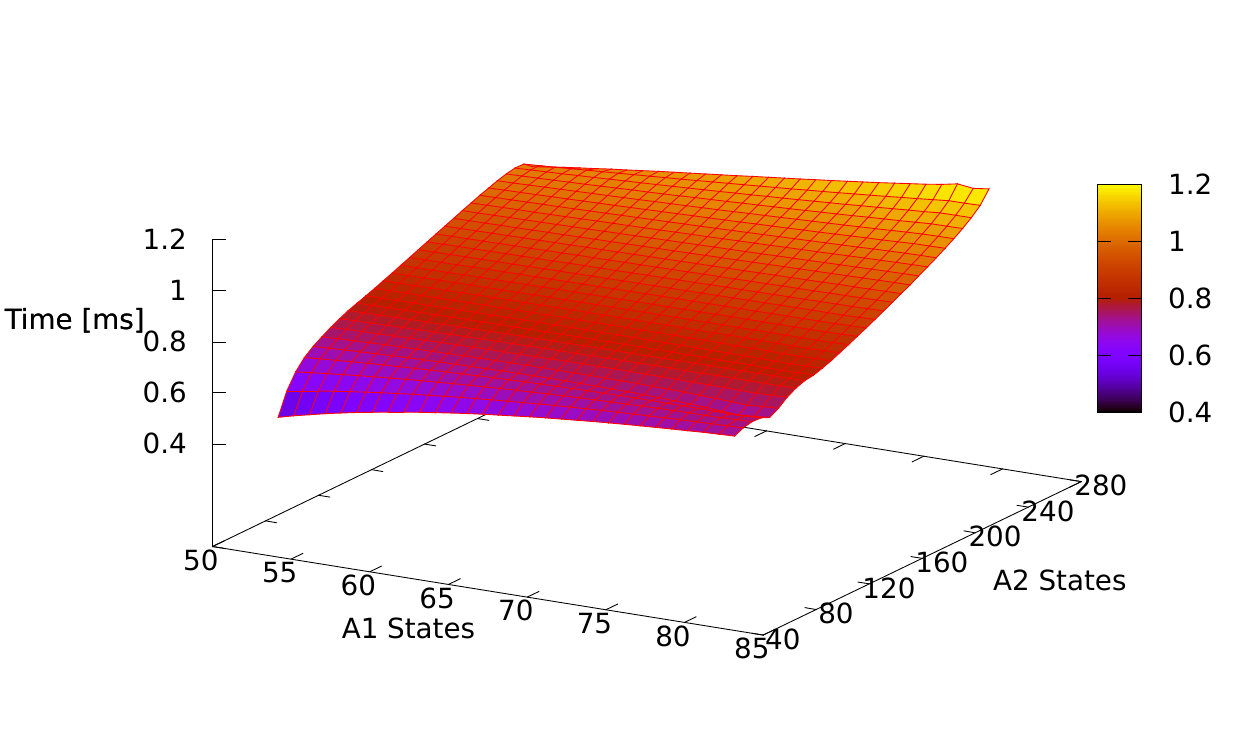}
    b) libSFTA
  \end{minipage}
  \caption{Performance comparison of language union for various state set size.}
  \label{fig_graph_union_states}
\end{figure}

We performed the tests on binary tree automata over an alphabet with 130 symbols and various size of the state set obtained from tree model checking of real systems.
It should be pointed out that the execution time for both libSFTA and Timbuk does not include the time necessary to load the automaton from a file.
This should give more valid results, since building an MTBDD for a transition function is not a trivial operation (note that the comparison is fair from a practical point of view since within a verification framework automata are built internally, not loaded from a file).
The results are given in Table~\ref{tab_union_states} (in a graphical form in Figure~\ref{fig_graph_union_states}).
It can be seen that libSFTA significantly outperforms Timbuk in all cases.

\vspace{5mm}
\begin{table}[h]
  \begin{center}
    \begin{tabular}{|c|c||r|r|}
      \hline
      \multicolumn{2}{|c||}{Automata}   &   Timbuk  &   libSFTA  \\
      \hline
      \hline
      A0053 & A0054 &   1.982\,s  &  0.0005\,s  \\
      \hline
      A0080 & A0082 &  37.645\,s  &  0.0007\,s  \\
      \hline
      A0080 & A0111 &  37.645\,s  &  0.0008\,s  \\
      \hline
      A0053 & A0246 & 414.104\,s  &  0.0010\,s  \\
      \hline
      A0080 & A0246 & 533.678\,s  &  0.0012\,s  \\
      \hline
      A0082 & A0246 & 542.069\,s  &  0.0012\,s  \\
      \hline
    \end{tabular}
  \end{center}
  \caption{Language union performance results.}
  \label{tab_union_states}
\end{table}

\section{Language Intersection} \label{sec_eval_intersection}

\begin{figure}[h]
  \hspace{-1.3cm}
  \begin{minipage}{8.5cm}
    \centering
    \includegraphics[width=8.5cm]{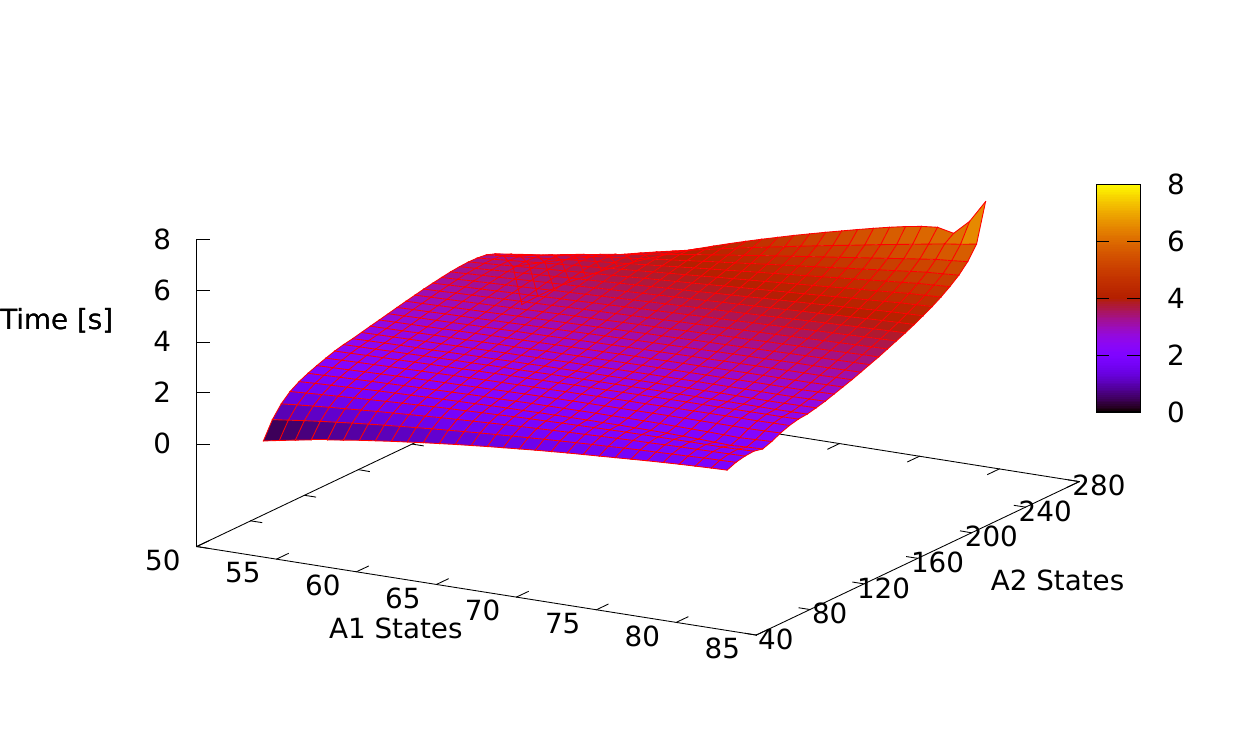}
    a) Timbuk
  \end{minipage}
  \hfill
  \begin{minipage}{8.5cm}
    \centering
    \includegraphics[width=8.5cm]{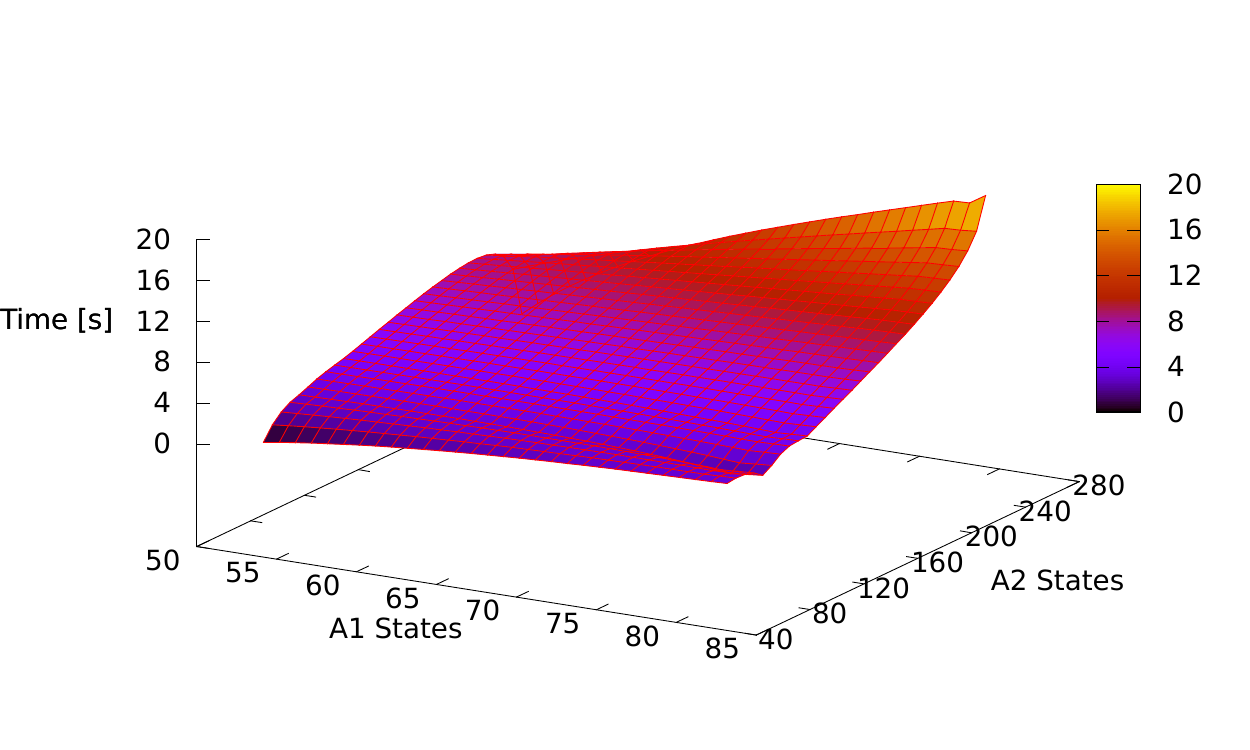}
    b) libSFTA
  \end{minipage}
  \caption{Performance comparison of language intersection for various state set size.}
    \vspace{-0.4cm}
  \label{fig_graph_intersection_states}
\end{figure}

The experiments with the language intersection operation on two input finite tree automata were also compared to Timbuk.
These experiments were conducted with the same set of test automata and the same testing conditions as the language union operation (see Section~\ref{sec_eval_union}).
The results are given in Table~\ref{tab_intersection_states} with the graph in Figure~\ref{fig_graph_intersection_states}.
The results show that for larger state sets Timbuk computes the finite tree automaton for language intersection slightly faster than libSFTA.

\begin{table}[h]
  \begin{center}
    \begin{tabular}{|c|c||r|r|}
      \hline
      \multicolumn{2}{|c||}{Automata}   &   Timbuk  &   libSFTA  \\
      \hline
      \hline
      A0053 & A0054 &  0.076\,s  &  0.057\,s  \\
      \hline
      A0053 & A0246 &  0.609\,s  &  0.617\,s  \\
      \hline
      A0080 & A0082 &  1.862\,s  &  1.675\,s  \\
      \hline
      A0080 & A0111 &  2.483\,s  &  3.765\,s  \\
      \hline
      A0080 & A0246 &  6.062\,s  & 18.320\,s  \\
      \hline
      A0082 & A0246 &  7.503\,s  & 19.355\,s  \\
      \hline
    \end{tabular}
  \end{center}
  \vspace{-0.4cm}
  \caption{Language intersection performance results.}
  \vspace{-0.4cm}
  \label{tab_intersection_states}
\end{table}

\section{Simulation Reduction} \label{sec_eval_simulation}

The performance of reduction of a finite tree automaton according to downward simulation relation was measured in the next series of tests.
The results were compared to SA~\cite{sa}, an OCaml tool implementing computation of downward simulation over labelled transition systems and finite tree automata.

The execution time for libSFTA comprises computing downward simulation over the input tree automaton, computing symmetric closure of the relation and reducing the automaton according to equivalence given by the simulation and its symmetric closure.
As SA cannot perform reduction of the automaton, the execution time of SA consists of the time it takes to load the automaton from a file (which should be negligible according to the authors) and compute the downward simulation.
Two different test cases were measured.

\begin{enumerate}

  \item  The first test case shows how the performance depends on the size of the \emph{state set} of the input finite tree automaton for a fixed small alphabet.
         The results are given in Table~\ref{tab_state_sets} and Figure~\ref{fig_graph_large_state_sets}.
         We can see that the performance of libSFTA is worse when compared to SA.
         The reason for this is that SA uses a more sophisticated algorithm for computation of simulation.
         In the future version of the library, we wish to focus on optimising the algorithm we use in order to be able to compete with the solution used in SA even for smaller alphabets.

  \begin{table}[h]
    \begin{center}
      \begin{tabular}{|c|c|r||r|r|}
        \hline
        Automaton & States & \multicolumn{1}{|c||}{Transitions} & \multicolumn{1}{|c|}{SA} & \multicolumn{1}{|c|}{libSFTA} \\
        \hline
        \hline
          A0053   & 53  & 159 &   0.04\,s    &       24.6\,s \\
        \hline
          A0054   & 54  & 241 &   0.04\,s    &       29.3\,s \\
        \hline
          A0063   & 63  & 571 &  0.10\,s    &       55.2\,s \\
        \hline
          A0070   & 70  & 622 &  0.07\,s    &       71.5\,s \\
        \hline
          A0080   & 80  & 672 &  0.11\,s    &      274.4\,s \\
        \hline
          A0082   & 82  & 713 &  0.09\,s    &      331.5\,s \\
        \hline
          A0089   & 89  & 1006 &  0.11\,s    &      226.1\,s \\
        \hline
      \end{tabular}
    \end{center}
    \vspace{-0.4cm}
    \caption{Experimental results of simulation reduction for various state set size.}
    \vspace{-0.4cm}
    \label{tab_state_sets}
  \end{table}

  \begin{figure}[h]
    \begin{center}
      \includegraphics[width=15cm]{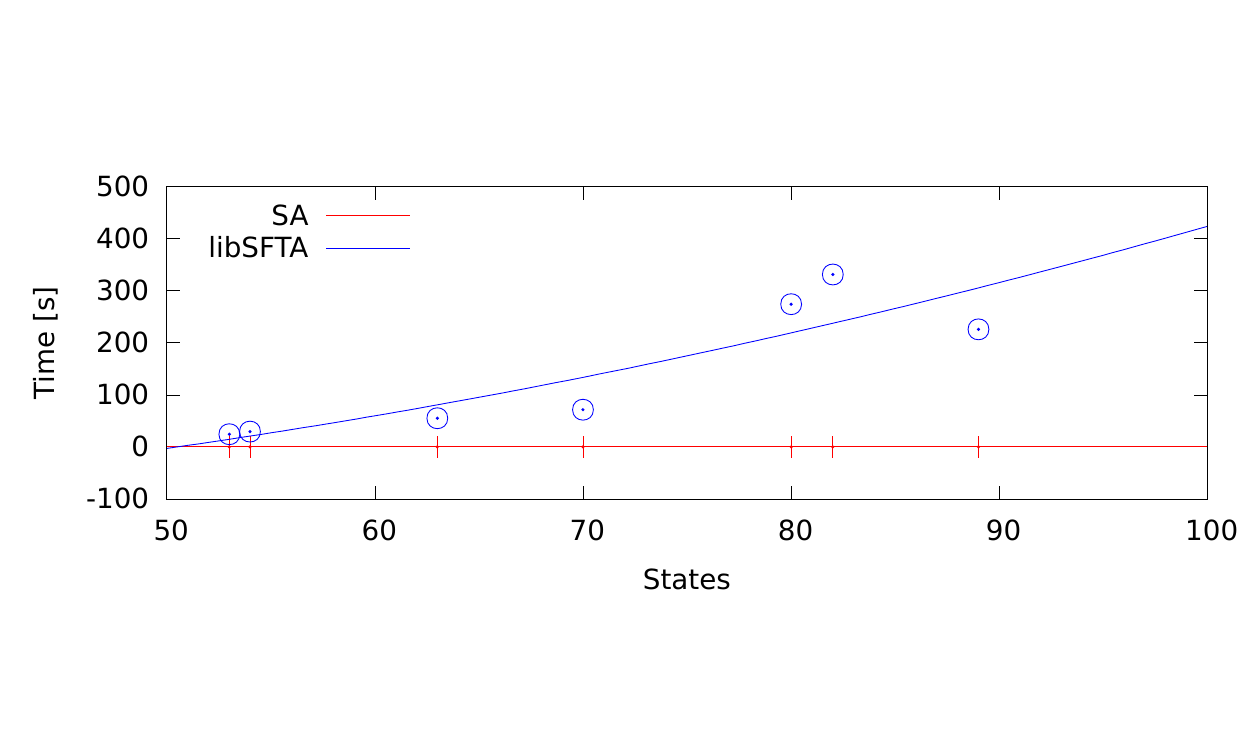}
    \end{center}
    \vspace{-2.7cm}
    \caption{Performance comparison of simulation reduction for various state set size.}
    \label{fig_graph_large_state_sets}
  \end{figure}

  \item  The second test case shows how the performance of libSFTA and SA relates to the size of the alphabet of the input finite tree automaton for a fixed set of states.
         We created a set of simple tree automata that perform transitions over symbols from alphabet of various size. 
         The number of transitions equals the number of symbols in the alphabet (we use one transition for every symbol).
         The results of the experiments with the size of the alphabet are in Table~\ref{tab_alphabets} and Figure~\ref{fig_graph_large_alphabets}.
         It is clear that the use of symbolic representation makes the performance of libSFTA far superior to the performance of SA.

  \begin{table}[h]
    \begin{center}
      \begin{tabular}{|r||r|r|}
        \hline
        Symbols      & \multicolumn{1}{|c|}{SA} & \multicolumn{1}{|c|}{libSFTA} \\
        \hline
        \hline
          1337       &           0.06\,s    &     0.0033\,s \\
        \hline
          3525       &           0.14\,s    &     0.0051\,s \\
        \hline
          7067       &           0.26\,s    &     0.0071\,s \\
        \hline
         15136       &           0.69\,s    &     0.0054\,s \\
        \hline
         31235       &           2.09\,s    &     0.0031\,s \\
        \hline
         65503       &           8.86\,s    &     0.0040\,s \\
        \hline
        130023       &          48.40\,s    &     0.0045\,s \\
        \hline
      \end{tabular}
    \end{center}
    \vspace{-0.4cm}
    \caption{Experimental results of simulation reduction for various alphabet size.}
    \vspace{-0.4cm}
    \label{tab_alphabets}
  \end{table}

  \begin{figure}[h]
    \begin{center}
      \includegraphics[width=15cm]{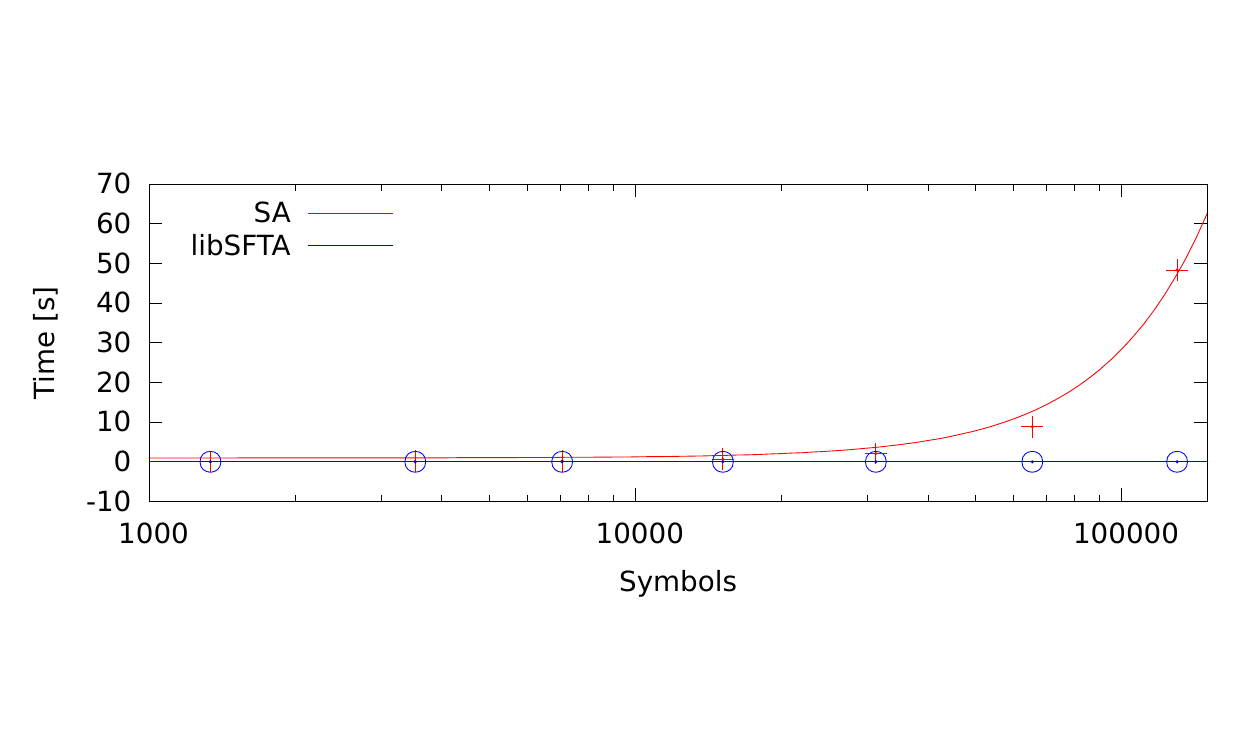}
    \end{center}
    \vspace{-2.7cm}
    \caption{Performance comparison of simulation reduction for various alphabet size.}
    \label{fig_graph_large_alphabets}
  \end{figure}

\end{enumerate}

\section{Discussion} \label{sec_discussion}

The experiments described in this chapter showed that libSFTA has a good potential to become an interesting tree automata library, especially for applications that need large alphabets and can exploit symbolic representation well.
The use of nondeterminism also considerably accelerates the computation of the automaton for language union and can keep automata small and clean.


\chapter{Conclusion} \label{chap_conclusion}

The aim of this Master's Thesis was to design and implement an efficient and flexible finite tree automata library for the use in symbolic formal verification, namely to be applicable for tree model checking techniques, such as regular tree model checking and abstract regular tree model checking.

The theoretical background of tree automata has been studied as well as formal verification methods for systems represented by tree automata.
Existing packages that implement tree automata have been surveyed and their advantages and disadvantages summarised.
An analysis of the aforementioned verification methods have yielded a list of necessary requirements for the library.

A representation of nondeterministic finite tree automata with symbolically represented transition functions has been proposed.
The representation is based on MTBDDs provided by an external package (which may be changed though by writing a simple wrapper with given interface for another library).
Algorithms that carry out standard as well as some verification-specific operations on tree automata using this representation have also been developed and described in this text.

An object-oriented modular design of the library based on design patterns and policies has been created.
A prototype implementation has been programmed, evaluated on testing data, and compared to other tools that provide the same functionality.
The results of experiments show that the concepts we employed are viable and that the library can complement currently available tree automata libraries, especially when used for applications with finite tree automata that make use of large alphabets and nondeterminism.

Future work will focus on redesigning the library according to our feedback from the implementation of the prototype and data collected from code profiling.
Further, we wish to create a library that supports both explicitly and symbolically represented finite tree automata.
We also plan to optimise the algorithms that are used by the library in order to give good performance even for small alphabets and large state sets.
Another direction of work then includes implementing a support of further algorithms for simulation reduction (upward and combined simulation based relations) and antichain-based inclusion checking (combination of antichains and simulations), including further research on still more advanced reduction and inclusion checking techniques.
A next step should then be to test the library as a part of various verification tools (for instance as a base of abstract regular tree model checking tools or with decision procedures for various logics, such as WS2S or separation logic).


\ifczech
  \bibliographystyle{czechiso}
\else 
  \bibliographystyle{unsrt}
\fi
  \begin{flushleft}
  \bibliography{literature} 
  \end{flushleft}
  \appendix
  
  \chapter{Storage Medium} \label{storage_medium}

A storage medium (DVD) containing an electronic version of the technical report
and source code of the prototype implementation including patched CUDD package
is enclosed to this thesis.
\end{document}